\documentclass[12pt,nofootinbib,superscriptaddress]{revtex4}

\usepackage{amssymb,color,paralist,amsmath,epsfig}
\usepackage{graphicx}
\usepackage{amsfonts}
\usepackage{nicefrac,esint}
\usepackage{verbatim}
\usepackage{float}
\usepackage{verbatim}
\allowdisplaybreaks
\usepackage{relsize}
 
\usepackage[colorlinks,citecolor=blue,linktoc=all,linkcolor=cyan]{hyperref}
\usepackage[section]{placeins}

\newcommand{\sla}[1]{{\not\! #1}} 

\newcommand{\beq}{\begin{equation}}
\newcommand{\eeq}{\end{equation}}

\newcommand{\ber}{\begin{eqnarray}} 
\newcommand{\eer}{\end{eqnarray}}

\usepackage{color}

\usepackage[normalem]{ulem}

\renewcommand\sout{\bgroup \color[rgb]{0,0.00,1.} \ULdepth=-.5ex \ULset}

\begin{document}

\title{Leading order corrections to the Bethe-Heitler process in the $\gamma p\rightarrow l^+l^-p$ reaction}
\author{Matthias Heller}
\affiliation{Institut f\"ur Kernphysik and $\text{PRISMA}^+$ Cluster of Excellence, Johannes Gutenberg Universit\"at, Mainz, Germany}
\author{Oleksandr Tomalak}
\affiliation{Institut f\"ur Kernphysik and $\text{PRISMA}^+$ Cluster of Excellence, Johannes Gutenberg Universit\"at, Mainz, Germany}
\affiliation{Department of Physics and Astronomy, University of Kentucky, Lexington, KY 40506, USA}
\affiliation{Fermilab, Batavia, IL 60510, USA}
\author{Marc Vanderhaeghen}
\affiliation{Institut f\"ur Kernphysik and $\text{PRISMA}^+$ Cluster of Excellence, Johannes Gutenberg Universit\"at, Mainz, Germany}
\author{Shihao Wu}
\affiliation{Memorial University of Newfoundland}

\date{\today}

\begin{abstract}
The ratio of di-lepton production cross sections on a proton, using the $\gamma p\rightarrow l^+ l^- p$ process, above and below di-muon production threshold allows to extract the effective lepton-proton interaction, which is required to be identical for electrons and muons if lepton universality is exact. To test for a scenario of broken universality at the percent level, of the size which could explain the different proton charge radii extracted from electron scattering and from muonic hydrogen spectroscopy, 
we evaluate all one-loop QED corrections to this process, including the full lepton mass dependencies. We furthermore show that two-photon exchange processes with both photons attached to the proton line vanish after averaging over di-lepton angles, and estimate the relatively small radiation off the proton. We compare the full one-loop calculation with a soft-photon approximation of the same order, and present estimates for a planned experiment. 

\end{abstract}

\maketitle

\tableofcontents

\vspace{-0.2cm}

\section{Introduction}
\label{sec1}

The proton radius puzzle, the discrepancy between extractions of the proton charge radius from electron scattering or electronic hydrogen spectroscopy on the one hand and from muonic hydrogen spectroscopy on the other hand, is not solved yet. The initial discrepancy amounted to around $5.6~\sigma$ when comparing both values: the extraction using elastic electron scattering, from which the A1@MAMI Collaboration reported the value $ R_E = 0.879(8) ~\mathrm{fm} $\cite{Bernauer:2010wm,Bernauer:2013tpr}, and the muonic hydrogen spectroscopy, which reported the value $ R_E = 0.84087(39) ~\mathrm{fm}$ \cite{Pohl:2010zza,Antognini:1900ns}, with more than an order of magnitude higher precision. This puzzle has spurred a lot of activity, resulting in a new round of experiments in the field, which are crucial to scrutinze and improve our understanding of systematic errors in such precision measurements. 
Recent measurements using electronic hydrogen spectroscopy 
\cite{Beyer:2017,Fleurbaey:2018fih} as well as new electron scattering experiments~\cite{Mihovilovic:2019jiz} have each reported support for both large and small values of the proton charge radius. Several further experiments~\cite{Vutha2012,Gasparian:2014rna} have reported preliminary results and are at the stage of final analysis.
 
Attempts to explain the discrepancy reach from systematic errors in the extraction of the radius, see Refs.~ \cite{Lorenz:2012tm,Lorenz:2014vha,Lorenz:2014yda,Lee:2015jqa,Arrington:2015ria,Arrington:2015yxa,Griffioen:2015hta,Higinbotham:2015rja,Alarcon:2018zbz}, to new physics models beyond the Standard Model of particle physics, see for example Refs.~ \cite{TuckerSmith:2010ra,Barger:2010aj,Barger:2011mt,Batell:2011qq,Brax:2010gp,Jentschura:2010ha,Carlson:2012pc,Wang:2013fma,Onofrio:2013fea,Karshenboim:2014tka} as well as 
Ref.~\cite{Carlson:2015jba} for an early review of the field. 

The possible explanations of electron versus muon discrepancy by a new physics scenario require to give up lepton universality, since the Standard Model has the same tree-level couplings for all leptons. In this context new experiments have been proposed to measure the proton form factors using muon beams: 
the MUSE@PSI experiment~\cite{Gilman:2013eiv,Gilman:2017hdr}, which is ongoing with the aim to compare low-energy elastic electron scattering versus elastic muon scattering off the proton, and the COMPASS@CERN~\cite{Denisov:2018unj} experiment, which plans to measure the elastic muon scattering off a proton using a 100~GeV muon beam. 

In Ref. \cite{Pauk:2015oaa}, the authors suggested to test lepton universality by comparing the cross sections for electron- and muon-pair production in the reaction  $\gamma\; p\rightarrow l^+l^-p$. According to the findings of Ref. \cite{Pauk:2015oaa}, 
a measurement of the cross-section ratio below and slightly above the di-muon production threshold can test lepton universality without having to rely on the precision which can be achieved for an  absolute cross section measurement. It was found that the difference between both values for the extracted proton charge radius amounts to an effect of 0.2~\% on this cross section ratio. Such measurement would therefore allow to test lepton universality at the $3\sigma$ significance level, if  one is able to measure such cross section ratio with a precision of around $7\times 10^{-4}$. An upcoming experiment at MAMI is presently conceived to perform such measurements \cite{MAMI_photopr}.

To make a conclusive statement from such kind of measurements, it is clearly necessary to include higher-order corrections. In our previous work \cite{Heller:2018ypa}, we have estimated the one-loop and higher-order QED corrections in the soft-photon approximation. We have found that the effect of radiative corrections is of the order 1~\% on the cross section ratio, and thus significantly larger than the 0.2~\% effect between both values for the radius. Therefore, the knowledge of the full one-loop corrections becomes imperative for an interpretation of an upcoming experiment.

In this work, we extend our calculation of the QED corrections in the soft-photon approximation and present a full one-loop QED calculation keeping all terms in the lepton mass. Moreover, we estimate the size of radiative effects on the proton side and indicate the vanishing two-photon exchange effect after the integration over the lepton-pair angles.

The outline of the paper is as follows. In Sec. \ref{sec2}, we introduce the Bethe-Heitler process at tree level and define the relevant kinematic variables.  In Sec. \ref{sec3}, we present details of the full one-loop QED calculation. We introduce the crossing relation to simplify the evaluation of the crossed diagrams from direct ones. We demonstrate that our full one-loop calculation reproduces the correct double-logarithmic behavior of the soft-photon approximation~\cite{Heller:2018ypa}. In Sec. \ref{sec4}, we present details of the numerical approach to this calculation, using the \texttt{Mathematcia} packages \texttt{FeynArts} and \texttt{FormCalc}, which serves to cross-check the analytic result. In Sec. \ref{sec5}, we present the calculation of the real corrections for the emission of a soft photon, as done in our previous work~\cite{Heller:2018ypa}. We verify that the infrared divergences in these contributions cancel with the infrared divergent part of the virtual one-loop corrections. In Sec. \ref{sec6}, we give the expression for the full one-loop QED corrections and exponentiate parts associated with the soft photon contributions. In Sec. \ref{proton_correction}, we present an estimate of the radiative corrections on the proton side. We prove that two-photon-exchange diagrams do not contribute on the level of the cross section after integrating over the di-lepton phase space. In Sec. \ref{results}, we present our numerical results, and show the effect on the absolute cross section as well as on the cross section ratio of muon- and electron-pair production. We conclude in Sec. \ref{conclusion}. Several technical details are discussed in five appendices. 

\section{Lepton-pair production at tree level}
\label{sec2}
\begin{figure}
	\includegraphics[scale=0.70]{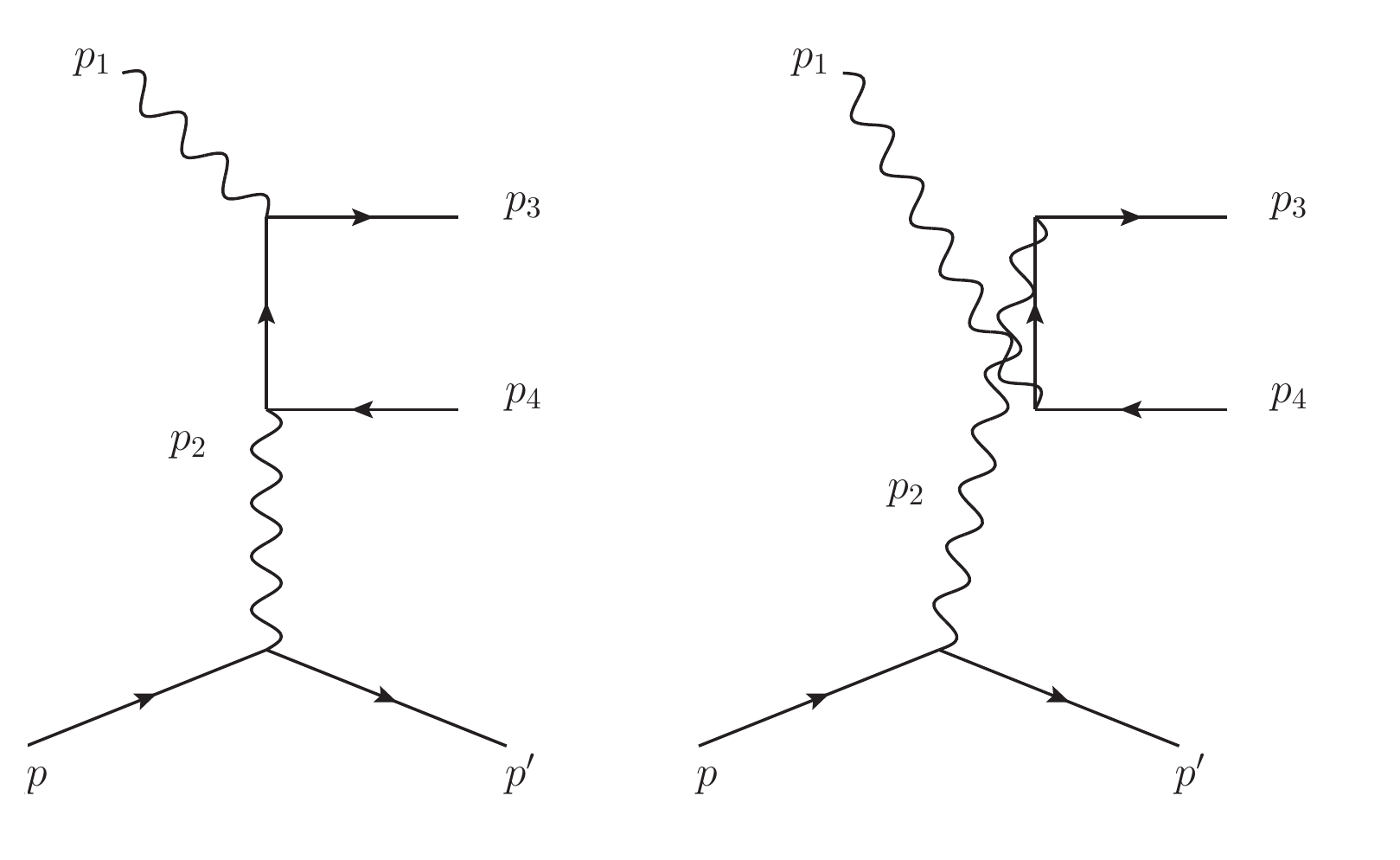}\hspace{0.cm}
	\caption{The Bethe-Heitler process at tree level.\label{tree}}
\end{figure}
The Bethe-Heitler process at tree level is described by two graphs, see Fig. \ref{tree}. We use $p$ ($p^{\prime}$) for the momenta of the initial (final) proton, and $p_3$ ($p_4$) for the momenta of leptons $l^-$ ($l^+$) respectively. The initial photon has momentum $p_1$, and the virtual photon momentum in the one-photon exchange graphs of Fig. \ref{tree} is defined as $p_2=p-p^{\prime}$. The Mandelstam variables for this process are defined as
\begin{align}
(p_3+p_4)^2&=s_{ll},\\
(p_3-p_1)^2&=t_{ll},\\
(p_3-p_2)^2&=u_{ll},\\
(p_1+p)^2&=s,\\
(p-p_3)^2&=u,\\
p_2^2=(p-p^{\prime})^2&=t.
\end{align}
The on-shell condition for external particles implies:
\begin{align}
p_3^2&=p_4^2=m^2,\\
p^2&=p^{\prime\; 2}=M^2,\\
p_1^2&=0.
\end{align}

At leading order, the scattering amplitude $\mathcal{M}_0$ is given by
\begin{align}
\mathcal{M}_0&=\bar{u}(p_3)(ie)\left[\gamma^{\mu}\frac{i(\sla{p}_3-\sla{p}_1+m)}{(p_3-p_1)^2-m^2}\gamma^{\nu}+\gamma^{\nu}\frac{i(\sla{p}_1-\sla{p}_4+m)}{(p_1-p_4)^2-m^2}\gamma^{\mu}\right](ie)v(p_4)\nonumber\\
&\times\frac{(-i)}{t}\varepsilon_{\mu}(p_1)\bar{u}(p^{\prime})(-ie)\Gamma_{\nu}(t)u(p),
\label{M0}\end{align}
where the electromagnetic vertex $\Gamma_{\nu}$ for the proton is expressed as
\begin{equation}
\Gamma_{\nu}(t)=F_D(t)\gamma_{\nu}-iF_P(t)\frac{\sigma_{\nu\alpha}(p_2)^{\alpha}}{2M},\label{protonVertex}
\end{equation} 
with the proton's Dirac and Pauli form factors $F_D$ and $F_P$, respectively.

The corresponding unpolarized differential cross section $d\sigma_0$ is given by
\begin{equation}
\left(\frac{d\sigma}{dt\,ds_{ll}\,d\Omega_{ll}^{CM_{l^+l^-}}}\right)_0=\frac{1}{(2\pi)^4}\frac{1}{64}\frac{\beta}{(2M E_{\gamma})^2}
\left[\overline{\sum_{i}}\sum_{f}\left(\mathcal{M}_0^{*}\;\mathcal{M}_0\right)\right],\label{CrossSectionSum}
\end{equation}
where $E_{\gamma}$ is the lab energy of the initial photon and $\Omega_{ll}^{CM_{l^+l^-}}$ is the solid angle of the lepton pair in their center-of-mass frame, in which the lepton velocity is denoted by
\begin{equation}
\beta=\sqrt{1-\frac{4m^2}{s_{ll}}}.
\end{equation}
In Eq. \eqref{CrossSectionSum}, we average over all polarizations in the initial state and sum over the polarizations in the final state. We express the cross section as a product of hadronic and leptonic parts as
\begin{equation}\label{CrossSection}
\left(\frac{d\sigma}{dt\,ds_{ll}\,d\Omega_{ll}^{CM_{l^+l^-}}}\right)_0=\frac{\alpha^3\beta}{16\pi(2M E_{\gamma})^2\;t^2}L_0^{\mu\nu}H_{\mu\nu},
\end{equation}
where the fine-structure constant is defined as $\alpha\equiv e^2/4\pi\approx 1/137$. Furthermore, the unpolarized leptonic tensor at leading order $L_0^{\mu\nu}$ (including the average over the initial photon polarization) is given by
\begin{align}
L_0^{\mu\nu}=-\frac{1}{2}\;\text{Tr}&\left[(\sla{p_3}+m)\left(\gamma^{\alpha}\frac{(\sla{p_3}-\sla{p_1}+m)}{(p_3-p_1)^2-m^2}\gamma^{\mu}+\gamma^{\mu}\frac{(\sla{p_1}-\sla{p_4}+m)}{(p_1-p_4)^2-m^2}\gamma^{\alpha}\right)\right.\nonumber\\
&\times\left.(\sla{p_4}-m)\left(\gamma^{\nu}\frac{(\sla{p_3}-\sla{p_1}+m)}{(p_3-p_1)^2-m^2}\gamma_{\alpha}+\gamma_{\alpha}\frac{(\sla{p_1}-\sla{p_4}+m)}{(p_1-p_4)^2-m^2}\gamma^{\nu}\right)\right],\label{L0Tensor}
\end{align}
and the unpolarized hadronic tensor $H^{\mu\nu}$ by
\begin{equation}
H^{\mu\nu}=\frac{1}{2}\text{Tr}\left[(\sla{p^{\prime}}+M)\;\Gamma^{\mu}\;(\sla{p}+M)\;(\Gamma^{\dagger})^{\nu}\right].
\end{equation}
Using \eqref{protonVertex}, the unpolarized hadronic tensor can be expressed as
\begin{align}
H^{\mu\nu}=(-g^{\mu\nu}+\frac{p_2^{\mu}p_2^{\nu}}{p_2^2})H_1+\tilde{p}^{\mu}\tilde{p}^{\nu}H_2,
\end{align}
where $\tilde{p}\equiv (p+p^{\prime})/2$. We have defined with $\tau\equiv -t/(4M^2)$
\begin{align}
H_1&=4M^2\tau G^2_M(t),\\
H_2&=\frac{4}{1+\tau}\left[G_E^2(t)+\tau G_M^2(t)\right],
\end{align}
and where the electric $(G_E)$ and magnetic $(G_M)$ form factors are defined as
\begin{align}
G_E&=F_D-\tau F_P,\\
G_M&=F_D+F_P,
\end{align}
which are functions of the spacelike momentum transfer $t$.

A very compact expression for the contraction of the tree-level lepton tensor $L_0^{\mu\nu}$ with the hadronic tensor $H_{\mu\nu}$ can be found in \cite{Drell:1963ej,Pauk:2015oaa}. We show the expression in our notation in appendix \ref{treecontraction}.

For the electric and magnetic proton form factors, which enter the total cross sections for lepton-pair production, we exploit the fit of Ref. \cite{Bernauer:2013tpr}, which is based on a global analysis of the electron-proton scattering data at $Q^2 < 10~\mathrm{GeV}^2$ with an empirical account of TPE corrections.

In the experimental setup, when only the recoil proton is measured, one has to integrate \eqref{CrossSection} over the lepton angles:
\begin{equation}
\left(\frac{d\sigma}{dt\,ds_{ll}}\right)_0=\frac{\alpha^3\beta}{16\pi (2ME_{\gamma})^2\;t^2}\cdot \int d\Omega_{ll}^{CM_{l^+l^-}} L_0^{\mu\nu}H_{\mu\nu}.\label{CrossSectionIntegrated}
\end{equation}
The kinematical invariant $t$ is in one-to-one relation with the recoiling proton lab momentum $\vec{p^\prime}$ (or energy $E^{\prime}$):
\begin{align}\label{TauLab}
|\vec{p^\prime}|&=2M\sqrt{\tau(1+\tau)},\\
E^{\prime}&=M(1+2\tau),
\end{align}
whereas the invariant $s_{ll}$ is then determined from the recoiling proton lab scattering angle:
\begin{equation}\label{ScatterAngleLab}
\text{cos}\;\theta_{p^{\prime}}=\frac{s_{ll}+2(s+M^2)\tau}{2(s-M^2)\sqrt{\tau(1+\tau)}},
\end{equation}
where $s$ can be expressed in terms of the initial photon-beam energy $E_{\gamma}$ as:
\begin{equation}
s=2E_{\gamma}M+M^2.
\end{equation}

In Ref. \cite{Pauk:2015oaa}, the authors calculated the cross section ratio $R$ between electron- and muon-pair production:
\begin{equation}
R(s_{ll},s^0_{ll})\equiv\frac{\left[\sigma_0(\mu^+\mu^-)\right](s_{ll})+[\sigma_0(e^+e^-)](s_{ll})}{[\sigma_0(e^+e^-)](s^0_{ll})},\label{ratio-def}
\end{equation}
which depends on the invariant mass of the lepton pair $s_{ll}$, and a reference point $s_{ll}^0$ to which the measurement is normalized. 

\begin{figure}
	\includegraphics[scale=0.9]{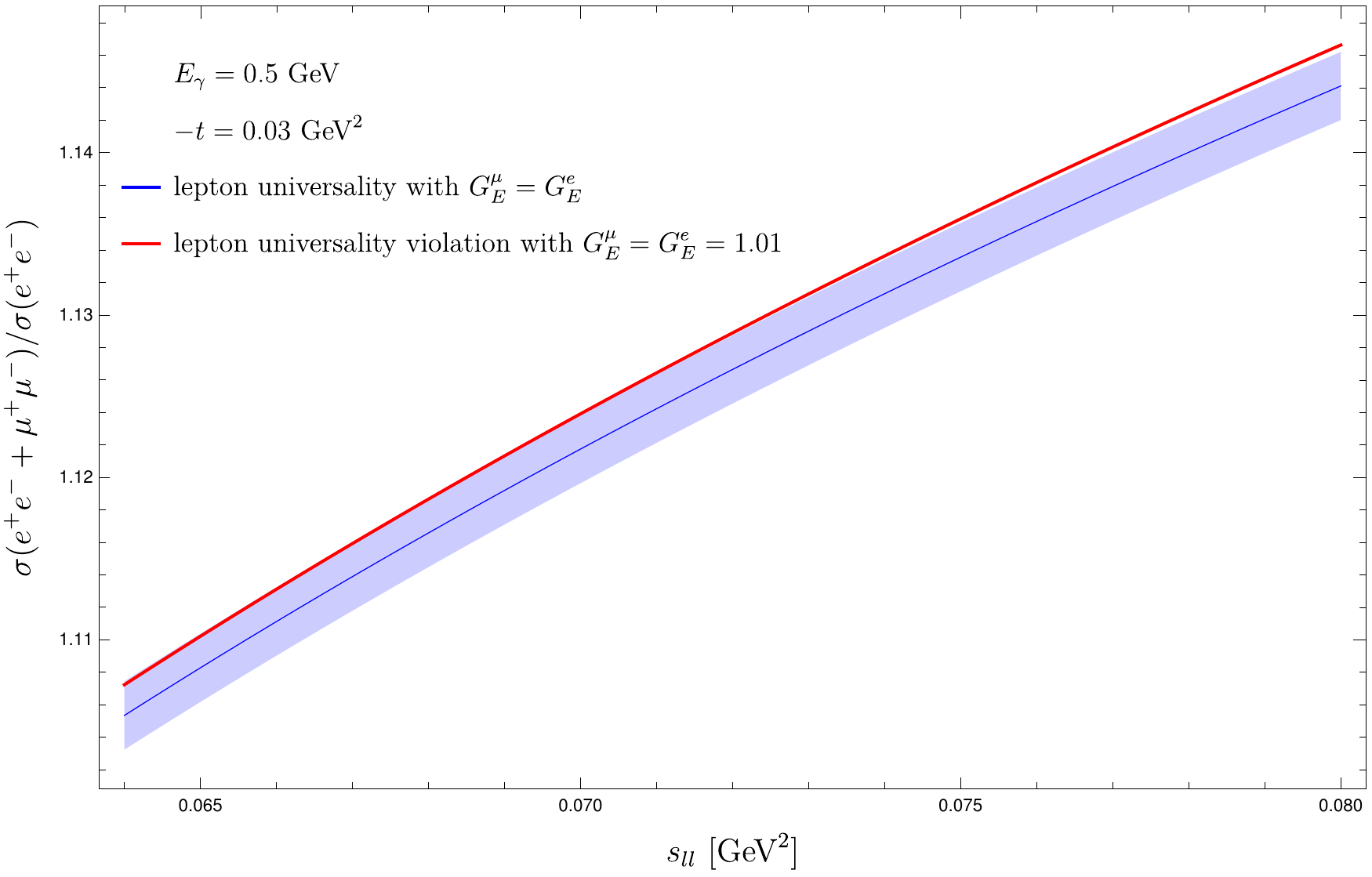}\hspace{0.cm}
	\caption{Ratio of the cross sections for $\gamma p\rightarrow (e^+e^-+\mu^+\mu^-)p$ vs $\gamma p\rightarrow (e^+e^-)p$. The blue band corresponds to a $3\sigma$ band around the lepton universality result, where $\sigma=7\times10^{-4}$.\label{RatioTree}}
\end{figure}

The corresponding plot for the kinematical range accessible at MAMI is shown in Fig. \ref{RatioTree}. The normalization is shown for the choice $s_{ll}^0=s_{ll}$, i.e., at each point above the muon-pair production threshold the sum of the cross sections for muon- and electron-pair production is divided by the corresponding cross section for electron-pair production. In this plot, the blue curve describes the scenario, when lepton universality holds, i.e., $G_E^{\mu}=G_E^{e}$, while the red curve corresponds to a case when lepton universality is broken by an amount of $1\%$, which would correspond with the difference in proton radii as extracted from muonic Hydrogen spectroscopy \cite{Pohl:2010zza,Antognini:1900ns} and from electron scattering \cite{Bernauer:2010wm,Bernauer:2013tpr}. The blue band describes the $3\sigma$ deviation if this observable is measured with an absolute accuracy of $7\times 10^{-4}$. We will show in this work that radiative corrections shift this curve by more than $3\sigma$, making their inclusion indispensable for a comparison with experiment.

\section{Calculation of the one-loop corrections}
\label{sec3}
In this work, we calculate all one-loop leptonic corrections contributing to the $\gamma p\rightarrow l^+l^-p$ in three independent setups:
\begin{itemize}
    \item[(i)] an analytic calculation using the techniques of Integration-By-Parts (IBP) identities,
    \item[(ii)]  a numerical calculation using the mathematica package \texttt{FormCalc}, which uses the Passarino-Veltman tensor reduction with numerical implementation of scalar integrals in \texttt{LoopTools},
    \item[(iii)] a calculation of the self-energy and vertex diagrams with the help of projection techniques.
\end{itemize}

\subsection{Crossing relations}
\label{SecCrossing}
\begin{figure}
	\includegraphics[scale=0.7]{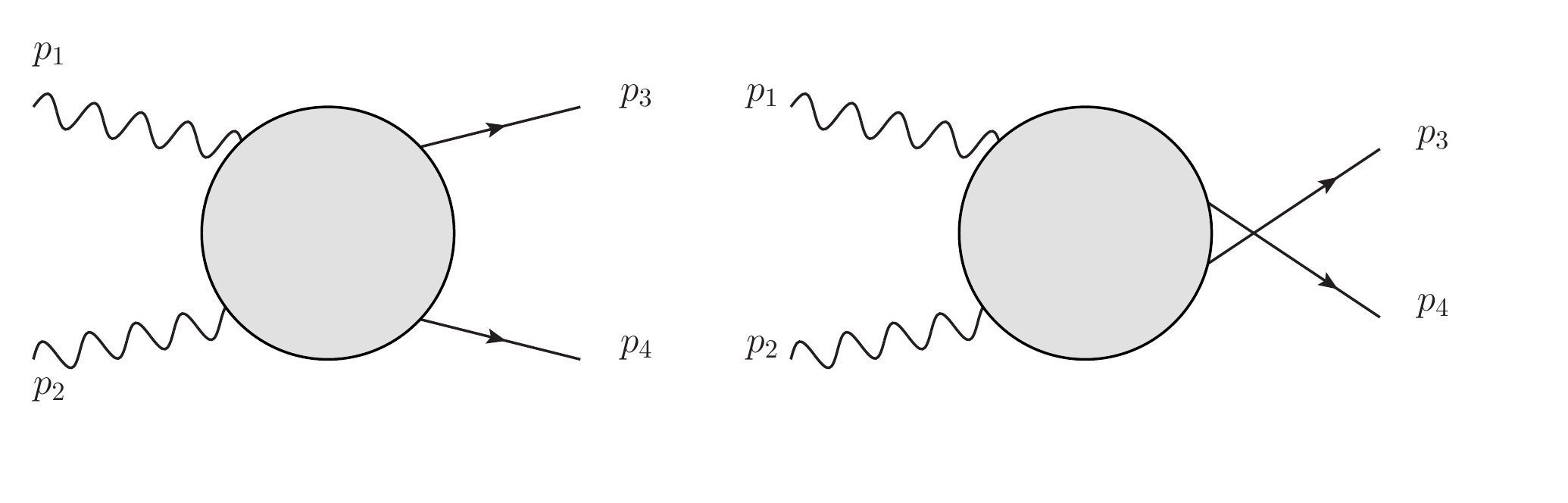}\hspace{0.cm}
	\caption{Generic one-loop leptonic diagram with its crossed counterpart.\label{CrossingRelation}}
\end{figure}
One can find relations between a given loop diagram and its crossed counterpart. Let us assume that the first diagram (direct diagram) in Fig. \ref{CrossingRelation} is given by
\begin{equation}
\mathcal{M}_\text{d}=\bar{u}(p_3)\;\Gamma_d(\sla{p}_1,\sla{p}_2,\sla{p}_3,\sla{p}_4,m)\; v(p_4).
\end{equation}
Then the crossed diagram is given by
\begin{equation}
\mathcal{M}_{c}=\bar{u}(p_3)\;\Gamma_{c}(\sla{p}_1,\sla{p}_2,\sla{p}_3,\sla{p}_4,m)\; v(p_4),
\end{equation}
where $\Gamma_{c}(\sla{p}_1,\sla{p}_2,\sla{p}_3,\sla{p}_4,m)$ can be related to $\Gamma_d(\sla{p}_1,\sla{p}_2,\sla{p}_3,\sla{p}_4,m)$ by reading $\Gamma_d(\sla{p}_1,\sla{p}_2,\sla{p}_3,\sla{p}_4,m)$ backwards and changing the direction of the momentum flow, i.e.:
\begin{equation}
\Gamma_{c}(\sla{p}_1,\sla{p}_2,\sla{p}_3,\sla{p}_4,m)=\gamma_0\Gamma_d^\dagger(-\sla{p}_1,-\sla{p}_2,-\sla{p}_4,-\sla{p}_3,m)\gamma_0.
\end{equation}
$\Gamma$ is always a product of an odd number of Fermion propagators, so we can pull out the minus sign and, instead, change $m$ to $-m$:
\begin{equation}
\Gamma_{c}(\sla{p}_1,\sla{p}_2,\sla{p}_3,\sla{p}_4,m)=-\gamma_0\Gamma_d^\dagger(\sla{p}_1,\sla{p}_2,\sla{p}_4,\sla{p}_3,-m)\gamma_0.
\end{equation}
This relation also holds for the tree level. Interfering the loop-diagrams with the tree diagrams we get therefore:
\begin{align}
\mathcal{M}_0^*\mathcal{M}_\text{d}=&\bar{v}(p_4)\left[\Gamma_{d,\text{tree}}(\sla{p}_1,\sla{p}_2,\sla{p}_3,\sla{p}_4,m)-\gamma_0\Gamma_{d,\text{tree}}^\dagger(\sla{p}_1,\sla{p}_2,\sla{p}_4,\sla{p}_3,-m)\gamma_0\right]^\dagger u(p_3)\nonumber\\
&\times\bar{u}(p_3)\left[\vphantom{\Gamma^{\dagger}_{d,\text{tree}}}\Gamma_\text{d}(\sla{p}_1,\sla{p}_2,\sla{p}_3,\sla{p}_4,m)\right] v(p_4).
\end{align}
Summing over spins, we find:

\begin{align}
\sum_s(\mathcal{M}_0^*\mathcal{M}_\text{d})=\text{Tr}&\left[(\sla{p_4}-m)\cdot\left(\Gamma_{d,\text{tree}}^\dagger(\sla{p}_1,\sla{p}_2,\sla{p}_3,\sla{p}_4,m)-\gamma_0\Gamma_{d,\text{tree}}(\sla{p}_1,\sla{p}_2,\sla{p}_4,\sla{p}_3,-m)\gamma_0\right)\right.\nonumber\\
&\times\left.(\sla{p}_3+m)\cdot\Gamma_\text{d}(\sla{p}_1,\sla{p}_2,\sla{p}_3,\sla{p}_4,m)\vphantom{\Gamma^{\dagger}_{d,\text{tree}}}\right].
\label{Interference}
\end{align}
The interference with the crossed counter-part gives:
\begin{align}
\mathcal{M}_0^*\mathcal{M}_{c}=&\bar{v}(p_4)\left[\Gamma_{d,\text{tree}}(\sla{p}_1,\sla{p}_2,\sla{p}_3,\sla{p}_4,m)-\gamma_0\Gamma_{d,\text{tree}}^\dagger(\sla{p}_1,\sla{p}_2,\sla{p}_4,\sla{p}_3,-m)\gamma_0\right]^\dagger u(p_3)\nonumber\\
&\times\bar{u}(p_3)\left[\vphantom{\Gamma^{\dagger}_{d,\text{tree}}}-\gamma_0\Gamma_\text{d}^\dagger(\sla{p}_1,\sla{p}_2,\sla{p}_4,\sla{p}_3,-m)\gamma_0\right] v(p_4),
\end{align}
and, summing over spins:
\begin{align}
\sum_s(\mathcal{M}_0^*\mathcal{M}_{c})=\text{Tr}&\left[(\sla{p_4}-m)\cdot\left(\Gamma_{d,\text{tree}}^\dagger(\sla{p}_1,\sla{p}_2,\sla{p}_3,\sla{p}_4,m)-\gamma_0\Gamma_{d,\text{tree}}(\sla{p}_1,\sla{p}_2,\sla{p}_4,\sla{p}_3,-m)\gamma_0\right)\right.\nonumber\\
&\times\left.(-\sla{p}_3-m)\cdot\gamma_0\Gamma_\text{d}^{\dagger}(\sla{p}_1,\sla{p}_2,\sla{p}_4,\sla{p}_3,-m)\gamma_0\vphantom{\Gamma^{\dagger}_{d,\text{tree}}}\right]\nonumber\\
=\text{Tr}&\left[\vphantom{\Gamma^{\dagger}_{d,\text{tree}}}(\sla{p}_3+m)\cdot\left(\Gamma^\dagger_{d,\text{tree}}(\sla{p}_1,\sla{p}_2,\sla{p}_4,\sla{p}_3,-m)-\gamma_0\Gamma_{d,\text{tree}}(\sla{p}_1,\sla{p}_2,\sla{p}_3,\sla{p}_4,m)\gamma_0\right)\right.\nonumber\\
&\times\left.\vphantom{\Gamma^{\dagger}_{d,\text{tree}}}(\sla{p}_4-m)\cdot\Gamma_\text{d}(\sla{p}_1,\sla{p}_2,\sla{p}_4,\sla{p}_3,-m)\right]^*,
\label{CrossedInterference}
\end{align}
where we used cyclicity and complex conjugation of the argument of the trace in the second step. Comparing Eqs. \eqref{Interference} and \eqref{CrossedInterference}, one can easily verify that:
\begin{equation}
\sum_f(\mathcal{M}_0^*\mathcal{M}_\text{d})=\left.\sum_f(\mathcal{M}_0^*\mathcal{M}_{c})^*\right|_{(p_3\rightarrow p_4,p_4\rightarrow p_3,m\rightarrow -m)}.\label{CrossingRelationEq}
\end{equation}
Since an odd number of gamma matrices does not contribute to the trace, only terms with an even number of powers of $m$ contribute. Therefore the substitution $m\rightarrow -m$ has no effect, and the crossing relation reads:
\begin{equation}
\sum_f(\mathcal{M}_0^*\mathcal{M}_\text{d})=\left.\sum_f(\mathcal{M}_0^*\mathcal{M}_{c})^*\right|_{(p_3\rightarrow p_4,p_4\rightarrow p_3)}.
\end{equation}
Having calculated any diagram, one can therefore easily obtain the expression for its crossed counterpart by only exchanging $p_3$ and $p_4$. Note that the complex conjugation in Eq. \eqref{CrossingRelationEq} only affects the algebraic, i.e. the trace over gamma matrices, and not the analytic part (the $i\epsilon$ prescription in the integrals) of the diagram. For the unpolarized cross section, which only involves the real part of such interference, this complex conjugation is of no relevance.

\subsection{Lepton self-energy}\label{SectionSelfEnergy}
\subsubsection{Lepton self-energy at first order}
\begin{figure}
	\includegraphics[scale=0.7]{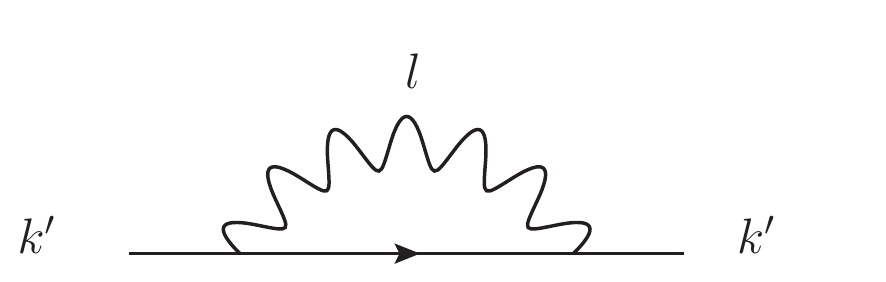}\hspace{0.cm}
	\caption{Lepton self-energy diagram with $k'^2=s'$.\label{SelfEnergyDiagram}}
\end{figure}
We show the first-order lepton self-energy diagram in Fig. \ref{SelfEnergyDiagram}. In the following, we use for the calculation of all Feynman diagrams the Feynman gauge and dimensional regularization for ultra-violet (UV) divergences ($\epsilon_{\text{UV}}=2-d/2>0$) and for infrared (IR) divergences ($\epsilon_{\text{IR}}=2-d/2<0$). The self-energy is then given by:
\begin{equation}
-i\Sigma(\sla{k'})=-e^2\mu^{4-d}\int\frac{d^dl}{(2\pi)^d}\frac{\gamma^{\alpha}(\sla{k}'+\sla{l}+m)\gamma_{\alpha}}{((k'+l)^2-m^2)\;l^2}.
\end{equation}
It can easily be reduced to
\begin{align}
\Sigma(\sla{k'})=&\sla{k}^{\prime} \frac{\alpha}{4\pi}\left\{-\left[\frac{1}{\epsilon_\text{UV}}-\gamma_E+\ln\left(4\pi \right)\right]+1-\frac{ \left(s^\prime + m^2\right)}{s^\prime}B_0(s',0, m^2)+\frac{1}{s^\prime}A_0(m^2)\right\}\nonumber\\
&+m\frac{\alpha}{4\pi}\left\{4\left[\frac{1}{\epsilon_\text{UV}}-\gamma_E+\ln\left(4\pi \right)\right]-2+ 4\;B_0(s',0, m^2)\right\},\label{SEOffshell}
\end{align}
where $s^\prime\equiv (k^\prime)^2$, and where the finite parts of the master integrals $A_0$ and $B_0$ are given by the expansions in Appendix \ref{master_integrals}.\\
For $\sla{k}'=m$, we find:
\begin{equation}
\Sigma(m)=m\frac{\alpha}{4\pi}\left\{3\left[\frac{1}{\epsilon_\text{UV}}-\gamma_E+\ln\left(4\pi\right)\right]+4-3\ln\left(\frac{m^2}{\mu^2}\right)\right\}.\label{SEOnshell}
\end{equation}
Note that we separate the UV-divergent part and imply only the regular part of the scalar integrals $A_0$, $B_0$ and $C_0$ in all expressions in the main part of this paper.

The on-shell renormalization condition fixes the pole at $(k^{\prime})^2=m^2$ with residue equal to one. This gives the renormalization constants $Z_2$ and $Z_m$:
\begin{align}
Z_2&=1+\left.\frac{d\;\Sigma (\sla{k'})}{d\sla{k'}}\right|_{\sla{\;k'}\;=m},\\
(1-Z_m)Z_2m&=\Sigma(m).
\end{align}
The evaluation of $\Sigma(k')$ and its derivative, results in the renormalization constants:
\begin{align}
Z_2&=1-\frac{\alpha}{4\pi}\left\{\left[\frac{1}{\epsilon_\text{UV}}-\gamma_E+\ln\left(\frac{4\pi\mu^2}{m^2}\right)\right]+2\left[\frac{1}{\epsilon_\text{IR}}-\gamma_E+\ln\left(\frac{4\pi\mu^2}{m^2}\right)\right]+4\right\},\\
Z_2Z_m&=1-\frac{\alpha}{4\pi}\left\{4\left[\frac{1}{\epsilon_\text{UV}}-\gamma_E+\ln\left(\frac{4\pi\mu^2}{m^2}\right)\right]+2\left[\frac{1}{\epsilon_\text{IR}}-\gamma_E+\ln\left(\frac{4\pi\mu^2}{m^2}\right)\right]+8\right\}.
\end{align}
The renormalized self-energy is then given by
\begin{equation}
\tilde{\Sigma}(k')=\Sigma(k')-(Z_2-1)\sla{p'}+(Z_2Z_m-1)m,
\end{equation}
which reads as:
\begin{align}
    \tilde{\Sigma}(k)&=-\frac{\alpha}{4\pi}\left\{(\sla{k}-m)\left[-2\left[\frac{1}{\epsilon_{\text{IR}}}-\gamma_E+\ln\left(\frac{4\pi\mu^2}{m^2}\right)\right]-3+\frac{m^2}{k^2}+\left(\frac{m^4}{k^4}-1\right)\ln\left(1-\frac{k^2}{m^2}\right)\right]\right.\nonumber\\
    &\left.+m\left[\left(\frac{m^2}{k^2}-1\right)+\left(3+\frac{m^4}{k^4}-4\frac{m^2}{k^2}\right)\ln\left(1-\frac{k^2}{m^2}\right)\right]\right\}\nonumber\\
    &\equiv(\sla{k}-m)\tilde{\Sigma}_1(k^2)+m\tilde{\Sigma}_2(k^2).\label{SelfEnergyRen}
\end{align}

\subsubsection{Self-energy diagram}
\begin{figure}
	\includegraphics[scale=0.7]{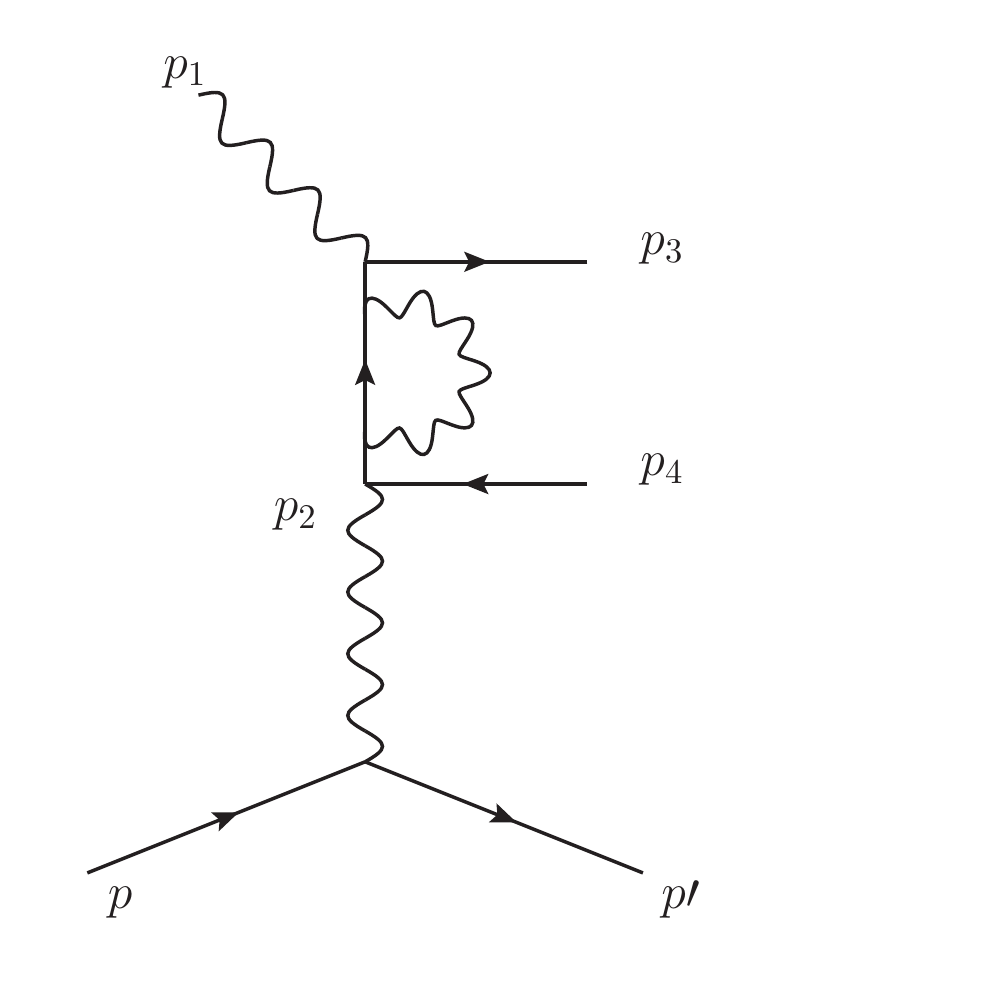}\hspace{0.cm}
	\caption{Lepton self-energy contributing to the Bethe-Heitler process. \label{SEDiagram}}
\end{figure}
In the on-shell scheme, only self-energy diagrams contribute with the virtual photon attached to an internal fermion line. In Fig. \ref{SEDiagram}, we show the corresponding diagram, contributing to the Bethe-Heitler process. The amplitude is given by
\begin{align}
\mathcal{M}_{\text{SE}}&=\bar{u}(p_3)(ie)\gamma^{\nu}\frac{i(\sla{p}_3-\sla{p}_1+m)}{(p_3-p_1)^2-m^2}(-i)\tilde{\Sigma}(p_3-p_1)\frac{i(\sla{p}_3-\sla{p}_1+m)}{(p_3-p_1)^2-m^2}\gamma^{\mu}(ie)v(p_4)\nonumber\\
&\times\frac{-i}{t}\varepsilon_{\nu}(p_1)\bar{u}(p^{\prime})(-ie)\Gamma_{\mu}(t)u(p),
\end{align}
where the renormalized self-energy $\tilde{\Sigma}$ is given by Eq. (\ref{SelfEnergyRen}).
The interference of the direct self-energy diagram with the lowest-order diagrams is given by:

\begin{align}
\overline{\sum_i}\sum_f\mathcal{M}^*_{0}\left(\mathcal{M}_{\text{SE}}\right)&=\frac{e^4}{t^2}\sum_{i=1,2} a^{\text{SE}}_i L_{i;\mu\nu}^{\text{SE}}H^{\mu\nu}
\end{align}
where    

\begin{align}
    L^{1;\mu\nu}_{\text{SE}}&=-\frac{1}{2}\text{Tr}\left[(\sla{p_4}-m)\left(\gamma^{\mu}\frac{(\sla{k_1}+m)}{k_1^2-m^2}\gamma_{\alpha}+\gamma_{\alpha}\frac{(\sla{k_2}+m)}{k_2^2-m^2}\gamma^{\mu}\right)(\sla{p_3}+m)\;\gamma^{\alpha}\frac{(\sla{k_1}+m)}{k_1^2-m^2}\gamma^{\nu}\right],\nonumber\\
    L^{2;\mu\nu}_{\text{SE}}&=-\frac{1}{2}\text{Tr}\left[(\sla{p_4}-m)\left(\gamma^{\mu}\frac{(\sla{k_1}+m)}{k_1^2-m^2}\gamma_{\alpha}+\gamma_{\alpha}\frac{(\sla{k_2}+m)}{k_2^2-m^2}\gamma^{\mu}\right)(\sla{p_3}+m)\;\gamma^{\alpha}\frac{1}{m}\gamma^{\nu}\right],
\end{align}
with
\begin{align}
k_1&=p_3-p_1,\\
k_2&=p_1-p_4,
\end{align}
and
\begin{align}
a_{\text{SE}}^{(1)}&=\tilde{\Sigma}_1(k_1^2)+2\frac{m^2}{k_1^2-m^2}\tilde{\Sigma}_2(k_1^2),\\
a_{\text{SE}}^{(2)}&=\frac{m^2}{k_1^2-m^2}\tilde{\Sigma}_2(k_1^2).
\end{align}
Evaluating these two traces, we find for the direct self-energy diagram:
\begin{align}
\overline{\sum_i}\sum_f\mathcal{M}^*_{0}\left(\mathcal{M}_{\text{SE}}\right)&=\frac{e^4}{t^2}\sum_{i=1}^2 a^{\text{SE}}_iT^{\text{SE}}_i,
\end{align}
where
\begin{align}
T^{\text{SE}}_1&=L_{0,d}^{\mu\nu}H_{\mu\nu},\\
T^{\text{SE}}_2&= \frac{4 H_1}{\left(m^2-t_{ll}\right) \left(-s_{ll}-t_{ll}+m^2+t\right)}\left\{-4 m^2 s_{ll}-4 m^2 t_{ll}-3 t s_{ll}+2 s_{ll} t_{ll}+s_{ll}^2+2 t_{ll}^2-5 t t_{ll}\right.\nonumber\\
&\left.+2 m^4+7 m^2 t+2 t^2\right\}\nonumber\\
&+ \frac{H_2}{\left(m^2-t_{ll}\right) \left(-s_{ll}-t_{ll}+m^2+t\right)}\left\{3 m^4 s_{ll}+17 m^2 M^2 s_{ll}+12 m^2 M^2 t_{ll}-6 m^2 t s_{ll}-4 m^2 s t_{ll}\right.\nonumber\\
&\left.-m^2 s_{ll} t_{ll}-5 m^2 u s_{ll}+m^2 s_{ll}^2-6 m^2 s s_{ll}-m^2 t t_{ll}+12 M^4 s_{ll}+6 M^4 t_{ll}-3 M^2 t s_{ll}-10 M^2 s t_{ll}\right.\nonumber\\
&\left.-7 M^2 s_{ll} t_{ll}-10 M^2 u s_{ll}-14 M^2 s s_{ll}-2 M^2 u t_{ll}-5 M^2 t_{ll}^2+10 M^2 t t_{ll}+4 s^2 t_{ll}+4 s^2 s_{ll}+4 t u s_{ll}\right.\nonumber\\
&\left.+2 s u t_{ll}+u s_{ll} t_{ll}+s t_{ll}^2+5 s t s_{ll}+s t t_{ll}+2 t s_{ll} t_{ll}+2 u^2 s_{ll}-u s_{ll}^2+6 s u s_{ll}-s   s_{ll}^2-3 t^2 t_{ll}-t u t_{ll}\right.\nonumber\\
&\left.+t t_{ll}^2-7 m^4 M^2+3 m^4 s-2 m^4 t-6 m^2 M^4+10 m^2 M^2 s-20 m^2 M^2 t+2 m^2 M^2 u-4 m^2 s^2\right.\nonumber\\
&\left.+5 m^2 s t-2 m^2 s u+6 m^2 t^2+5 m^2 t u-12 M^4 t+14 M^2 s t+3 M^2 t^2+10 M^2 t u-4 s^2 t-4 s t^2\right.\nonumber\\
&\left.-6 s t u-3 t^2 u-2 t u^2\right\},
\end{align}
where $L_{0,d}^{\mu\nu}$ denotes the direct part of the tree-level lepton tensor, which is given by $L_{0,d}^{\mu\nu}= L^{1;\mu\nu}_{\text{SE}}$, and where $T^{\text{SE}}_1$ is given by Eq. \eqref{DirectTreeContraction}.

In the limit for small lepton masses, the expression for the direct and crossed lepton self-energy contributions keeping only terms with logarithmic terms in the lepton mass scaled by $s_{ll}$ reduces to:
\begin{align}
\overline{\sum_i}\sum_f\mathcal{M}^*_{0}\left(\mathcal{M}_{\text{SE}}\right)\approx \frac{e^4}{t^2}\left(\frac{\alpha}{\pi}\right)L_0^{\mu\nu}H_{\mu\nu}\left\{\frac{1}{2}\left[\frac{1}{\epsilon_{\text{IR}}}-\gamma_E+\ln\left(\frac{4\pi \mu^2}{s_{ll}}\right)\right]-\frac{3}{4}\ln\left(\frac{m^2}{s_{ll}}\right)\right\}.
\label{SeApp}\end{align}

\subsection{Vertex corrections}\label{SectionVertex}
\subsubsection{Decomposition of the half-off-shell vertex}
\begin{figure}
	\includegraphics[scale=0.9]{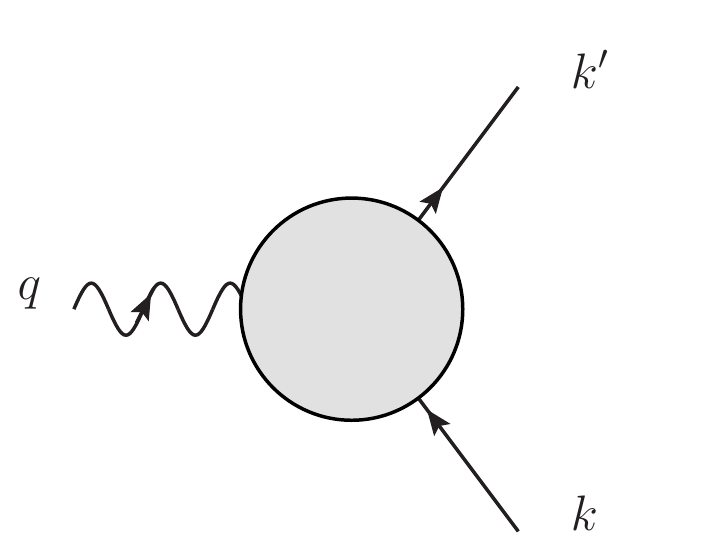}\hspace{0.cm}
	\caption{Half-off-shell vertex with $k^2=m^2$, $k^{\prime}$ off-shell, and $q=k^{\prime}-k$.\label{HOVertex} }
\end{figure}
In order to calculate the one-loop vertex corrections on the lepton side, we need to evaluate the half-off-shell vertex $\tilde{\Gamma}^{\mu}$, shown in Fig. \ref{HOVertex}. It can be constructed with the following set of Lorentz structures:
\begin{equation}
\gamma^{\mu},q^{\mu},(k+k^{\prime})^{\mu},\sla{k},\sla{k}^{\prime},
\end{equation}
with 
\begin{equation}
q=k^{\prime}-k.
\end{equation}
Using the Dirac equation:
\begin{equation}
\sla{k}\;u(k)=m\;u(k),
\end{equation}
we see, that $\sla{k}$ does not appear in the decomposition. All in all, we find $6$ independent Lorentz structures in the decomposition of the vertex:
\begin{align}
\tilde{\Gamma}^{\mu}(k^{\prime},k)=&\left\{\Lambda^{(+)}(k')\left[F_{1^+}(s',q^2)\gamma^{\mu}+F_{2^+}(s',q^2)\frac{(k+k^{\prime})^{\mu}}{2m}-F_{3^+}(s',q^2)\frac{q^{\mu}}{2m}\right]\right.\nonumber\\
&+\left.\Lambda^{(-)}(k')\left[F_{1^-}(s',q^2)\gamma^{\mu}+F_{2^-}(s',q^2)\frac{(k+k^{\prime})^{\mu}}{2m}-F_{3^-}(s',q^2)\frac{q^{\mu}}{2m}\right]\right\},\label{VertexDec}
\end{align}
with scalar form factors $F_{i^\pm}$, $s^{\prime}=(k^{\prime})^2$ and projectors to the on-shell states:
\begin{equation}
\Lambda^{(\pm)}(k')=\frac{\sla{k}^{\prime}\pm m}{2m}.
\end{equation}
The structure proportional to $F_{3^\pm}$ does not contribute to the cross section, since $q^{\mu}$ gets either contracted with the photon momenta or with the hadron tensor; both contributions give zero due to gauge invariance. However, for the proof of gauge invariance, we need to consider this structure.\\
One can construct projectors to extract the six form factors by
\begin{equation}
\hat{P}_{\mu}^{(i,\pm)}=(\sla{k}^{\prime}+m)\sum_{\sigma=\pm}\left(a_{(1,\sigma)}^{(i,\pm)}\gamma_{\mu}\Lambda^{(\sigma)}(k')+a^{(i,\pm)}_{(2,\sigma)}\frac{(k+k^{\prime})_{\mu}}{2m}\Lambda^{(\sigma)}(k')+a^{(i,\pm)}_{(3,\sigma)}\frac{q_{\mu}}{2m}\Lambda^{(\sigma)}(k')\right),
\end{equation}
where the coefficients $a_{(j,\pm)}^{(i,\pm)}$ are chosen, such that the $\hat{P}_{\mu}^{(i,\pm)}$ project out the different Lorentz structures of the form factor:
\begin{equation}
\text{Tr}\left(\hat{P}_{\mu}^{(i,\pm)}\tilde{\Gamma}^{\mu}\right)=F_{i^\pm}.
\end{equation}

For the form factors, we find:

\begin{align}
F_{1^-}&(s',q^2)=\left(\frac{\alpha}{4\pi}\right)\left\{-\left[\frac{1}{\epsilon_\text{UV}}-\gamma_E+\ln\left(\frac{4\pi\mu^2}{m^2}\right)\right]-\frac{\left(m^2-s'-2 q^2\right) \left(m^2+s'-q^2\right)}{m^4-2 m^2 \left(s'+q^2\right)+\left(s'-q^2\right)^2} B_0\left(s',0,m^2\right)\right.\nonumber\\
&\left.+\frac{q^2
   \left(-m^2-3 s'+3 q^2\right) }{m^4-2 m^2 \left(s'+q^2\right)+\left(s'-q^2\right)^2}B_0\left(q^2,m^2,m^2\right)\right.\nonumber\\
&\left.+\frac{2 \left(q^2
   \left(s'^2-m^4\right)+\left(m^3-m s'\right)^2-2 q^4 \left(m^2+s'\right)+q^6\right)}{m^4-2 m^2
   \left(s'+q^2\right)+\left(s'-q^2\right)^2} C_0\left(m^2,q^2,s',0,m^2,m^2\right)\right.\nonumber\\
 &\left.+\frac{ \left(2 m^2 s'-2 \left(s'-q^2\right)^2\right)}{m^2 \left(m^4-2 m^2
   \left(s'+q^2\right)+\left(s'-q^2\right)^2\right)}A_0\left(m^2\right)+\frac{2 m^2 \left(m^2-s'-2 q^2\right)}{m^4-2 m^2 \left(s'+q^2\right)+\left(s'-q^2\right)^2}\right\},\label{F1-}
\end{align}

\begin{align}
F_{1^+}&(s',q^2)=\left(\frac{\alpha}{4\pi}\right)\left\{\left[\frac{1}{\epsilon_\text{UV}}-\gamma_E+\ln\left(\frac{4\pi\mu^2}{m^2}\right)\right] + \frac{3 q^2 \left(3 m^2+s'-q^2\right) }{m^4-2 m^2 \left(s'+q^2\right)+\left(s'-q^2\right)^2}B_0\left(q^2,m^2,m^2\right)\right.\nonumber\\
&\left.+\left(\frac{3 \left(m^2-s'\right)
   \left(m^2+s'-q^2\right)}{m^4-2 m^2 \left(s'+q^2\right)+\left(s'-q^2\right)^2}+2\right) B_0\left(s',0,m^2\right)\right.\nonumber\\
 &\left.+\left(4 m^2-\frac{6 \left(m^3-m
   s'\right)^2}{m^4-2 m^2 \left(s'+q^2\right)+\left(s'-q^2\right)^2}-2 q^2\right) C_0\left(m^2,q^2,s',0,m^2,m^2\right)\right.\nonumber\\
&\left.
  + \left(\frac{2}{m^2}+\frac{6 s'-6 m^2}{m^4-2 m^2 \left(s'+q^2\right)+\left(s'-q^2\right)^2}\right)A_0\left(m^2\right) +\frac{6 m^2 s'-6 m^4}{m^4-2 m^2
   \left(s'+q^2\right)+\left(s'-q^2\right)^2}\right\},\label{F1+}
\end{align}

\begin{align}
F_{2^-}&(s',q^2)=\left(\frac{\alpha}{4\pi}\right)\frac{1}{s' \left(m^4-2 m^2 \left(s'+q^2\right)+\left(s'-q^2\right)^2\right)^2}\nonumber\\
&\times\left\{\vphantom{\frac{1}{2}}4 m^2 s' q^2 \left(5 m^4+m^2 \left(2 s'+5 q^2\right)-\left(7 s'-4 q^2\right) \left(s'-q^2\right)\right) B_0\left(q^2,m^2,m^2\right)\right.\nonumber\\
&\left.+2 m^2 \left[\vphantom{\frac{1}{2}}q^6
   \left(m^2+5 s'\right)-q^4 \left(3 m^4+18 m^2 s'+11 s'^2\right)-\left(m^2-s'\right)^2 \left(m^4-5 s'^2\right)+q^2\times\right.\right.\nonumber\\
&\left.\left. \left(3 m^6+11 m^4 s'+17 m^2 s'^2+s'^3\right)\vphantom{\frac{1}{2}}\right] B_0\left(s',0,m^2\right)-4 m^2 s' \left[-q^6 \left(9 m^2+7 s'\right)+8 q^4 \left(m^2+s'\right)^2\right.\right.\nonumber\\
&\left.\left.+q^2
   \left(m^2-s'\right) \left(3 m^2+s'\right) \left(m^2+3 s'\right)+2 m^2 \left(m^2-s'\right)^3+2 q^8\right] C_0\left(m^2,q^2,s',0,m^2,m^2\right)\right.\nonumber\\
&\left.+2
   \left[\vphantom{\frac{1}{2}}m^8-m^6 \left(2 s'+3 q^2\right)+m^4 \left(-4 s'^2-21 s' q^2+3 q^4\right)+m^2 \left(10 s'^3-21 s'^2 q^2+8 s'
   q^4-q^6\right)\right.\right.\nonumber\\
&\left.\left.-s' \left(5 s'-3 q^2\right) \left(s'-q^2\right)^2\right]A_0\left(m^2\right) \right.\nonumber\\
&\left.-4 m^2 s' \left(m^6-m^4 \left(s'-9 q^2\right)-m^2 \left(s'^2-10
   s' q^2+3 q^4\right)+\left(s'-q^2\right)^3\right)\vphantom{\frac{1}{2}}\right\},\label{F2-}
\end{align}

\begin{align}
F_{2^+}&(s',q^2)=\left(\frac{\alpha}{4\pi}\right)\left\{\frac{4 m^2 q^2 \left(2 m^4+m^2 \left(5 q^2-4 s'\right)+\left(s'-q^2\right) \left(2 s'+q^2\right)\right)}{\left(m^4-2 m^2
   \left(s'+q^2\right)+\left(s'-q^2\right)^2\right)^2} B_0\left(q^2,m^2,m^2\right)\right.\nonumber\\
&\left.+\frac{2 m^2 \left(q^6 \left(m^2+s'\right)-q^4 \left(3 m^2-s'\right) \left(m^2+3 s'\right)+3 q^2
   \left(m^2-s'\right) \left(m^2+s'\right)^2-\left(m^2-s'\right)^4\right) }{s' \left(m^4-2 m^2
   \left(s'+q^2\right)+\left(s'-q^2\right)^2\right)^2}\right.\nonumber\\
&\times B_0\left(s',0,m^2\right)\left.-\frac{4 m^2 q^2 \left(m^2-s'\right) \left(5 m^4+2 m^2 \left(q^2-2
   s'\right)-\left(s'-q^2\right)^2\right)}{\left(m^4-2 m^2
   \left(s'+q^2\right)+\left(s'-q^2\right)^2\right)^2} C_0\left(m^2,q^2,s',0,m^2,m^2\right)\right.\nonumber\\
&\left.+\frac{2 \left(m^2-s'\right) \left(m^6-3 m^4 \left(s'+q^2\right)+m^2
   \left(s'-3 q^2\right) \left(3 s'-q^2\right)-\left(s'-q^2\right)^2 \left(s'+q^2\right)\right)}{s' \left(m^4-2 m^2
   \left(s'+q^2\right)+\left(s'-q^2\right)^2\right)^2}A_0\left(m^2\right) \right.\nonumber\\
&\left.+\frac{24 m^4 q^2 \left(s'-m^2\right)}{\left(m^4-2 m^2
   \left(s'+q^2\right)+\left(s'-q^2\right)^2\right)^2}\right\}.\label{F2+}
\end{align}
The corresponding expressions for the form factors $F_{3^+}$ and $F_{3^-}$, which do not contribute to the observables due to gauge invariance, are given in Appendix \ref{F3}.

\subsubsection{The on-shell limit and renormalization}
\label{onshell_limit}

In this section we show, that the form factors $F_{1^+}$, $F_{2^+}$ and $F_{3^+}$ reproduce the correct on-shell limit. The on-shell form factor can be decomposed into two structures, according to:
\begin{align}
\bar{u}(k^{\prime})\Gamma^{\mu}u(k)=\bar{u}(k^{\prime})\left[F_1(q^2)\gamma^{\mu}+iF_2(q^2)\sigma^{\mu\nu}\frac{q_{\nu}}{2m}\right]u(k).
\end{align} 
For $s' = m^2$, we find for the on-shell expressions of the form factors:

\begin{align}
F_{1^+}(m^2,q^2)=&\left(\frac{\alpha}{4\pi}\right)\left\{\left[\frac{1}{\epsilon_\text{UV}}-\gamma_E+\ln\left(\frac{4\pi\mu^2}{m^2}\right)\right]
+\frac{1}{m^2}\left[2 A_0(m^2)+2m^2 B_0(m^2,0,m^2)\right.\right.\nonumber\\
&\left.\left.-3 m^2 B_0(q^2,m^2,m^2)+(4 m^4-2 m^2 q^2)\;C_0(m^2,q^2,m^2,0,m^2,m^2)\right]\vphantom{\frac{1}{2}}\right\}\nonumber\\
=&\left(\frac{\alpha}{4\pi}\right)\left\{\left[\frac{1}{\epsilon_\text{UV}}-\gamma_E+\ln\left(\frac{4\pi\mu^2}{m^2}\right)\right]\right.\nonumber\\
&\left.+\left[\frac{1}{\epsilon_\text{IR}}-\gamma_E+\ln\left(\frac{4\pi\mu^2}{m^2}\right)\right]\frac{1+v^2}{v}\ln\left(\frac{v+1}{v-1}\right)\right.\nonumber\\
&\left.+\frac{1+v^2}{v}\left[\text{Li}_2\left(\frac{v+1}{2v}\right)-\text{Li}_2\left(\frac{v-1}{2v}\right) - \ln^2\left(\frac{v-1}{2v}\right)\right.\right.\nonumber\\
&\left.\left.+\ln\left(\frac{v-1}{2v}\right)\ln\left(\frac{v+1}{2v}\right)+\frac{1}{2}\ln^2\left(\frac{v-1}{v+1}\right)\right]-3v\ln\left(\frac{v-1}{v+1}\right)\right\}, \label{F1ponshell}\\
F_{2^+}(m^2,q^2)=&\left(\frac{\alpha}{4\pi}\right)\frac{4m^2}{4m^2-q^2}\left\{B_0(q^2,m^2,m^2)-B_0(m^2,0,m^2)\right\}\nonumber\\
=&\frac{\alpha}{4\pi}\frac{v^2-1}{v}\ln\left(\frac{v-1}{v+1}\right),\label{renormalized_QED_F2} \\
F_{3^+}(m^2,q^2)=&\left(\frac{\alpha}{4\pi}\right)\frac{4}{q^2}\left\{-m^2-A_0(m^2)+m^2 B_0(m^2,0,m^2)\right\}\nonumber\\
=&0,
\end{align}
where we defined 
\begin{equation}
v^2\equiv1-\frac{4m^2}{q^2}.\label{EqDef:v}
\end{equation}
$F_{1^+}(m^2,q^2)$ has an infrared divergence, which arises from $C_0(m^2,q^2,m^2,0,m^2,m^2)$.

We reproduce the on-shell Dirac form factor which is given by
\begin{align}
F_1(q^2)=F_{1^+}(m^2,q^2)+F_{2^+}(m^2,q)&=\left(\frac{\alpha}{4\pi}\right)\left\{\left[\frac{1}{\epsilon_\text{UV}}-\gamma_E+\ln\left(\frac{4\pi\mu^2}{m^2}\right)\right]\right.\nonumber\\
&\left.+\left[\frac{1}{\epsilon_\text{IR}}-\gamma_E+\ln\left(\frac{4\pi\mu^2}{m^2}\right)\right]\frac{1+v^2}{v}\ln\left(\frac{v+1}{v-1}\right)\right.\nonumber\\
&\left.+\frac{2v^2+1}{v}\ln\left(\frac{v+1}{v-1}\right)+\frac{v^2+1}{2v}\ln\left(\frac{v+1}{v-1}\right)\ln\left(\frac{v^2-1}{4v^2}\right)\right.\nonumber\\
&\left.+\frac{1+v^2}{v}\left[\text{Li}_2\left(\frac{v+1}{2v}\right)-\text{Li}_2\left(\frac{v-1}{2v}\right)\right]\right\}. \label{F1onshell}
\end{align}
Moreover, we reproduce the Schwinger correction to the electron magnetic moment given by
\begin{equation}
\kappa=-F_{2^+}(m^2,0)=\frac{\alpha}{2\pi}.
\end{equation}

In the on-shell subtraction scheme, the vertex counter term is defined to fix the electron charge $e$ at $q^2=0$.
Only the Dirac form factor $F_1(q^2)$ is UV divergent, and one finds at $q^2=0$ the renormalization constant:
\begin{align}
Z_1&=1-F_1(0)=\nonumber\\
&=1-\frac{\alpha}{4\pi}\left\{\left[\frac{1}{\epsilon_\text{UV}}-\gamma_E+\ln\left(\frac{4\pi\mu^2}{m^2}\right)\right]+2\left[\frac{1}{\epsilon_\text{IR}}-\gamma_E+\ln\left(\frac{4\pi\mu^2}{m^2}\right)\right]+4\right\}. \label{onshell_vertex_counterterm}
\end{align}
This leads to the renormalized (on-shell) form factor:
\begin{equation}
\tilde{F}_1(q^2)=F_1(q^2)-F_1(0). \label{renormalized_QED_F1}
\end{equation}

\subsubsection{Ward-Takahashi identity}
As a further check of our one-loop expression for the half-off-shell vertex we check the Ward-Takahashi identity, which reads:
\ber
 q_\mu \Gamma^\mu \left(k',~k\right) = - \left(\vphantom{\frac{1}{2}}\Sigma(k') - \Sigma(k) \right),\label{WardIdentity}
 \eer
where $\Sigma(k)$ denotes the Fermion propagator.
Contracting the vertex of Eq. \eqref{VertexDec} with the photon momentum $q_\mu$, we find:
\ber
q_\mu \Gamma^\mu &=& \Lambda^{(+)} \left[ F_{1^+} \left( \sla{k}' - m \right) +\left(\frac{s^\prime-m^2}{2m^2}\right)m F_{2^+}   -\frac{q^2}{2m}  F_{3^+}  \right] \nonumber \\
&+& \Lambda^{(-)} \left[ F_{1^-} \left( \sla{k}' - m \right) + \left(\frac{s^\prime-m^2}{2m^2}\right) m F_{2^-} - \frac{q^2}{2m} F_{3^-}  \right], \\
q_\mu \Gamma^\mu&=& \frac{\sla{k}'}{2} \left[ \left(\frac{s^\prime-m^2}{2m^2}\right) \left( F_{2^+} + F_{2^-} \right) +  \frac{-q^2}{2m^2} \left( F_{3^+} + F_{3^-} \right)   - 2 F_{1^-} \right] -\frac{m}{2} \left[ F_{1^+} - F_{1^-} \right] \nonumber \\
&+&\frac{m}{2} \left[ \left(\frac{s^\prime-m^2}{2m^2}\right) \left( F_{2^+} - F_{2^-} \right)+ \frac{-q^2}{2m^2} \left( F_{3^+} - F_{3^-} \right) + \frac{s^\prime}{m^2}\left( F_{1^+} + F_{1^-} \right) \right].
\eer
Plugging in the form factors (Eqs. \eqref{F1-}-\eqref{F2+},~\eqref{F3+},~\eqref{F3-}), we find:
\begin{align}
q_\mu \Gamma^\mu \left(k',~k\right) &=\sla{k}^{\prime} \frac{\alpha}{4\pi}\left\{\left[\frac{1}{\epsilon_\text{UV}}-\gamma_E+\ln\left(4\pi\right)\right]-1+\frac{s^\prime+m^2}{s^\prime}B_0(s',0, m^2)-\frac{1}{s^\prime}A_0(m^2)\right\}\nonumber\\
&+m\frac{\alpha}{4\pi}\left\{-\left[\frac{1}{\epsilon_\text{UV}}-\gamma_E+\ln\left(4\pi\right)\right]+3+\frac{3}{ m^2}A_0(m^2)- 4\;B_0(s',0, m^2)\right\}.
\end{align}
This is indeed equivalent to the right-hand side of Eq. \eqref{WardIdentity}:
\begin{equation}
\Sigma(k) - \Sigma(k') ,
\end{equation}
considering the one-loop self-energy of the fermion, as can be seen by using Eqs. \eqref{SEOffshell} and \eqref{SEOnshell}.

\subsubsection{Vertex diagrams}
\begin{figure}
	\includegraphics[scale=0.6]{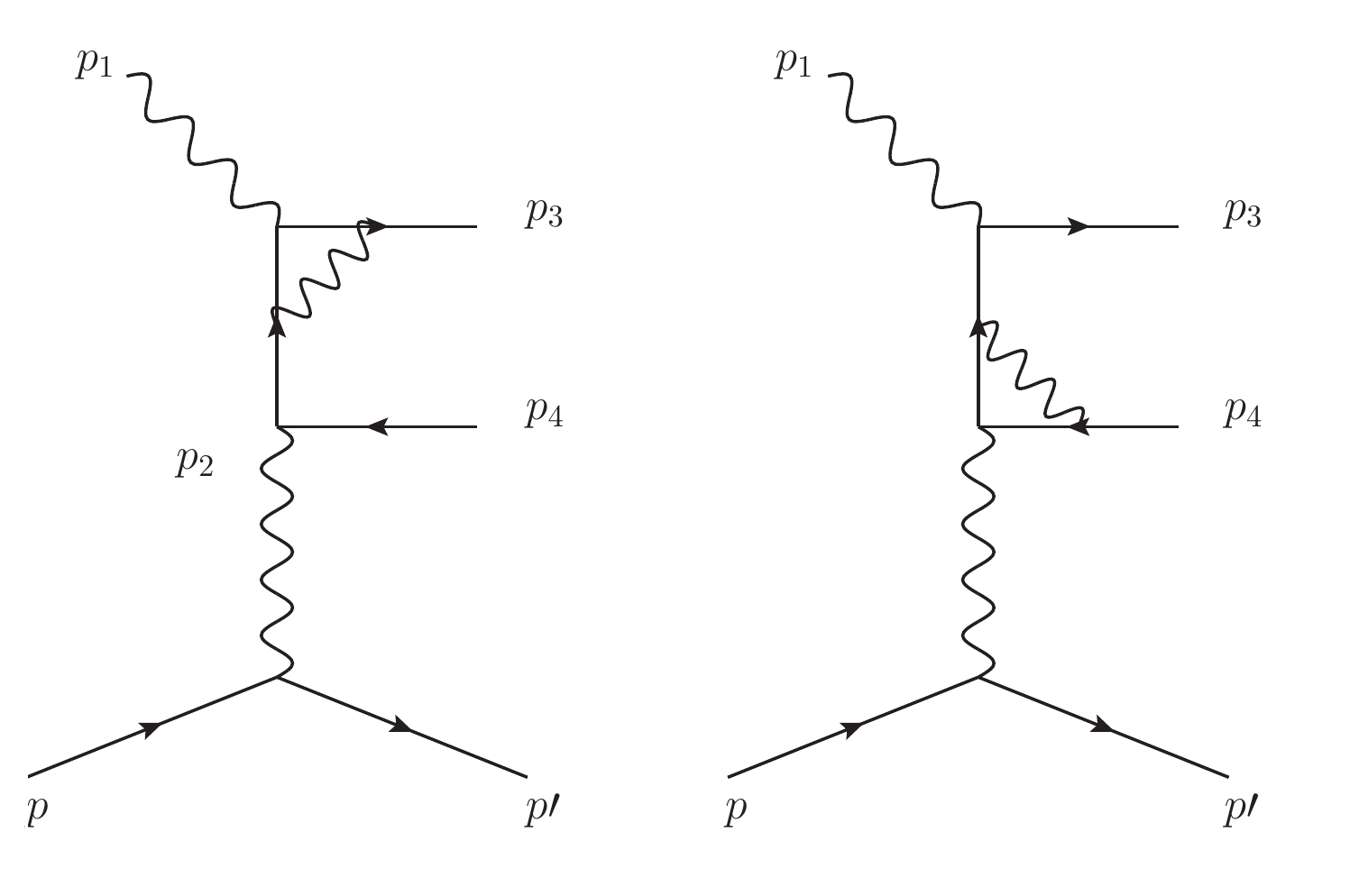}\hspace{0.cm}
	\caption{Vertex diagrams contributing to the Bethe-Heitler process. \label{VertexDiagrams}}
\end{figure}

Having determined the half-off shell one-loop expressions for the vertex, we can now evaluate the the two vertex diagrams contributing to the Bethe-Heitler process, as shown in Fig. \ref{VertexDiagrams}. The first diagram of Fig. \ref{VertexDiagrams} is given by
\begin{align}
\mathcal{M}_{\text{V}_1}&=\bar{u}(p_3)(ie)^2i\left\{\left[F_{1^+}(t_{ll},0)\gamma^{\mu}+F_{2^+}(t_{ll},0)\frac{(-p_1+2\;p_3)^{\mu}}{2m}\right]\left[\frac{1}{2m}+\frac{\sla{p_3}-\sla{p_1}+m}{(p_3-p_1)^2-m^2}\right]\right.\nonumber\\
&\left.+\frac{1}{2m}\left[F_{1^-}(t_{ll},0)\gamma^{\mu}+F_{2^-}(t_{ll},0)\frac{(-p_1+2\;p_3)^{\mu}}{2m}\right]\right\}\gamma^{\nu} v(p_4)\nonumber\\
&\times\frac{-i}{t}\varepsilon_\mu(p_1)\bar{u}(p^{\prime})(-ie)\Gamma_{\nu}(t)u(p).
\end{align}
Comparing this to the expression of $\mathcal{M}_0$, given by Eq. \eqref{M0}, $3$ additional Lorentz structures appear.
The interference of the first vertex diagram with the two tree level diagrams is given by
\begin{align}
\bar{\sum_i}\sum_f\mathcal{M}^*_{0}(\mathcal{M}_{\text{V}_1})&=\frac{e^4}{t^2}\sum_{i=1}^4 a_{\text{V}_1}^{(i)}L_{V_1}^{i;\mu\nu}H_{\mu\nu},
\end{align}
where 
\begin{align}
a_{\text{V}_1}^{(1)}&=F_{1^+}(t_{ll},0),\\
a_{\text{V}_1}^{(2)}&=\frac{F_{1^+}(t_{ll},0)+F_{1^-}(t_{ll},0)}{2},\\
a_{\text{V}_1}^{(3)}&=F_{2^+}(t_{ll},0),\\
a_{\text{V}_1}^{(4)}&=\frac{F_{2^+}(t_{ll},0)+F_{2^-}(t_{ll},0)}{2},
\end{align}
and
\begin{align}
L^{1;\mu\nu}_{\text{V}_1}&=-\frac{1}{2}\text{Tr}\left[(\sla{p_4}-m)\left(\gamma^{\mu}\frac{(\sla{k_1}+m)}{k_1^2-m^2}\gamma_{\alpha}+\gamma_{\alpha}\frac{(\sla{k_2}+m)}{k_2^2-m^2}\gamma^{\mu}\right)(\sla{p_3}+m)\;\gamma^{\alpha}\frac{(\sla{k_1}+m)}{k_1^2-m^2}\gamma^{\nu}\right],\nonumber\\
L^{2;\mu\nu}_{\text{V}_1}&=-\frac{1}{2}\text{Tr}\left[(\sla{p_4}-m)\left(\gamma^{\mu}\frac{(\sla{k_1}+m)}{k_1^2-m^2}\gamma_{\alpha}+\gamma_{\alpha}\frac{(\sla{k_2}+m)}{k_2^2-m^2}\gamma^{\mu}\right)(\sla{p_3}+m)\;\gamma^{\alpha}\frac{1}{m}\gamma^{\nu}\right],\nonumber\\
L^{3;\mu\nu}_{\text{V}_1}&=-\frac{1}{2}\text{Tr}\left[(\sla{p_4}-m)\left(\gamma^{\mu}\frac{(\sla{k_1}+m)}{k_1^2-m^2}\gamma_{\alpha}+\gamma_{\alpha}\frac{(\sla{k_2}+m)}{k_2^2-m^2}\gamma^{\mu}\right)(\sla{p_3}+m)\;\frac{p_3^{\alpha}}{m}\frac{(\sla{k_1}+m)}{k_1^2-m^2}\gamma^{\nu}\right],\nonumber\\
L^{4;\mu\nu}_{\text{V}_1}&=-\frac{1}{2}\text{Tr}\left[(\sla{p_4}-m)\left(\gamma^{\mu}\frac{(\sla{k_1}+m)}{k_1^2-m^2}\gamma_{\alpha}+\gamma_{\alpha}\frac{(\sla{k_2}+m)}{k_2^2-m^2}\gamma^{\mu}\right)(\sla{p_3}+m)\;\frac{p_3^{\alpha}}{m}\frac{1}{m}\gamma^{\nu}\right].
\end{align}
Evaluating these $4$ traces, we find for the first vertex diagram:
\begin{align}
\overline{\sum_i}\sum_f\mathcal{M}^*_{0}\left(\mathcal{M}_{\text{V}_1}\right)&=\frac{e^4}{t^2}\sum_{i=1}^4 a^{\text{V}_1}_iT^{\text{V}_1}_i,
\end{align}
where
\begin{align}
T^{\text{V}_1}_1&=L_{0,d}^{\mu\nu}H_{\mu\nu},\\
T^{\text{V}_1}_2&= T^{\text{SE}}_2,\\
T^{\text{V}_1}_3&=\frac{2\; H_1}{(m^2-t_{ll})^2(-s_{ll}-t_{ll}+m^2+t)}\left\{4 m^6-2 t m^4+s_{ll} m^4-8 t_{ll} m^4-3 t^2 m^2+4 t_{ll}^2 m^2\nonumber\right.\\
&\left.+t s_{ll} m^2-8 t t_{ll} m^2+6 s_{ll} t_{ll} m^2+2 t t_{ll}^2+s_{ll} t_{ll}^2-t^2 t_{ll}+3 t s_{ll} t_{ll}\right\}\nonumber\\
&+\frac{H_2}{4(m^2-t_{ll})^2(-s_{ll}-t_{ll}+m^2+t)}\left\{18 M^2 m^6-18 s m^6+36 t m^6-18 s_{ll} m^6+22 M^4 m^4\right.\nonumber\\
&\left.+10 s^2 m^4-18 t^2 m^4+s_{ll}^2 m^4-32 M^2 s m^4+81 M^2 t m^4-35 s t m^4-12 M^2 u m^4+12 s u m^4\nonumber\right.\\
&\left.-54 t u m^4-59 M^2 s_{ll} m^4+29 s s_{ll} m^4+7 t s_{ll} m^4+30 u s_{ll} m^4-18 M^2 t_{ll} m^4+18 s t_{ll} m^4\nonumber\right.\\
&\left.-30 t t_{ll} m^4-15 M^2 t^2 m^2+12 s t^2 m^2+20 t u^2 m^2+2 M^2 s_{ll}^2 m^2-s s_{ll}^2 m^2-u s_{ll}^2 m^2-2 M^2 t_{ll}^2 m^2\nonumber\right.\\
&\left.+2 s t_{ll}^2 m^2+12 t t_{ll}^2 m^2+2 s_{ll} t_{ll}^2 m^2+62 M^4 t m^2+12 s^2 t m^2-54 M^2 s t m^2+15 t^2 u m^2-70 M^2 t u m^2\nonumber\right.\\
&\left.+30 s t u m^2-44 M^4 s_{ll} m^2-10 s^2 s_{ll} m^2-12 u^2 s_{ll} m^2+42 M^2 s s_{ll} m^2+9 M^2 t s_{ll} m^2-7 s t s_{ll} m^2\nonumber\right.\\
&\left.+46 M^2 u s_{ll} m^2-22 s u s_{ll} m^2-6 t u s_{ll} m^2-12 M^4 t_{ll} m^2-4 s^2 t_{ll} m^2+t^2 t_{ll} m^2-s_{ll}^2 t_{ll} m^2\nonumber\right.\\
&\left.+16 M^2 s t_{ll} m^2-20 M^2 t t_{ll} m^2+8 s t t_{ll} m^2+8 M^2 u t_{ll} m^2-8 s u t_{ll} m^2+28 t u t_{ll} m^2-8 M^2 s_{ll} t_{ll} m^2\nonumber\right.\\
&\left.+4 s s_{ll} t_{ll} m^2+12 t s_{ll} t_{ll} m^2+4 u s_{ll} t_{ll} m^2+2 M^2 t_{ll}^3-2 s t_{ll}^3-2 t t_{ll}^3-10 M^4 t_{ll}^2-6 s^2 t_{ll}^2+t^2 t_{ll}^2\nonumber\right.\\
&\left.+16 M^2 s t_{ll}^2+3 M^2 t t_{ll}^2-5 s t t_{ll}^2+4 M^2 u t_{ll}^2-4 s u t_{ll}^2-6 t u t_{ll}^2+3 M^2 s_{ll} t_{ll}^2-s s_{ll} t_{ll}^2-3 t s_{ll} t_{ll}^2\nonumber\right.\\
&\left.-2 u s_{ll} t_{ll}^2-M^2 t^2 t_{ll}+4 s t^2 t_{ll}-4 t u^2 t_{ll}-2 M^2 s_{ll}^2 t_{ll}+s s_{ll}^2 t_{ll}+u s_{ll}^2 t_{ll}+2 M^4 t t_{ll}+4 s^2 t t_{ll}\nonumber\right.\\
&\left.-10 M^2 s t t_{ll}+t^2 u t_{ll}+6 M^2 t u t_{ll}+2 s t u t_{ll}-20 M^4 s_{ll} t_{ll}-6 s^2 s_{ll} t_{ll}-4 u^2 s_{ll} t_{ll}+22 M^2 s s_{ll} t_{ll}\nonumber\right.\\
&\left.+7 M^2 t s_{ll} t_{ll}-9 s t s_{ll} t_{ll}+18 M^2 u s_{ll} t_{ll}-10 s u s_{ll} t_{ll}-10 t u s_{ll} t_{ll}\right\},\\
T^{\text{V}_1}_4&=\frac{-2 H_1}{m^2(m^2-t_{ll})(-s_{ll}-t_{ll}+m^2+t)}\left\{2 m^6-3 t m^4+2 s_{ll} m^4-4 t_{ll}
   m^4-t^2 m^2+2 t_{ll}^2 m^2\nonumber\right.\\
&\left.-t t_{ll} m^2+s_{ll} t_{ll} m^2+s_{ll} t_{ll}^2+s_{ll}^2 t_{ll}\right\}\nonumber\\
&+\frac{H_2}{4m^2(m^2-t_{ll})(-s_{ll}-t_{ll}+m^2+t)} \left\{-9 M^2 m^6+9 s m^6-18 t m^6+9 s_{ll} m^6-10 M^4 m^4\nonumber\right.\\
&\left.-4 s^2 m^4+8 t^2 m^4-s_{ll}^2 m^4+14 M^2 s m^4-41 M^2 t m^4+14 s t m^4+6 M^2 u m^4-6 s u m^4\nonumber\right.\\
&\left.+27 t u m^4+33 M^2 s_{ll} m^4-14 s s_{ll} m^4-2 t s_{ll} m^4-15 u s_{ll} m^4+9 M^2 t_{ll} m^4-9 s t_{ll} m^4+15 t t_{ll} m^4\nonumber\right.\\
&\left.+6 M^2 t^2 m^2-4 s t^2 m^2-10 t u^2 m^2-2 M^2 s_{ll}^2 m^2+s s_{ll}^2 m^2+u s_{ll}^2 m^2+M^2 t_{ll}^2 m^2-s t_{ll}^2 m^2\nonumber\right.\\
&\left.-6 t t_{ll}^2 m^2-s_{ll} t_{ll}^2 m^2-26 M^4 t m^2-4 s^2 t m^2+20 M^2 s t m^2-6 t^2 u m^2+32 M^2 t u m^2-12 s t u m^2\nonumber\right.\\
&\left.+20 M^4 s_{ll} m^2+4 s^2 s_{ll} m^2+6 u^2 s_{ll} m^2-18 M^2 s s_{ll} m^2-3 M^2 t s_{ll} m^2+2 s t s_{ll} m^2-22 M^2 u s_{ll} m^2\nonumber\right.\\
&\left.+10 s u s_{ll} m^2+t u s_{ll} m^2+8 M^4 t_{ll} m^2+4 s^2 t_{ll} m^2-2 t^2 t_{ll} m^2-s_{ll}^2 t_{ll} m^2-12 M^2 s t_{ll} m^2\nonumber\right.\\
&\left.+16 M^2 t t_{ll} m^2-2 s t t_{ll} m^2-4 M^2 u t_{ll} m^2+4 s u t_{ll} m^2-14 t u t_{ll} m^2-6 M^2 s_{ll} t_{ll} m^2-3 t s_{ll} t_{ll} m^2\nonumber\right.\\
&\left.-2 u s_{ll} t_{ll} m^2-M^2 t_{ll}^3+s t_{ll}^3+t t_{ll}^3+2 M^4 t_{ll}^2-2 M^2 s t_{ll}^2-3 M^2 t t_{ll}^2-2 M^2 u t_{ll}^2+2 s u t_{ll}^2+3 t u t_{ll}^2\nonumber\right.\\
&\left.+M^2 s_{ll} t_{ll}^2+2 s s_{ll} t_{ll}^2+t s_{ll} t_{ll}^2+u s_{ll} t_{ll}^2+2 t u^2 t_{ll}+2 M^2 s_{ll}^2 t_{ll}+s s_{ll}^2 t_{ll}+u s_{ll}^2 t_{ll}+2 M^4 t t_{ll}\nonumber\right.\\
&\left.-4 M^2 t u t_{ll}+4 M^4 s_{ll} t_{ll}+2 u^2 s_{ll} t_{ll}-2 M^2 s s_{ll} t_{ll}-3 M^2 t s_{ll} t_{ll}-6 M^2 u s_{ll} t_{ll}+2 s u s_{ll} t_{ll}\nonumber\right.\\
&\left.+3 t u s_{ll} t_{ll}\right\}.
\end{align}

The second diagram of Fig. \ref{VertexDiagrams} is given by
\begin{align}
\mathcal{M}_{\text{V}_2}&=\bar{u}(p_3)(ie)^2\gamma^{\mu}i\left\{\left[\frac{1}{2m}+\frac{\sla{p_3}-\sla{p_1}+m}{(p_3-p_1)^2-m^2}\right]\left[F_{1^+}(t_{ll},t)\gamma^{\nu}+F_{2^+}(t_{ll},t)\frac{(p_2-\;2 p_4)^{\nu}}{2m}\right]\right.\nonumber\\
&\left.+\frac{1}{2m}\left[F_{1^-}(t_{ll},t)\gamma^{\nu}+F_{2^-}(t_{ll},t)\frac{(p_2 -2\;p_4 )^{\nu}}{2m}\right]\right\} v(p_4)\times (-i)\varepsilon_\mu(p_1)\bar{u}(p^{\prime})(-ie)\Gamma_{\nu}(t)u(p). \label{second_vertex}
\end{align}
The interference of the second vertex diagram with the two Born diagrams is given by
\begin{align}
\bar{\sum_i}\sum_f\mathcal{M}^*_{0}(\mathcal{M}_{\text{V}_2})&=\frac{e^4}{t^2}\sum_{i=1}^4 a^{\text{V}_2}_iL_{V_2}^{i;\mu\nu}H_{\mu\nu},
\end{align}
where 
\begin{align}
a_{\text{V}_2}^{(1)}&=F_{1^+}(t_{ll},t),\\
a_{\text{V}_2}^{(2)}&=\frac{F_{1^+}(t_{ll},t)+F_{1^-}(t_{ll},t)}{2},\\
a_{\text{V}_2}^{(3)}&=F_{2^+}(t_{ll},t),\\
a_{\text{V}_2}^{(4)}&=\frac{F_{2^+}(t_{ll},t)+F_{2^-}(t_{ll},t)}{2},
\end{align}
and
\begin{align}
L^{1;\mu\nu}_{\text{V}_2}&=-\frac{1}{2}\text{Tr}\left[(\sla{p_4}-m)\left(\gamma^{\mu}\frac{(\sla{k_1}+m)}{k_1^2-m^2}\gamma_{\alpha}+\gamma_{\alpha}\frac{(\sla{k_2}+m)}{k_2^2-m^2}\gamma^{\mu}\right)(\sla{p_3}+m)\;\gamma^{\alpha}\frac{(\sla{k_1}+m)}{k_1^2-m^2}\gamma^{\nu}\right],\nonumber\\
L^{2;\mu\nu}_{\text{V}_2}&=-\frac{1}{2}\text{Tr}\left[(\sla{p_4}-m)\left(\gamma^{\mu}\frac{(\sla{k_1}+m)}{k_1^2-m^2}\gamma_{\alpha}+\gamma_{\alpha}\frac{(\sla{k_2}+m)}{k_2^2-m^2}\gamma^{\mu}\right)(\sla{p_3}+m)\;\gamma^{\alpha}\frac{1}{m}\gamma^{\nu}\right],\nonumber\\
L^{3;\mu\nu}_{\text{V}_2}&=-\frac{1}{2}\text{Tr}\left[(\sla{p_4}-m)\left(\gamma^{\mu}\frac{(\sla{k_1}+m)}{k_1^2-m^2}\gamma_{\alpha}+\gamma_{\alpha}\frac{(\sla{k_2}+m)}{k_2^2-m^2}\gamma^{\mu}\right)(\sla{p_3}+m)\gamma^{\alpha}\frac{(\sla{k_1}+m)}{k_1^2-m^2}\;\frac{(-p_4)^{\nu}}{m}\right],\nonumber\\
L^{4;\mu\nu}_{\text{V}_2}&=-\frac{1}{2}\text{Tr}\left[(\sla{p_4}-m)\left(\gamma^{\mu}\frac{(\sla{k_1}+m)}{k_1^2-m^2}\gamma_{\alpha}+\gamma_{\alpha}\frac{(\sla{k_2}+m)}{k_2^2-m^2}\gamma^{\mu}\right)(\sla{p_3}+m)\gamma^{\alpha}\;\frac{1}{m}\frac{(-p_4)^{\nu}}{m}\right].
\end{align}
These lepton tensors enter the calculation like before:
\begin{align}
\overline{\sum_i}\sum_f\mathcal{M}^*_{0}\left(\mathcal{M}_{\text{V}_2}\right)&=\frac{e^4}{t^2}\sum_{i=1}^4 a^{\text{V}_2}_iT^{\text{V}_2}_i,
\end{align}
where
\begin{align}
T^{\text{V}_2}_1&=T^{\text{V}_1}_1,\\\
T^{\text{V}_2}_2&=T^{\text{V}_1}_2,\\
T^{\text{V}_2}_3&=\frac{2 H_1}{\left(m^2-t_{ll}\right){}^2 \left(-s_{ll}-t_{ll}+m^2+t\right)} \left\{m^4 s_{ll}-8 m^4 t_{ll}-m^2 t s_{ll}+6 m^2 s_{ll} t_{ll}+4 m^2 t_{ll}^2\nonumber\right.\\
&\left.-4 m^2 t t_{ll}+s_{ll} t_{ll}^2-t s_{ll} t_{ll}+4 m^6-4 m^4 t+2 m^2 t^2\right\}\nonumber\\
&+\frac{H_2}{2\left(m^2-t_{ll}\right){}^2 \left(-s_{ll}-t_{ll}+m^2+t\right)}  \left\{-t_{ll}+3 m^2+4 M^2-2 s-t-2 u
\right\}\left\{-3 m^4 s_{ll}\nonumber\right.\\
&\left.-6 m^2 M^2 s_{ll}-2 m^2 M^2 t_{ll}+m^2 t s_{ll}+2 m^2 s t_{ll}-m^2 s_{ll} t_{ll}+3 m^2 u s_{ll}+3 m^2 s s_{ll}-3 m^2 t t_{ll}\nonumber\right.\\
&\left.-2 M^2 s_{ll} t_{ll}-M^2 t_{ll}^2-M^2 t t_{ll}+u s_{ll} t_{ll}+s t_{ll}^2+s s_{ll} t_{ll}+t s_{ll} t_{ll}+t u t_{ll}+t t_{ll}^2+3 m^4 M^2\nonumber\right.\\
&\left.-3 m^4 s+6 m^4 t+9 m^2 M^2 t-4 m^2 s t-2 m^2 t^2-5 m^2 t u\right\},\\
T^{\text{V}_2}_4&=\frac{-2 H_1}{m^2 \left(m^2-t_{ll}\right) \left(-s_{ll}-t_{ll}+m^2+t\right)} \left\{2 m^4 s_{ll}-4 m^4 t_{ll}-2 m^2 t s_{ll}+m^2 s_{ll} t_{ll}+2 m^2 t_{ll}^2\nonumber\right.\\
&\left.-3 m^2 t t_{ll}+s_{ll} t_{ll}^2+s_{ll}^2 t_{ll}-t s_{ll} t_{ll}+2 m^6-m^4 t+2 m^2 t^2\right\}\nonumber\\
&+\frac{-H_2}{4m^2 \left(m^2-t_{ll}\right) \left(-s_{ll}-t_{ll}+m^2+t\right)} \left\{-t_{ll}+3 m^2+4 M^2-2 s-t-2 u\right\} \left\{-4 m^4 s_{ll}\nonumber\right.\\
&\left.-7 m^2 M^2 s_{ll}-4 m^2 M^2 t_{ll}+5 m^2 t s_{ll}+4 m^2 s t_{ll}+4 m^2 u s_{ll}-m^2 s_{ll}^2+3 m^2 s s_{ll}+3 M^2 t s_{ll}\nonumber\right.\\
&\left.-M^2 s_{ll} t_{ll}-2 M^2 s_{ll}^2-M^2 t t_{ll}-2 t u s_{ll}-s t s_{ll}+s t t_{ll}+s s_{ll} t_{ll}-t s_{ll} t_{ll}+u s_{ll}^2+s s_{ll}^2+t^2 t_{ll}\nonumber\right.\\
&\left.+4 m^4 M^2-4 m^4 s+4 m^4 t+9 m^2 M^2 t-5 m^2 s t-4 m^2 t^2-4 m^2 t u-M^2 t^2+t^2 u\right\}.
\end{align}
The contributions of the crossed vertex diagrams can be calculated by using the crossing relations derived in Sec. \ref{SecCrossing}. Note that the replacement $t_{ll}\rightarrow u_{ll}$ also affects the scalar form factors.

\subsubsection{Leading contribution for small lepton masses}
Taking the limit of small lepton masses $m^2\rightarrow 0$, keeping only terms with either double-logarithmic dependence or proportional to $\ln(m^2/s_{ll})$, we find that we can rewrite the interference of the sum of all $4$ vertex diagrams with the tree-level diagrams, as:

\begin{align}
\overline{\sum_i}\sum_f\mathcal{M}^*_{0}\left(\mathcal{M}_{\text{Vertex}}\right)&\approx \frac{e^4}{t^2}\left(\frac{\alpha}{\pi}\right)\left\{-L_0^{\mu\nu}H_{\mu\nu}\left[\ln\left(\frac{4\pi \mu^2}{s_{ll}}\right)+\frac{1}{\epsilon_{\text{IR}}}-\gamma_E\right]+(B+\tilde{B})\;\ln\left(\frac{m^2}{s_{ll}}\right)\nonumber\right.\\
&\left.+C\left[\ln^2\left(\frac{-t_{ll}}{s_{ll}}\right)-2\ln\left(\frac{-t_{ll}}{s_{ll}}\right)\ln\left(\frac{m^2}{s_{ll}}\right)\right]\nonumber\right.\\
&\left.+\tilde{C}\left[\ln^2\left(\frac{-u_{ll}}{s_{ll}}\right)-2\ln\left(\frac{-u_{ll}}{s_{ll}}\right)\ln\left(\frac{m^2}{s_{ll}}\right)\right]\nonumber\right.\\
&\left.-(C+\tilde{C})\left[\ln^2\left(\frac{-t}{s_{ll}}\right)-2\ln\left(\frac{-t}{s_{ll}}\right)\ln\left(\frac{m^2}{s_{ll}}\right)\right]\right\},
\label{VertexApp}\end{align}
where

\begin{align}
B&=\frac{2H_1}{(t-t_{ll})t_{ll}(-s_{ll}+t-t_{ll})}\left\{-t^2 s_{ll}+2 t s_{ll}^2+5 t s_{ll} t_{ll}-4 s_{ll} t_{ll}^2-3 s_{ll}^2 t_{ll}-3 t^2
   t_{ll}+3 t t_{ll}^2-t_{ll}^3+t^3\right\}\nonumber\\
&-\frac{H_2}{4(t-t_{ll})(-s_{ll}+t-t_{ll})t_{ll}}\left\{12 M^4 t s_{ll}-4 M^4 s_{ll} t_{ll}-8 M^4 s_{ll}^2-20 M^4 t t_{ll}-7 M^2 t^2 s_{ll}\right.\nonumber\\
&\left.-18 M^2 t u s_{ll}+6 M^2 u s_{ll} t_{ll}+6 M^2 t s_{ll}^2-2 M^2 s t_{ll}^2-7 M^2 s_{ll} t_{ll}^2-6 M^2 s t s_{ll}-2 M^2 s_{ll}^2 t_{ll}\right.\nonumber\\
&\left.+22 M^2 s t t_{ll}+2 M^2 s s_{ll} t_{ll}+6 M^2 t s_{ll} t_{ll}+8 M^2 u s_{ll}^2+8 M^2 s s_{ll}^2-7 M^2 t^2 t_{ll}+2 M^2 u t_{ll}^2\right.\nonumber\\
&\left.+18 M^2 t u t_{ll}-3 M^2 t_{ll}^3+20 M^2 t t_{ll}^2+2 s^2 t_{ll}^2-8 s^2 t t_{ll}-2 s^2 s_{ll}^2+5 t^2 u s_{ll}-2 s t^2 s_{ll}-6 s t^2 t_{ll}\right.\nonumber\\
&\left.+t^2 s_{ll} t_{ll}+6 t u^2 s_{ll}-2 u^2 s_{ll} t_{ll}+t u s_{ll}^2-2 s u t_{ll}^2-u s_{ll} t_{ll}^2+6 s t u s_{ll}-3 u s_{ll}^2 t_{ll}-6 s t u t_{ll}-2 s u s_{ll} t_{ll}\right.\nonumber\\
&\left.+2 t u s_{ll} t_{ll}-s t_{ll}^3+s t s_{ll}^2-s t t_{ll}^2-4 s s_{ll} t_{ll}^2+t s_{ll} t_{ll}^2-3 s s_{ll}^2 t_{ll}+8 s t s_{ll} t_{ll}-2 u^2 s_{ll}^2-4 s u s_{ll}^2+2 t^3 t_{ll}\right.\nonumber\\
&\left.+t^2 u t_{ll}-3 t^2 t_{ll}^2-6 t u^2 t_{ll}-7 t u t_{ll}^2-t t_{ll}^3+12 M^4 t^2-12 M^2 s t^2-2 M^2 t^3-12 M^2 t^2 u+4 s^2 t^2\right.\nonumber\\
&\left.+4 s t^3+4 s t^2 u+2 t^3 u+4 t^2 u^2\right\}, \\
C&=\frac{H_1 t}{(t-t_{ll})^2(-s_{ll}+t-t_{ll})t_{ll}}\left\{-t^2 s_{ll}+t s_{ll}^2+2 t s_{ll} t_{ll}-s_{ll} t_{ll}^2-3 t^2 t_{ll}+3 t t_{ll}^2-t_{ll}^3+t^3\right\}\nonumber\\
&-\frac{H_2 t}{4(t-t_{ll})^2(-s_{ll}+t-t_{ll})t_{ll}}\left\{-8 M^4 t s_{ll}+4 M^4 s_{ll} t_{ll}+4 M^4 s_{ll}^2+M^4 t_{ll}^2-6 M^4 t t_{ll}+M^2 t^2 s_{ll}\right.\nonumber\\
&\left.+8 M^2 t u s_{ll}-2 M^2 u s_{ll} t_{ll}-2 M^2 s t_{ll}^2-4 M^2 s_{ll} t_{ll}^2+8 M^2 s t s_{ll}-2 M^2 s_{ll}^2 t_{ll}+6 M^2 s t t_{ll}\right.\nonumber\\
&\left.-6 M^2 s s_{ll} t_{ll}+7 M^2 t s_{ll} t_{ll}-4 M^2 u s_{ll}^2-4 M^2 s s_{ll}^2-8 M^2 t^2 t_{ll}+6 M^2 t u t_{ll}-2 M^2 t_{ll}^3+8 M^2 t t_{ll}^2\right.\nonumber\\
&\left.+s^2 t_{ll}^2-2 s^2 t s_{ll}-2 s^2 t t_{ll}+2 s^2 s_{ll} t_{ll}+s^2 s_{ll}^2-t^2 u s_{ll}-2 s t^2 s_{ll}-t^2 s_{ll} t_{ll}-2 t u^2 s_{ll}+t u s_{ll}^2\right.\nonumber\\
&\left.-4 s t u s_{ll}-2 s t u t_{ll}+2 s u s_{ll} t_{ll}-2 t u s_{ll} t_{ll}+s t  s_{ll}^2+s t s_{ll} t_{ll}+u^2 s_{ll}^2+2 s u s_{ll}^2+2 t^3 t_{ll}+2 t^2 u t_{ll}\right.\nonumber\\
&\left.-t^2 t_{ll}^2-2 t u^2 t_{ll}-2 t u t_{ll}^2+9 M^4 t^2-8 M^2 s t^2-2 M^2 t^3-10 M^2 t^2 u+2 s^2 t^2+2 s t^3+4 s t^2 u\right.\nonumber\\
&\left.+2 t^3 u+3 t^2 u^2\right\},
\end{align}
$\tilde{B}$ ($\tilde{C}$), are given by making the replacements $t_{ll}\rightarrow u_{ll}$ and $u \rightarrow -u - s - t + 3M^2$ in $B$ ($C$), respectively.

\subsection{Vacuum polarization}
\subsubsection{Vacuum polarization at first order}
\begin{figure}
	\includegraphics[scale=0.7]{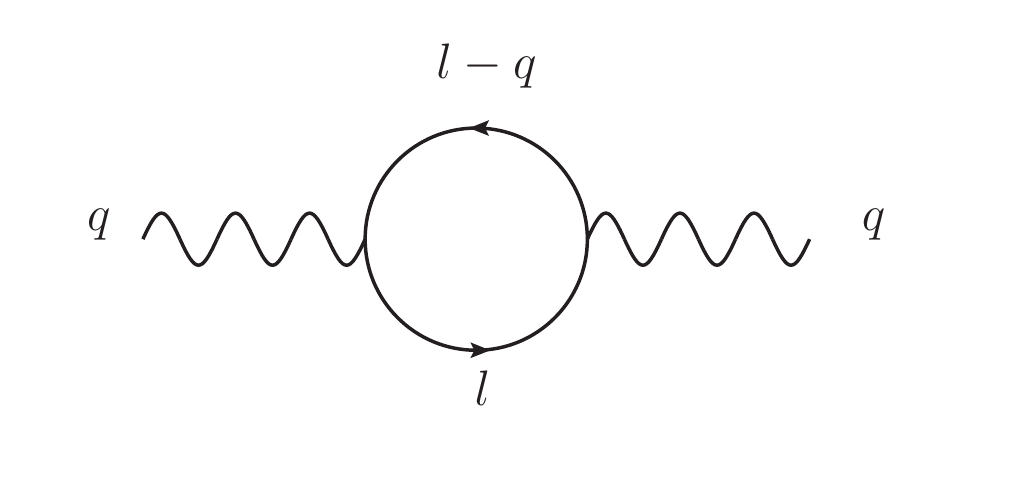}\hspace{0.cm}
	\caption{Vacuum polarization diagram. The fermion loop can be either electrons or muons.\label{Vacuumdiagram}}
\end{figure}
We show the first-order vacuum polarization diagram in Fig. \ref{Vacuumdiagram}. The photon propagator can be written as:
\begin{equation}
    D^{\mu\nu}(q)=D_0^{\mu\nu}(q)+D_0^{\mu\alpha}(q)\Pi_{\alpha\beta}(q)D_0^{\beta\nu}(q),
\end{equation}
where $D_0^{\mu\nu}$ is the leading order photon propagator
\begin{equation}
    D_0^{\mu\nu}=\frac{-g^{\mu\nu}}{q^2},
\end{equation}
and $\Pi_{\alpha\beta}$ is the vacuum polarization, which is given at first order in $\alpha$ by:
\begin{equation}
    -i\Pi^{\mu\nu}(q)=-e^2\mu^{2\epsilon}\int\frac{d^dl}{(2\pi)^d}\frac{\text{Tr}\left[\gamma^{\mu}(\sla{l}-\sla{k}+m)\gamma^{\nu}(\sla{l}+m)\right]}{[(l-k)^2-m^2][]l^2-m^2]}.
\end{equation}
Due to gauge-invariance, $q_{\mu}\Pi^{\mu\nu}=q_{\nu}\Pi^{\mu\nu}=0$, and the vacuum polarization can be decomposed as:
\begin{equation}
    \Pi^{\mu\nu}(q)=(-g^{\mu\nu}q^2+q^{\mu}q^{\nu})\Pi(q^2),
\end{equation}
where $\Pi(q^2)$ is given by \cite{Vanderhaeghen:2000ws}:
\begin{equation}
    \Pi(q^2)=-\frac{\alpha}{3\pi}\left[\frac{1}{\epsilon_{\text{UV}}}-\gamma_E+\ln\left(\frac{4\pi\mu^2}{m^2}\right)-\left(v^2-\frac{8}{3}\right)+\frac{v}{2}(v^2-3)\ln\left(\frac{v+1}{v-1}\right)\right],\label{vacuumunren}
\end{equation}
where $v$ is defined in Eq. \eqref{EqDef:v}.

The UV-divergence in Eq. \eqref{vacuumunren} is removed by the renormalization constant $Z_3$:
\begin{equation}
    \tilde{\Pi}(q^2)=\Pi(q^2)-(Z_3-1),
\end{equation}
which is fixed by requiring, that the renormalized vacuum polarization $\tilde{\Pi}(q^2)$ has a pole with residue $1$ at $q^2=0$:
\begin{equation}
    Z_3=1+\Pi(q^2=0).
\end{equation}
The renormalized vacuum polarization is then given by:
\begin{equation}
    \tilde{\Pi}(q^2)=\frac{\alpha}{3\pi}\left[\left(v^2-\frac{8}{3}\right)+\frac{v}{2}(3-v^2)\ln\left(\frac{v+1}{v-1}\right)\right].\label{EqVPRen}
\end{equation}
The renormalized photon propagator is therefore given by
\begin{equation}
    \tilde{D}^{\mu\nu}(q)=\frac{-g^{\mu\nu}}{q^2}\left[1+\tilde{\Pi}(q^2)\right]+\frac{q^{\mu}q^{\nu}\Pi(q^2)}{q^4}.
\end{equation}
Note that, due to gauge invariance, only the term proportional to $g^{\mu\nu}$ contributes to a physical cross section.

\begin{figure}
	\includegraphics[scale=0.7]{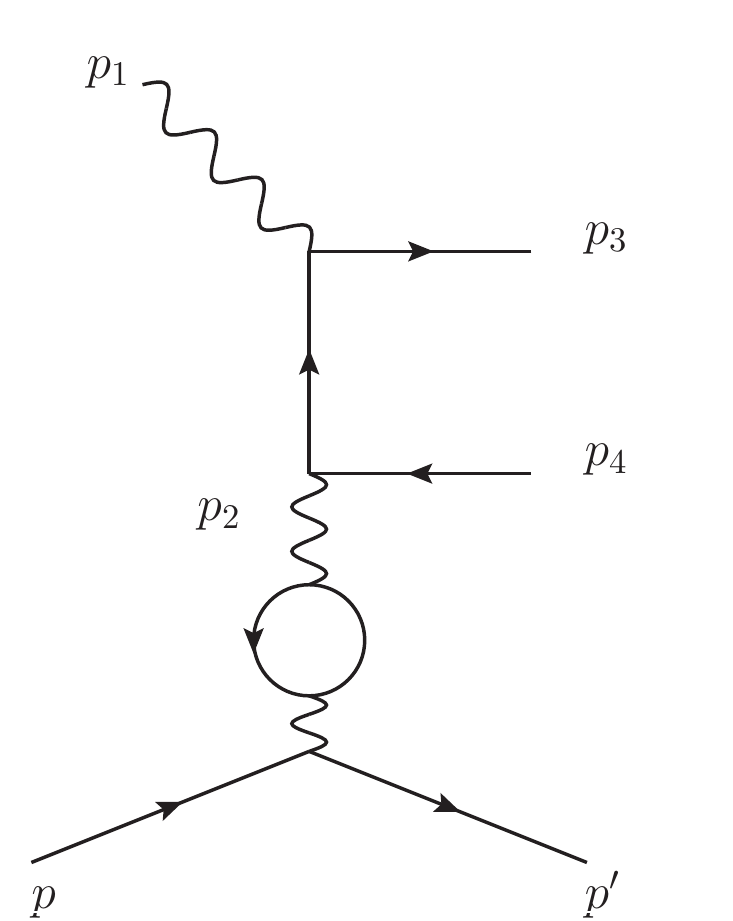}\hspace{0.cm}
	\caption{Vacuum polarization diagram contributing to the Bethe-Heitler process. In the ratio of electron versus muon pair production cross sections, this contribution drops out.\label{VacuumBH}}
\end{figure}
We show the contributing vacuum polarization diagrams in Fig. \ref{VacuumBH}. The amplitude for the sum of both diagrams is given by
\begin{align}
    \mathcal{M}_\text{VP}&=\tilde{\Pi}(t)\;\mathcal{M}_0,
\end{align}
where $\tilde{\Pi}(t)$ is given by Eq. \eqref{EqVPRen} and $\mathcal{M}_0$ by Eq. \eqref{M0}.

Note that the correction due to the vacuum polarization is the same for muon- and electron-pair production. It therefore drops out in the cross section ratio of lepton-pair production, considered in this paper. This remains also true when one considers hadronic vacuum polarization.

\subsection{Lepton box diagrams and reproduction of soft-photon result}
\begin{figure}
	\includegraphics[scale=0.7]{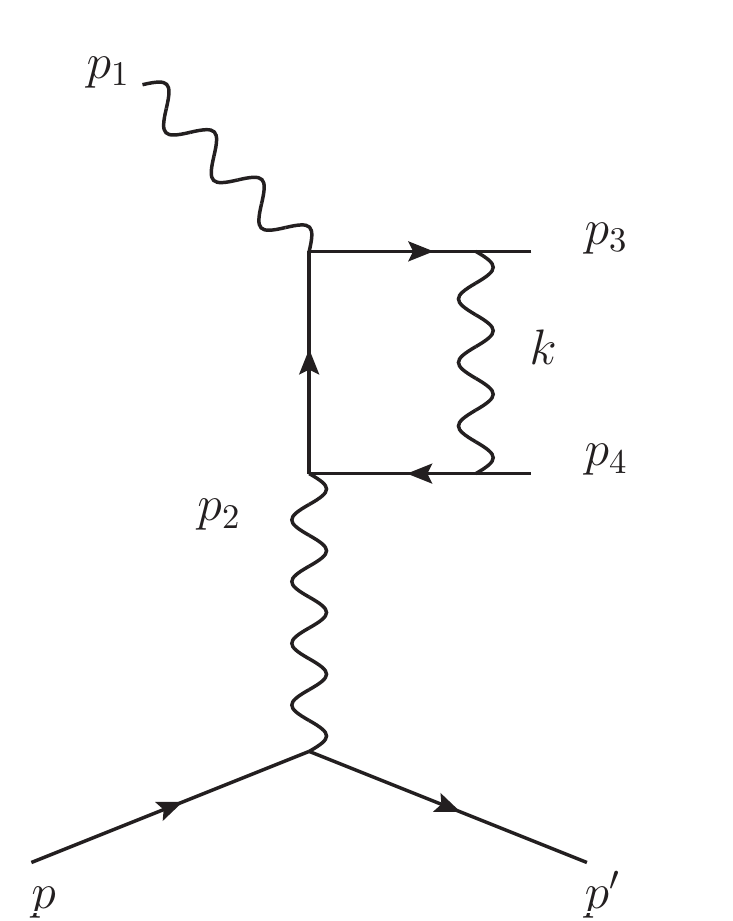}\hspace{0.cm}
	\caption{Lepton box diagrams contributing to the Bethe-Heitler process. \label{Boxdiagram}}
\end{figure}

In this section, we will present the calculation of the lepton box diagrams, as shown in Fig \ref{Boxdiagram}.
For the analytic calculation, we use \texttt{QGRAF} \cite{Nogueira:1991ex} to generate all diagrams in a representation, which can be processed with the algebra program \texttt{Form} \cite{Kuipers:2012rf}. To generate Integration-By-Part (IBP) identities, which are used to express a set of integrals in terms of a smaller set of so-called master integrals, we use the program \texttt{Reduze 2} \cite{vonManteuffel:2012np}. These master integrals are either known analytically in the literature, or could be evaluated with the help of Feynman parameters. We give a list of all master integrals in Appendix \ref{master_integrals}.

In the general case with a finite lepton mass, the calculation of the lepton box diagrams leads to very large expressions. Nevertheless, we were able to extract the analytical expressions with the setup described above.

Having calculated the leptonic tensor $L^{\mu\nu}$, we were able to check gauge invariance by contracting with the external photon momenta, i.e.,
\begin{align}
    p_1^{\mu}L_{\mu\nu}= p_2^{\nu}L_{\mu\nu}=0.
\end{align}
It turns out, that this identity is only fulfilled, once all diagrams (self-energy, vertex and box) and all counter-terms are taken into account. Therefore, only the renormalized lepton tensor is gauge invariant. 

Here we give the leading contribution stemming from double logarithms, for the limit of small lepton masses $m^2\rightarrow 0$ and show that we reproduce the correct asymptotic behavior for $s_{ll}>>4m^2$, which was derived in Ref. \cite{Heller:2018ypa}. In this case, the contribution to the cross section can be written as

\begin{align}
\overline{\sum_i}\sum_f\mathcal{M}^*_{0}\left(\mathcal{M}_{\text{Box}}\right)&\approx \frac{e^4}{t^2}\left(\frac{\alpha}{\pi}\right)\left\{(C+\tilde{C})\left[\ln^2\left(\frac{-t}{s_{ll}}\right)-2\ln\left(\frac{-t}{s_{ll}}\right)\ln\left(\frac{m^2}{s_{ll}}\right)\right]+D\ln^2\left(\frac{-t_{ll}}{s_{ll}}\right)\right.\nonumber\\
&\left.+2C\ln\left(\frac{-t_{ll}}{s_{ll}}\right)\ln\left(\frac{m^2}{s_{ll}}\right)+\tilde{D}\ln^2\left(\frac{-u_{ll}}{s_{ll}}\right)+2\tilde{C}\ln\left(\frac{-u_{ll}}{s_{ll}}\right)\ln\left(\frac{m^2}{s_{ll}}\right)\right.\nonumber\\
&\left.+\left(\frac{L_0^{\mu\nu}H_{\mu\nu}}{2}-B-\tilde{B}\right)\ln\left(\frac{m^2}{s_{ll}}\right)+L_0^{\mu\nu}H_{\mu\nu}\left[\frac{1}{4}\ln^2\left(\frac{m^2}{s_{ll}}\right)-\frac{1}{2}\ln\left(\frac{m^2}{s_{ll}}\right)\right.\right.\nonumber\\
&\times\left.\left.\left[\frac{1}{\epsilon_\text{IR}}-\gamma_E+\ln\left(\frac{4\pi \mu^2}{s_{ll}}\right)\right]\right]\right\},
\label{BoxApp}\end{align}
where $B$ and $C$ are defined in Eq. \eqref{VertexApp}, while the other coefficients are defined as:
\begin{align}
D&=\frac{H_1 \left(t^2-2 t_{ll} t+s_{ll}^2+t_{ll}^2\right)}{\left(t-s_{ll}-t_{ll}\right) t_{ll}}+\frac{-H_2 }{4\left(t-s_{ll}-t_{ll}\right){}^3 t_{ll}}\left\{9 t^3 M^4+10 t s_{ll}^2 M^4+9 t t_{ll}^2 M^4-18 t^2 s_{ll} M^4\right.\nonumber\\
&\left.-14 t^2 t_{ll} M^4+18 t s_{ll} t_{ll} M^4-2 t^4 M^2+2 s_{ll}^4 M^2+2 t_{ll}^4 M^2-8 s t^3 M^2-4 t  s_{ll}^3 M^2-12 t t_{ll}^3 M^2\right.\nonumber\\
&\left.+4 s_{ll} t_{ll}^3 M^2+t^2 s_{ll}^2 M^2-10 s t s_{ll}^2 M^2-10 t u s_{ll}^2 M^2+14 t^2 t_{ll}^2 M^2+4 s_{ll}^2 t_{ll}^2 M^2-10 s t t_{ll}^2 M^2\right.\nonumber\\
&\left.-8 t u t_{ll}^2 M^2-21 t s_{ll} t_{ll}^2 M^2-10 t^3 u M^2+3 t^3 s_{ll} M^2+16 s t^2 s_{ll} M^2+20 t^2 u s_{ll} M^2-6 t^3 t_{ll} M^2\right.\nonumber\\
&\left.+4 s_{ll}^3 t_{ll} M^2+14 s t^2 t_{ll} M^2-13 t s_{ll}^2 t_{ll} M^2+14 t^2 u t_{ll} M^2+14 t^2 s_{ll} t_{ll} M^2-20 s t s_{ll} t_{ll} M^2\right.\nonumber\\
&\left.-16 t u s_{ll} t_{ll} M^2+2 s t^4+2 s^2 t^3-s t s_{ll}^3+t u s_{ll}^3+2 t^2 t_{ll}^3+2 s t t_{ll}^3+2 t u t_{ll}^3+3 t^3 u^2+3 s t^2 s_{ll}^2\right.\nonumber\\
&\left.+3 t u^2 s_{ll}^2+3 s^2 t s_{ll}^2+4 s t u s_{ll}^2.-3 t^3 t_{ll}^2+2 t u^2 t_{ll}^2+3 s^2 t t_{ll}^2-2 t^2 u t_{ll}^2+4 s t u t_{ll}^2+3 t^2 s_{ll} t_{ll}^2\right.\nonumber\\
&\left.+3 s t s_{ll} t_{ll}^2+4 t u s_{ll} t_{ll}^2+2 t^4 u+4 s t^3 u-4 s t^3 s_{ll}-4 s^2 t^2 s_{ll}-6 t^2 u^2 s_{ll}-3 t^3 u s_{ll}-8 s t^2 u s_{ll}\right.\nonumber\\
&\left.+2 t^4 t_{ll}-2 s t^3 t_{ll}-4 s^2 t^2 t_{ll}-4 t^2 u^2 t_{ll}+t^2 s_{ll}^2 t_{ll}+3 t u s_{ll}^2 t_{ll}-6 s t^2 u t_{ll}-3 t^3 s_{ll} t_{ll}+3 s t^2 s_{ll} t_{ll}\right.\nonumber\\
&\left.+4 t u^2 s_{ll} t_{ll}+6 s^2 t s_{ll} t_{ll}-3 t^2 u s_{ll} t_{ll}+8 s t u s_{ll} t_{ll}\right\}-C,
\end{align}
and $\tilde{D}$ is given by by making the replacements $t_{ll}\rightarrow u_{ll}$ and $u \rightarrow -u - s - t + 3M^2$ in $D$.

By adding the contribution of self-energy, vertex and box diagrams, i.e Eqs. \eqref{SeApp}, \eqref{VertexApp} and \eqref{BoxApp}, all mixed double-logarithmic terms involving the factor $\ln(m^2/s_{ll})$ drop out in the sum. Furthermore, the contribution proportional to $\ln(m^2/s_{ll})$ and $\ln^2(m^2/s_{ll})$ factorizes in terms of the tree-level amplitude, such that we can write:

\begin{align}
\left(\frac{d\sigma}{dtds_{ll}}\right)_{\text{V}}\approx\left(\frac{d\sigma}{dtds_{ll}}\right)_{0}\left(\frac{-\alpha}{\pi}\right)&\left\{\left[\ln\left(\frac{m^2}{s_{ll}}\right)+1\right]\left[\frac{1}{\epsilon_\text{IR}}-\gamma_E+\ln\left(\frac{4\pi \mu^2}{m^2}\right)\right]\right.\nonumber\\
&\left.+\frac{1}{2}\ln^2\left(\frac{m^2}{s_{ll}}\right)+\frac{1}{2}\ln\left(\frac{m^2}{s_{ll}}\right)\right\},\label{LeadingDoubleLogs}
\end{align}
where $\mu$ is the scale associated to the infrared divergence of the soft-photon loops. The dependence on $\mu$ cancels, once one takes real photon corrections into account.

We are thus able to reproduce the correct double-logarithmic behavior for the virtual corrections, which we derived in \cite{Heller:2018ypa} in Eqs. $(46,47)$, in the limit of $\beta\rightarrow 1$, corresponding to $s_{ll}>>4m^2$.

Furthermore, in Ref. \cite{Huld:1968zz} the radiative corrections to the dilepton pair photoproduction cross section were studied in the approximation of a small lepton mass. The correction is shown in Eq. (B1) of that paper. We compared our analytical result after expansion in $m^2$ and find exact agreement. To make the comparison, we had to make the identifications (kinematical quantities on left were introduced in Ref. \cite{Huld:1968zz}):
\begin{align}
    Q^2&\rightarrow s_{ll}\nonumber\\
    q^2&\rightarrow t\nonumber\\
    \beta_-&\rightarrow -t_{ll}+m^2\nonumber\\
    \beta_+&\rightarrow -u_{ll}+m^2.
\end{align}
Note that the $L$ and $H$ functions of Ref. \cite{Huld:1968zz} are given analytically by:
\begin{align}
    L(x)&=-\text{Li}_2(-x)-\frac{\pi^2}{12}\nonumber\\
    H_0(Q^2)&=-\frac{2}{3}\pi^2+\frac{1}{2} \ln^2\left(\frac{m^2}{s_{ll}}\right) + \ln\left(\frac{m^2}{s_{ll}}\right) \left[\frac{1}{\epsilon_\text{IR}}-\gamma_E+\ln\left(\frac{4\pi\mu^2}{m^2}\right)\right].
\end{align}

\subsection{Integration over lepton angles}
For an experiment which only measures the recoiling proton, we have to integrate the differential cross section over the lepton angles $d\Omega_{ll}^{CM_{ll}}$, as shown in Eq. \eqref{CrossSectionIntegrated}. 
In order to perform the integration we have to express the kinematic invariants $t_{ll}$ and $u$ in terms of the lepton angles $\theta_{ll}$ and $\phi_{ll}$ defined in the rest frame of the di-lepton pair:
\begin{align}
    t_{ll}&=\frac{1}{2}(2m^2+t-s_{ll})+\frac{\beta}{2}(b_1\cos\theta_{ll}+b_2\cos\phi_{ll}\sin\theta_{ll}),\nonumber\\
    u&=\frac{1}{2}(2m^2+3M^2-s-t)-\frac{\beta}{2}(a\cos\theta_{ll}+b_1\cos\theta_{ll}+b_2\cos\phi_{ll}\sin\theta_{ll}),\label{leptonangles}
\end{align}
where
\begin{align}
    a&=\sqrt{(s-s_{ll}-M^2)^2-4M^2s_{ll}},\nonumber\\
    b_1&=\frac{(s-M^2-s_{ll})s_{ll}+(s-M^2+s_{ll})t}{a},\nonumber\\
    b_2&=\sqrt{(s_{ll}-t)^2-b_1^2}.
\end{align}
The integration over the lepton angles $\theta_{ll}$ and $\phi_{ll}$ was done numerically in \texttt{Mathematica}.

\section{Numerical Calculation of virtual photon corrections}
\label{sec4}
The numerical calculation of the virtual photon corrections were mostly done in \texttt{Mathematica}, using the packages \texttt{FeynArts}, \texttt{FormCalc}, \texttt{FeynCalc} and  \texttt{Looptools} \cite{Mertig:1990an,Hahn:2010zi}.

To obtain the leptonic tensor, we calculated the cross section of a truncated two-to-two scattering with an off-shell photon and an on-shell photon as incoming particles in dilepton production. This simplifies the calculation significantly since it can use the automated techniques to generate cross sections in \texttt{FeynArts} and \texttt{FormCalc}. Furthermore, one can directly apply the existing regularization and renormalization scheme from regular two-to-two scattering processes provided by \texttt{FormCalc}.

After obtaining the expression for the leptonic cross section, one can use the polarization vectors for the photons to construct the leptonic tensor structure.
The renormalization constants are provided numerically within the \texttt{FormCalc} in the on-shell scheme, and are given in terms of one- and two-point functions and derivatives of these. After renormalization, the result can be tested to be UV finite by varying $\mu^2$.

After obtaining the expression for the leptonic cross section, one can use the polarization vectors for the photons to construct the leptonic tensor. We created a python script to replace all the photon polarization vectors into tensorial form in the format of \texttt{FeynCalc} notation.  Note that after the contraction with the hadronic tensor, the scalar result is still infrared divergent. One has then to include the soft photon bremsstrahlung. For the regularization of infrared divergences, \texttt{FormCalc} introduces a photon mass. Cancellation of infrared divergences between the real and virtual corrections was tested by varying the photon mass.

Using the numerical result, we were able to check the analytic expression at a given phase-space point.

\section{Soft-photon bremsstrahlung}
\label{sec5}
\begin{figure}
	\includegraphics[scale=0.70]{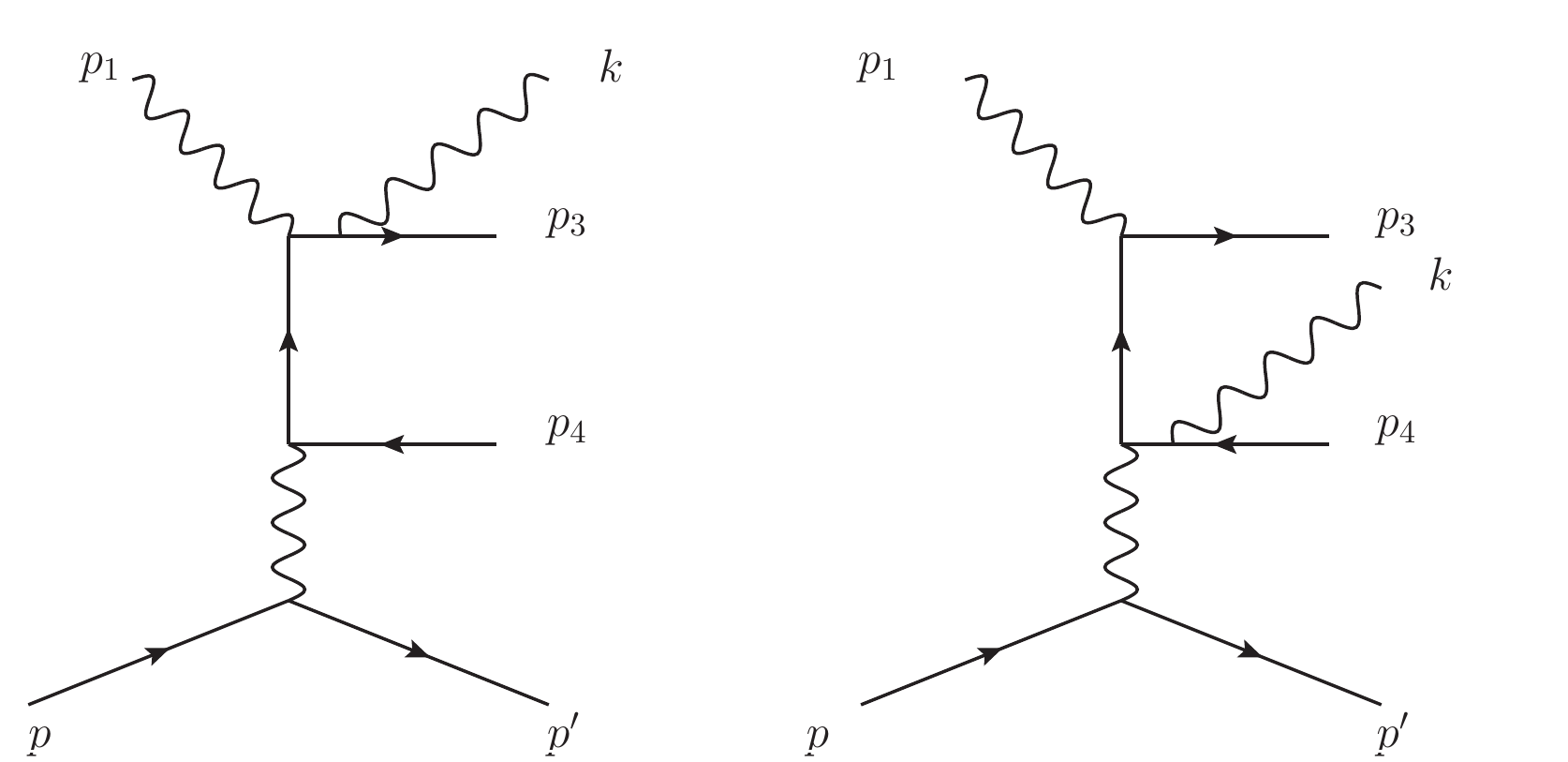}\hspace{0.cm}
	\caption{Diagrams with real photon emission from the lepton lines of the Bethe-Heitler process. In the soft-photon limit, the diagram with the photon attached to the internal (off-shell) fermion line does not contribute.\label{SoftBrems}}
\end{figure}
Having evaluated all one-loop virtual corrections, we have to take into account the radiation of real undetected photons to get an infrared finite cross section. We calculate real radiation only in the soft-photon limit and show the contributing diagrams in Fig. \ref{SoftBrems}. Note that the diagram, where the photon is attached to the internal lepton line, does not contribute in the soft-photon limit. Denoting the momentum of the photon by $k$, the squared matrix element is given by:
\begin{align}
&\left|\mathcal{M}(\gamma p\rightarrow\gamma_\text{s}\;l^+l^-p)\right|^2=\left|\mathcal{M}(\gamma p\rightarrow l^+l^-p)\right|^2\;(-e^2)\left[ \frac{p_3^{\mu}}{p_3\cdot k}-\frac{p_4^{\mu}}{p_4\cdot k}\right]\cdot\left[ \frac{p_{3\mu}}{p_3\cdot k}-\frac{p_{4\mu}}{p_4\cdot k}\right].
\end{align}
To calculate the contribution to the cross section, one then has to integrate over the undetected soft-photon energy up to a small value $\Delta E_s$, determined by the experimental resolution. 

Due to the energy-momentum conserving $\delta$-function, $\delta^4(p_1+p-p_3-p_4-p^{\prime}-k)$, the integration domain has a complicated shape in the lab system. The integration can be carried out in the rest frame $\mathcal{S}$ of the real ($p_1)$ and virtual ($p_2$) photons, which is also the rest frame of the di-lepton pair and soft photon, defined by
\begin{equation}
\vec{p}_1+\vec{p}_2=\vec{p}_3+\vec{p}_4+\vec{k}=0.
\end{equation}
In such frame, the dependence of the integral with respect to the soft-photon momentum becomes isotropic.
For the differential cross section, we then need to evaluate:
\begin{equation}
\left(\frac{d\sigma}{dt ds_{ll}}\right)_\text{s;R}=-\left(\frac{d\sigma}{dt ds_{ll}}\right)_0\frac{e^2}{(2\pi)^3}\int_{|\vec{k}|<\Delta E_s}\frac{d^3\vec{k}}{2k^0}\left[\frac{m^2}{(p_3k)^2}+\frac{m^2}{(p_4k)^2}-\frac{2(p_3p_4)}{(p_3k)(p_4k)}\right ],\label{SoftPhotonIntegral}
\end{equation}
where the integration is performed in the frame $\mathcal{S}$. 

The integrals are infrared divergent and can be carried out analytically after dimensional regularization. They have been worked out, e.g., in Ref. \cite{tHooft:1978jhc}.
For the kinematics in system $\mathcal{S}$, where the soft-photon momentum:
\begin{equation}
|\vec{k}|\ll\left|\vec{p}_3\right|,\left|\vec{p}_4\right|,
\end{equation}
with the lepton four-momenta:
\begin{align}
p_3^0=p_4^0=\frac{\sqrt{s_{ll}}}{2},\ \ \vec{p}_3=-\vec{p}_4,
\end{align}
we obtain:
\begin{align}
\left(\frac{d\sigma}{dt ds_{ll}}\right)_\text{s;R}&=\left(\frac{d\sigma}{dtds_{ll}}\right)_{0}\left(\vphantom{\frac{1}{2}}\delta_\text{s;R}^\text{IR}+\delta_\text{s;R}\right),
\end{align}
where $\delta_\text{s;R}^\text{IR}$ is the infrared-divergent contribution due to real photon emission:
\begin{align}
\delta^\text{IR}_\text{s;R}=\left(\frac{-\alpha}{\pi}\right)\left[\left(\frac{1+\beta^2}{2\beta}\right)\ln\left(\frac{1+\beta}{1-\beta}\right)-1\right]\left[\frac{1}{\epsilon_\text{IR}}-\gamma_E+\ln\left(\frac{4\pi\mu^2}{m^2}\right)\right], \label{IRReal}
\end{align}
and $\delta_\text{s;R}$ is the corresponding finite part:
\begin{align}
\delta_\text{s;R}=\left(\frac{-\alpha}{\pi}\right)\left\{\ln\left(\frac{4\Delta E_s^2}{m^2}\right)\right.&\left.\left[1+\left(\frac{1+\beta^2}{2\beta}\right)\ln\left(\frac{1-\beta}{1+\beta}\right)\right]+\frac{1}{\beta}\ln\left(\frac{1-\beta}{1+\beta}\right)+\right.\nonumber\\
\phantom{=}+\left(\frac{1+\beta^2}{2\beta}\right)&\left.\left[2\;\text{Li}_2\left(\frac{2\beta}{1+\beta}\right)+\frac{1}{2}\ln^2\left(\frac{1-\beta}{1+\beta}\right)\right]\right\}.\label{FiniteReal}
\end{align}

The maximum value of the undetected soft-photon energy $\Delta E_s$ is defined in the system $\mathcal{S}$. One can re-express it in terms of the detector resolutions. We consider the case of detecting the recoil proton only. The energy $E^{\prime}$ and angle $\theta_{p^{\prime}}$ of the scattered proton are measured in the lab frame.
The missing mass $ M_\text{miss} $ of the system is defined by
\ber
M^2_\text{miss} & = & (p_3 + p_4 + k )^2 = s_{ll} + 2 M_\text{miss}  E_s,  \\
E_s & = & \frac{M^2_\text{miss} - s_{ll}}{2 M_\text{miss}},\label{Reso}
\eer
where $E_s$ denotes the soft-photon energy.

The missing mass $M_\text{miss}$ is experimentally determined from the quantity:
\begin{align}
M_{\text{miss}}^2&=(p_1+p-p^{\prime})^2 \nonumber\\
&=4M\tau \left(E_{\gamma}\sqrt{\frac{1+\tau}{\tau}}\text{cos}\;\theta_{p^{\prime}}-E_{\gamma}-M\right),\label{MissMassLab}
\end{align}
where $\tau$ is determined from the lab proton momentum by Eq. \eqref{TauLab}, and $\theta_{p^{\prime}}$ is the experimentally measured recoil proton scattering angle in the laboratory frame.

For the process without radiation, this angle is given by Eq. \eqref{ScatterAngleLab}, which can be equivalently obtained from Eq. \eqref{MissMassLab} by the replacement $M_{\text{miss}}^2 \rightarrow s_{ll}$:
\begin{equation}
s_{ll}=4M\tau\left( E_{\gamma}\sqrt{\frac{1+\tau}{\tau}}\text{cos}\;\theta_{p^{\prime}}|_{\text{no rad}}-E_{\gamma}-M\right)\label{sllLab}.
\end{equation}

Combining Eqs. \eqref{MissMassLab} and \eqref{sllLab}, we can express the soft-photon energy of Eq. \eqref{Reso} approximately as:
\begin{equation}
E_\text{s}=\frac{2ME_{\gamma}\sqrt{\tau(1+\tau)}}{\sqrt{s_{ll}}}\left[\vphantom{\frac{1}{2}}\text{cos}\;\theta_{p^{\prime}}-\text{cos}\;\theta_{p^{\prime}}|_{\text{no rad}}\right].
\end{equation}

Consequently, the experimental recoiling proton angular resolution, denoted as $\Delta\theta_{p^\prime}$, determines the maximum value $\Delta E_s$ of the undetected soft-photon energy, which enters the radiative correction of Eq. \eqref{FiniteReal}, as
\begin{equation}
\Delta E_s=\frac{2ME_{\gamma}\sqrt{\tau(1+\tau)}}{\sqrt{s_{ll}}}\;\text{sin}\;\theta_{p^{\prime}}\;\Delta\theta_{p^{\prime}}.\label{AngularReso}
\end{equation}

\section{Total result and exponentiation of the soft-photon contribution}
\label{sec6}
Adding the virtual one-loop corrections and the real soft-photon correction, we find a cancellation of all infrared divergences. The full result can be written as:

\begin{align}
\left(\frac{d\sigma}{dt ds_{ll}}\right)&=\left(\frac{d\sigma}{dt ds_{ll}}\right)_0\left\{1+\left[\delta_{\text{1-loop}}+\delta_\text{s;R}\right]\right\}\nonumber\\
&\equiv\left(\frac{d\sigma}{dtds_{ll}}\right)_0\left(1+\delta_{\text{exp}}\right),\label{CrossSec-oneloop}
\end{align}
where $\delta_{\text{1-loop}}$ is the finite part of the one-loop virtual corrections. In the limit $s_{ll}>>4m^2$, one obtains from Eq. \eqref{LeadingDoubleLogs} the approximate result to double logarithmic accuracy:
\begin{equation}
\delta_{\text{1-loop}}\approx -\frac{\alpha}{\pi}\left\{\frac{1}{2}\ln^2\left(\frac{m^2}{s_{ll}}\right)+\frac{1}{2}\ln\left(\frac{m^2}{s_{ll}}\right)\right\}.
\end{equation}

As shown in Ref. \cite{Heller:2018ypa}, we can exponentiate the double logarithmic part of $\delta_{\text{exp}}$, which in the soft-photon limit is given by:

\begin{equation}
\delta_{\text{exp}}\approx-\frac{\alpha}{\pi}\left[\ln\left(\frac{4\Delta E_s^2}{m^2}\right)+\ln\left(\frac{1-\beta}{1+\beta}\right)\right]\left[1+\left(\frac{1+\beta^2}{2\beta}\right)\ln\left(\frac{1-\beta}{1+\beta}\right)\right].
\end{equation}
For $s_{ll}>>4m^2$, this becomes

\begin{equation}
\delta_{\text{soft}}\approx -\left(\frac{\alpha}{\pi}\right)\left\{\ln\left(\frac{4\Delta E_s^2}{s_{ll}}\right)\left[1+\ln\left(\frac{m^2}{s_{ll}}\right)\right]\right\}.
\end{equation}
By accounting for the exponentiation of large double-logarithms in $m^2/s_{ll}$, the full one-loop correction of the cross section is therefore given by:

\begin{align}
\left(\frac{dtd\sigma}{dtds_{ll}}\right)=\left(\frac{d\sigma}{dtds_{ll}}\right)_0\cdot F e^{\delta_{\text{soft}}}\times&\left\{1+\left[\delta_{\text{1-Loop}}+\left(\frac{\alpha}{\pi}\right)\frac{1+\beta^2}{4\beta}\ln^2\left(\frac{1-\beta}{1+\beta}\right)\right.\right.\nonumber\\
&\left.\left.-\frac{\alpha}{\pi}\left(\frac{1+\beta^2}{\beta}\text{Li}_2\left(\frac{2\beta}{1+\beta}\right)+\frac{1-\beta}{\beta}\ln\left(\frac{1-\beta}{1+\beta}\right)\right)\right]\right\}\nonumber\\
\equiv\left(\frac{d\sigma}{dtds_{ll}}\right)_0\left(1+\delta_{\text{exp}}\right),&\label{CrossSec-oneloop-exp}
\end{align}
Note that, in Eq. \eqref{CrossSec-oneloop-exp} we have to subtract the leading double-logarithmic term $-\alpha/\pi(1+\beta^2)/(4\beta)\ln^2((1-\beta)/(1+\beta))$ from the virtual one-loop corrections and add the terms from the real soft-photon corrections that cannot be exponentiated, c.f. Ref. \cite{Heller:2018ypa}. In the limit $s_{ll}>>4m^2$, Eq. \eqref{CrossSec-oneloop-exp} is given by:
\begin{align}
\left(\frac{dtd\sigma}{dtds_{ll}}\right)&=\left(\frac{d\sigma}{dtds_{ll}}\right)_0\cdot F e^{\delta_{\text{soft}}}\times\left\{1+\left[\delta_{\text{1-Loop}}+\left(\frac{\alpha}{\pi}\right)\left(\frac{1}{2}\ln^2\left(\frac{m^2}{s_{ll}}\right)-\frac{\pi^2}{3}\right)\right]\right\},
\end{align}
in which the correct leading logarithm dependence from Eq. \eqref{LeadingDoubleLogs} is reproduced. 

Furthermore, it was shown in Ref. \cite{Yennie:1961ad} that the normalization factor $F$ arises due to the physical assumption that in an experiment the sum of all soft-photon energies is smaller than $\Delta E_s$, instead of requiring that each soft-photon energy is individually smaller than $\Delta E_s$. Its leading correction from unity is given by:
\begin{equation}
F=1-\frac{\alpha^2}{3}\left[1+\left(\frac{1+\beta^2}{2\beta}\right)\ln\left(\frac{1-\beta}{1+\beta}\right)\right]^2+...
\end{equation}
Although we account for the factor $F$ explicitly, its deviation from unity is quite small: for $s_{ll}=0.077\text{ GeV}^2$ approximately $-2.4 \times 10^{-3}$ for electrons and $-8.5 \times 10^{-6}$ for muons.

\section{Proton-line corrections}
\label{proton_correction}
\begin{figure}
	\includegraphics[scale=0.6]{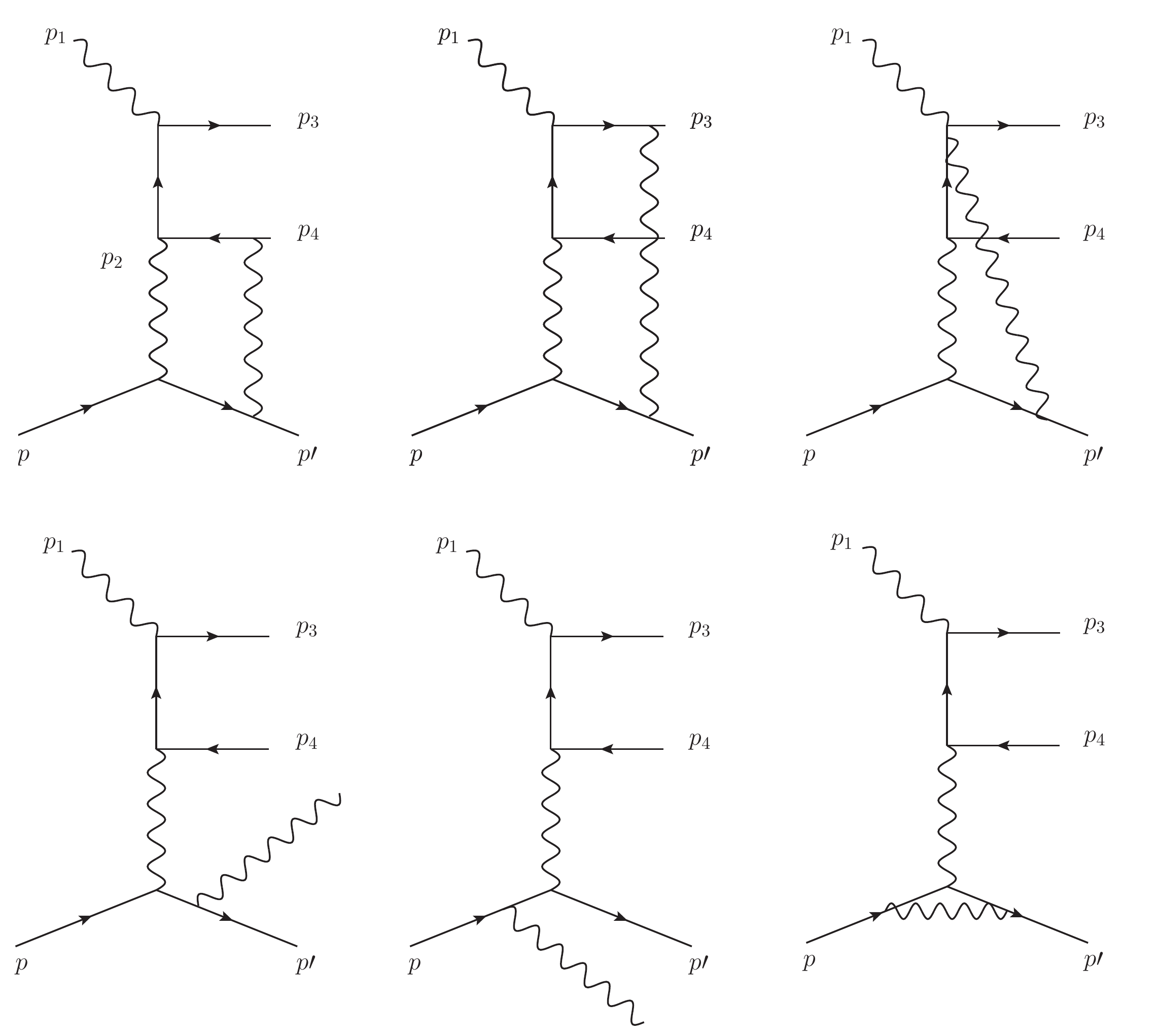}\hspace{0.cm}
	\caption{One-loop radiative corrections to the $\gamma p\rightarrow l^+l^-p$ process. Upper diagrams: box graphs with proton, lower right diagram: the proton vertex correction, two lower left diagrams: soft bremsstrahlung from the proton line. The graphs with an interchange of final leptons are not shown.\label{protonside}}
\end{figure}

Besides the one-loop leptonic corrections, we also estimate the one-loop hadronic corrections. As the latter are much smaller than their leptonic counterparts at low energies, due to the much larger proton mass, we will estimate the hadronic corrections in the soft-photon approximation. We have to account for the soft bremsstrahlung from the proton line, box graphs with protons and proton vertex correction, as shown by the diagrams in Fig. \ref{protonside}.

First, we consider the interference of graphs with bremsstrahlung from the proton and lepton lines. The parity transformation: 
\ber
\theta \to \pi -\theta, \qquad \phi \to \phi + \pi, \label{parity_lepton}
\eer
corresponds to the interchange of lepton and anti-lepton. When calculating the cross section correction due to the interference of graphs with the radiation from the proton line and the soft bremsstrahlung from the lepton lines, the parity transformation of Eq. (\ref{parity_lepton}) swaps the bremsstrahlung vertex between final particles in the di-lepton pair. Such transformation results in the overall sign change due to the opposite electric charge of the lepton and anti-lepton. Consequently, the cross section contribution from the interference of lepton and proton bremsstrahlung has a symmetry property:
\ber
\mathrm{d} \sigma_\text{p}^{\gamma p \to \gamma p l_+ l_-} \left( \pi -\theta,~ \phi + \pi \right) = - \mathrm{d} \sigma_\text{p}^{\gamma p \to \gamma p l_+ l_-} \left( \theta,~\phi \right). 
\eer
The same arguments are valid for the interference of the tree-level graphs with the contribution of the corresponding box proton diagrams: 
\ber
\mathrm{d} \sigma^{\gamma p \to p l_+ l_-}_{\mathrm{p Box}} \left( \pi -\theta,~ \phi + \pi \right) = - \mathrm{d} \sigma^{\gamma p \to p l_+ l_-}_{\mathrm{p Box}} \left( \theta,~\phi \right).
\eer
Integrating over the lepton-pair angles, the leading proton box contributions and the bremsstrahlung interference between lepton and proton lines exactly yields zero. \footnote{This property was first mentioned in Ref. \cite{Huld:1968zz}.}

In the following, we estimate the proton vertex correction as the difference between the calculation with proton form factors corrected by the 1-loop QED renormalized on-shell vertex of Eqs. (\ref{renormalized_QED_F2}, \ref{renormalized_QED_F1}) and the tree level result.

As the vertex diagrams has an IR divergence, we also need to account for the soft bremsstrahlung from the proton line. We denote the momentum of the outgoing photon by $k$. The corresponding matrix element has the following form:
\begin{align}
&\left|\mathcal{M}(\gamma p\rightarrow\gamma_\text{s}\;l^+l^-p)\right|^2=\left|\mathcal{M}(\gamma p\rightarrow l^+l^-p)\right|^2\;(-e^2)\left[ \frac{p^{\mu}}{p \cdot k}-\frac{p'^{\mu}}{p' \cdot k}\right]\cdot\left[ \frac{p_{\mu}}{p\cdot k}-\frac{p'_{\mu}}{p' \cdot k}\right].
\end{align}

In the rest frame of the dilepton pair, where the dependence of the phase-space integral on the photon momentum direction is isotropic,  the soft-photon contribution factorizes in terms of the BH cross section as
\ber
\left(\frac{d\sigma}{dt ds_{ll}}\right)_\text{s;pR} &=& - \left(\frac{d\sigma}{dt ds_{ll}}\right)_0  \frac{e^2}{(2\pi)^3}\int_{|\vec{k}|<\Delta E_s}\frac{d^3\vec{k}}{2 k^0}   \left[ \frac{p^{\mu}}{p \cdot k}-\frac{p'^{\mu}}{p' \cdot k}\right]\cdot\left[ \frac{p_{\mu}}{p\cdot k}-\frac{p'_{\mu}}{p' \cdot k}\right], 
\eer
where the integration is performed up to a small value of the soft-photon energy $\Delta E_s$, determined by the experimental resolution. The resulting correction is evaluated in the rest frame of the dilepton pair \cite{Heller:2018ypa} and can be expressed as \cite{Vanderhaeghen:2000ws,Arbuzov:2007ct}
\ber
\left(\frac{d\sigma}{dt ds_{ll}}\right)_\text{s;R} &=& \left(\frac{d\sigma}{dt ds_{ll}}\right)_0\left(\vphantom{\frac{1}{2}}\delta_\text{s;pR}^\text{IR}+\delta_\text{s;pR}\right),
\eer
with the infrared-divergent contribution $\delta_\text{s;pR}^\text{IR}$:
\begin{align}
\delta^\text{IR}_\text{s;pR}=\left(\frac{-\alpha}{\pi}\right)\left[\frac{1}{2 \bar{v}} \ln \left(\frac{1+\bar{v}}{1-\bar{v}}\right)-1\right]\left[\frac{1}{\epsilon_\text{IR}}-\gamma_E+\ln\left(\frac{4\pi\mu^2}{M^2}\right)\right], \label{IRReal_proton}
\end{align}
where $ \bar{v}= 2 \sqrt{\tau \left( 1 + \tau \right)}/\left( 1+2 \tau \right)$, and the corresponding finite part $\delta_\text{s;pR}$:
\ber
\delta_\text{s;pR} &=& \left(\frac{-\alpha}{\pi}\right) \left\{   \ln\left(\frac{4\Delta E_s^2}{M^2}\right) \left[1 + \frac{1}{2 \bar{v}} \ln \left(\frac{1-\bar{v}}{1+\bar{v}} \right)\right] - I\left(\beta_p,~\beta_{p'},~\bar{v}\right) \right \} \nonumber \\
 &+& \left(\frac{-\alpha}{\pi}\right) \left \{ \frac{1}{2 \beta_p}\ln\left(\frac{1-\beta_p}{1+\beta_p}\right)+\frac{1}{2 \beta_{p'}}\ln\left(\frac{1-\beta_{p'}}{1+\beta_{p'}}\right) \right \}.\label{equation_soft_bremsstrahlung}
\eer
The initial ($\beta_p$) and final ($\beta_{p'}$) proton velocities in the lepton-pair c.m. frame are given by
\ber
\beta_p &=& \sqrt{1-\frac{4M^2 s_{ll}}{\left(s+t-M^2\right)^2}},\\
\beta_{p'} &=& \sqrt{1-\frac{4M^2 s_{ll}}{\left(s -s_{ll}-M^2\right)^2}}.
\eer
The soft-photon integral $I\left(\beta_p,~\beta_{p'},~\bar{v}\right) $ can be expressed as
\ber
I \left(\beta_p,~\beta_{p'},~\bar{v}\right) &=& g \left(\beta_p,~\beta_{p'},~\bar{v}\right) + g \left(\beta_{p'},~\beta_p,~\bar{v}\right)  + \left[ \bar{v} \leftrightarrow -\bar{v} \right] ,
\eer
with
\ber
g \left(\beta_p,~\beta_{p'},~\bar{v}\right) &=&\frac{1}{2\bar{v}}\mathrm{Li}_2 \left( \frac{1}{1+\beta_p} \left(\beta_p -\frac{1}{\bar{v}} + \frac{\sqrt{1-\bar{v}^2}}{\bar{v}} \frac{\sqrt{1-\beta_p^2}}{\sqrt{1-\beta_{p'}^2}} \right)\right) + \left[ \beta_p \leftrightarrow -\beta_p \right].
\eer
We provide an alternative expression for the integral $I$ in the Appendix \ref{soft_photon_integral}. 

Note the exact cancellation of the infrared divergences between the soft bremsstrahlung of Eq. (\ref{IRReal_proton}) and the infrared part of the proton vertex correction coming from $ 2 \tilde{F}_1 \left( t \right)$ with the renormalized form factor of Eq. (\ref{renormalized_QED_F1}), using the identification
\begin{equation}
    \bar{v}=\frac{2v}{v^2+1},
\end{equation}
with $v$ defined in Eq. \eqref{EqDef:v}.

\section{Results and discussions}
\label{results}
\begin{figure}[H]
	\includegraphics[scale=0.95]{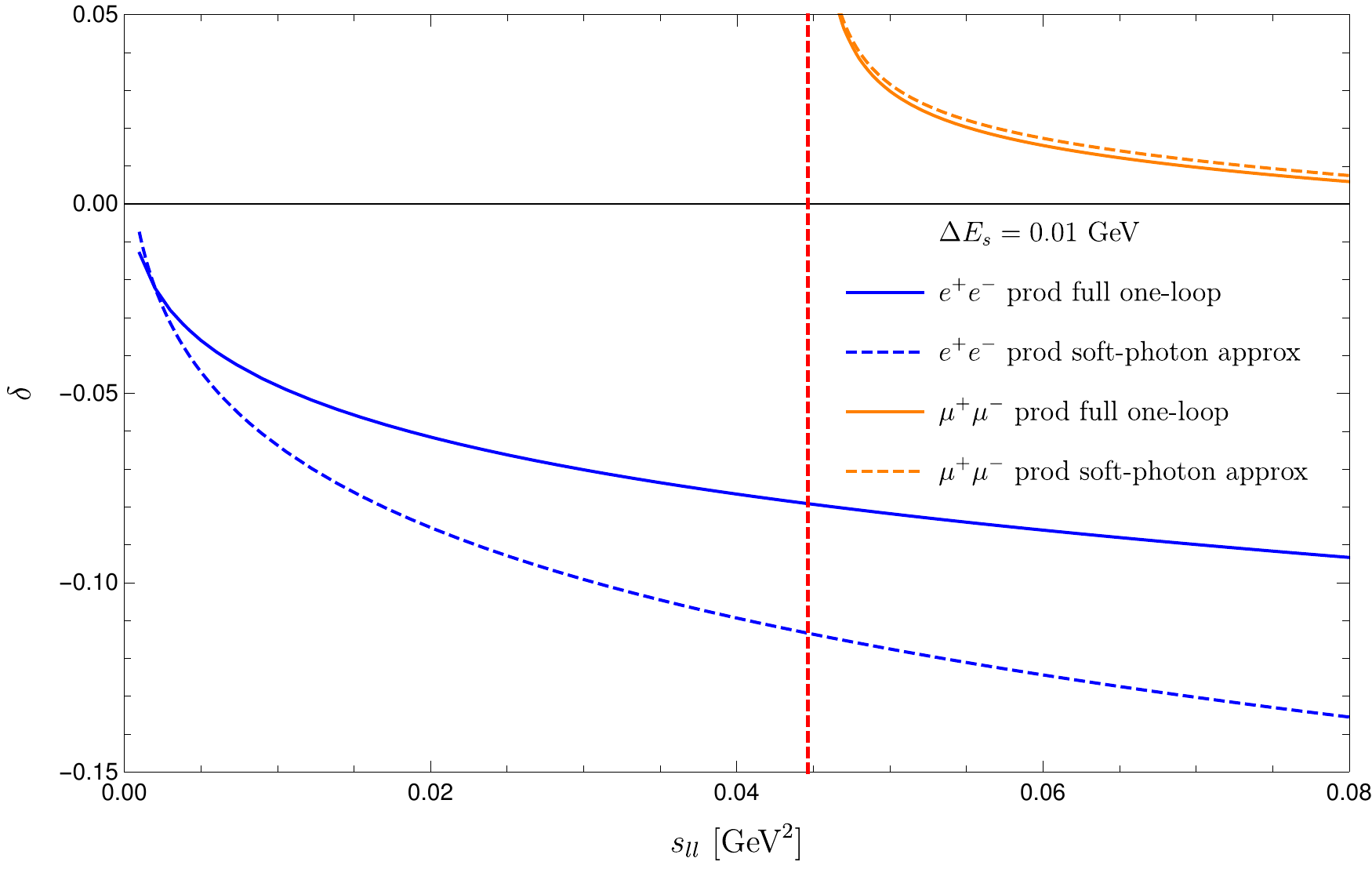}\hspace{0.cm}
	\caption{Comparison of first-order QED corrections to the cross section, including soft-photon bremsstrahlung with $\Delta E_s=0.01\text{ GeV}$ (solid lines), with the calculation in the soft-photon approximation (dashed lines). The vertical dashed red line indicates the muon-pair production threshold at $s_{ll}\approx 0.045\text{ GeV}^2$.\label{PlotCorrections}}
\end{figure}

In Fig. \ref{PlotCorrections} we show the radiative corrections to the cross section in the kinematical range of $s_{ll}$ between $0$ and $0.08\ \text{GeV}^2$ and compare with our previous result in the soft-photon approximation. The muon threshold is at $s_{ll}=4m_{\mu}^2\approx 0.045\ \text{GeV}^2$ (vertical dashed red line in Fig \ref{PlotCorrections}). We observe that the corrections for electrons are negative of order $10$ percent, while the corrections for muons are positive of order $1$ percent. The difference between the full one-loop calculation and the soft-photon approximation comes from terms which are not proportional to the double logarithmic form $\ln^2(m^2/s_{ll})^2$. 

\begin{figure}[H]
	\includegraphics[scale=0.95]{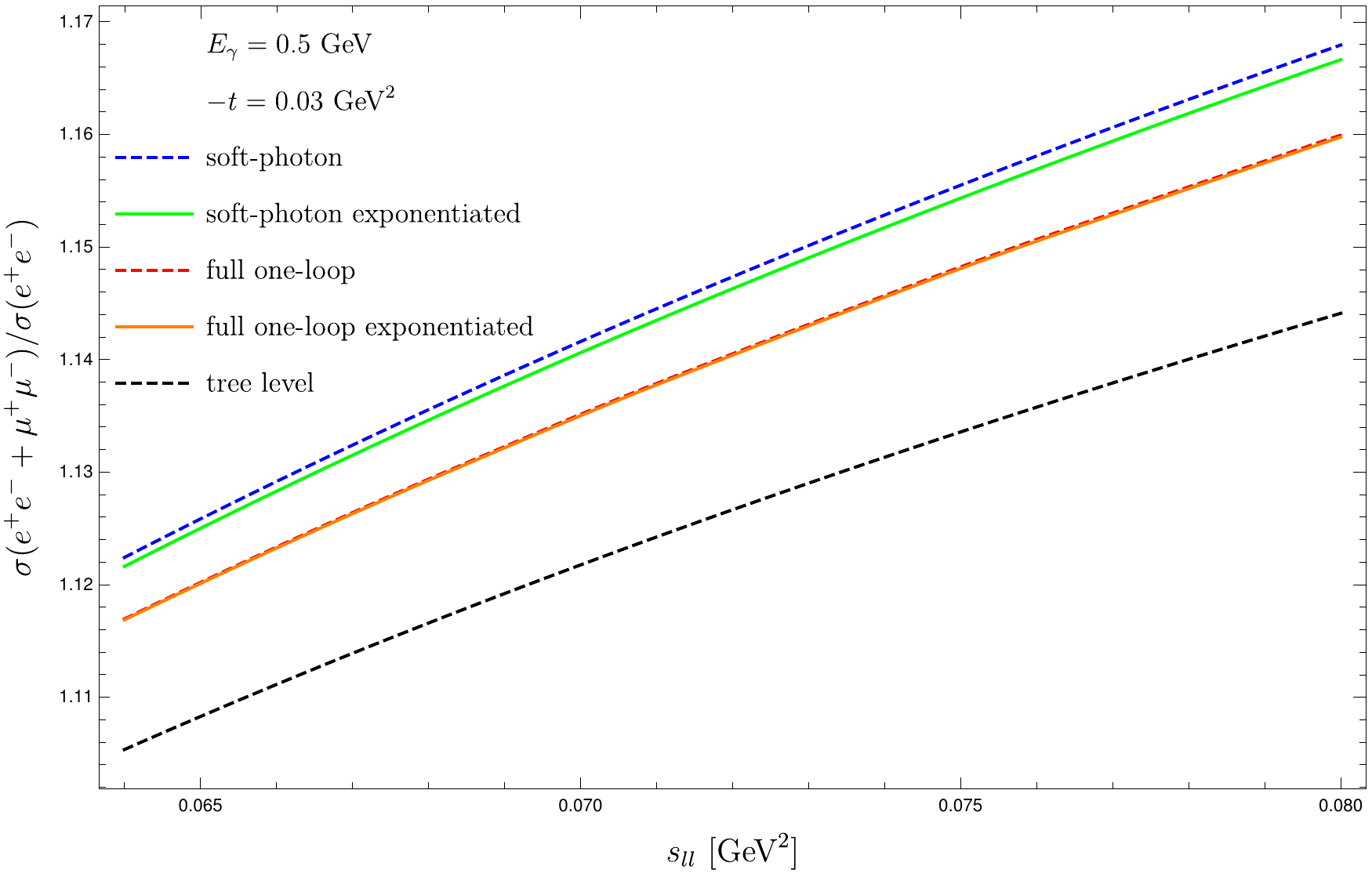}\hspace{0.cm}
	\caption{Effect of the one-loop radiative correction on the ratio $(d\sigma(e^+e^-)+d\sigma(\mu^+\mu^-))/d\sigma(e^+e^-)$, comparing with the soft-photon result of Ref. \cite{Heller:2018ypa}.\label{Plot-NLO-ratio}}
\end{figure}

In Fig. \ref{Plot-NLO-ratio} we show the effect of the full one-loop radiative corrections on the ratio defined in Eq. \eqref{ratio-def}. The exponentiation has a considerably smaller effect on the full one-loop calculation, than on the soft-photon approximation. Furthermore, the soft-photon approximation clearly overestimates the effect of radiative corrections in this calculation.

Taking radiative corrections into account, the ratio of Eq. \eqref{ratio-def} is now given by
\begin{equation}
R(s_{ll},s^0_{ll})\equiv\frac{\left[\sigma_0(\mu^+\mu^-)(1+\delta^{\mu})\right](s_{ll})+[\sigma_0(e^+e^-)(1+\delta^e)](s_{ll})}{[\sigma_0(e^+e^-)(1+\delta^e)](s^0_{ll})},
\end{equation}
which depends on the measured invariant lepton mass $s_{ll}$ and the reference point $s^0_{ll}$, to which the cross section is normalized. $\delta^e$ and $\delta^\mu$ are given by Eq. \eqref{CrossSec-oneloop-exp}. One chooses $s_{ll}^0<4m_{\mu}^2$, such that the reference measurement is below the muon-pair-production threshold, and only electron pairs are created.
\begin{figure}[H]
	\includegraphics[scale=0.95]{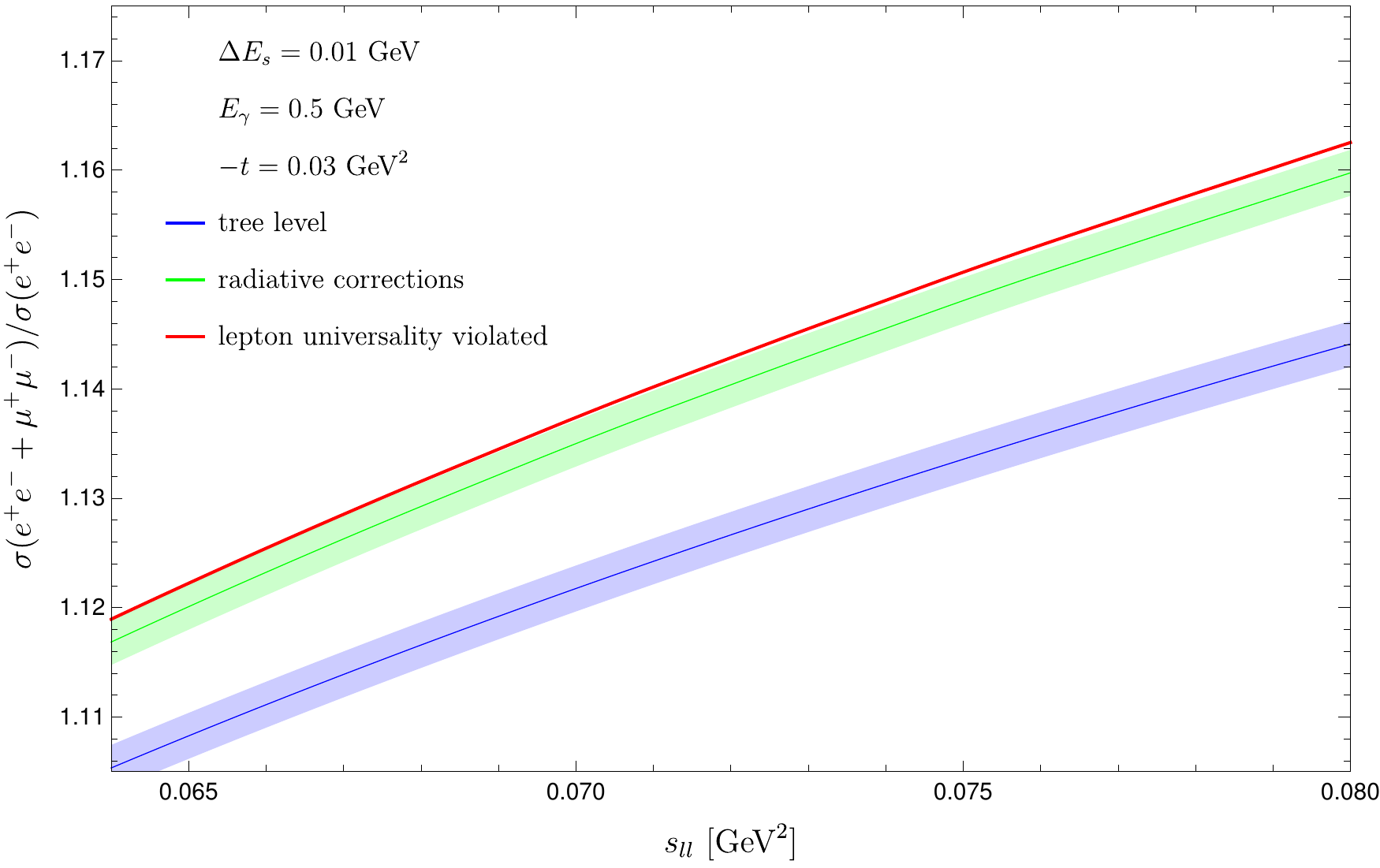}\hspace{0.cm}
	\caption{Ratio of cross sections between electron- and muon-pair production at tree level (blue curve) with account of full one-loop QED corrections estimated using $\Delta E_s=0.01 \text{ GeV}$ (green curve) with corresponding $3\sigma$ error bands. The red curve denotes the scenario when lepton universality is broken with $G^{\mu}_E/G^e_E=1.01$, including the one-loop radiative corrections.\label{Ratio1}}
\end{figure}

In Fig. \ref{Ratio1} we show the differential cross section ratio $R$ of Eq. \eqref{ratio-def}, including full one-loop QED corrections with $\Delta E_s=0.01\text{ GeV}$. The radiative corrections to $R$ are of the order of $1\%$. The red curve in Fig. \ref{Ratio1} shows the scenario when lepton universality is violated by $G^{\mu}_E/G^e_E=1.01$, which is an effect of order $0.2\%$. Following Ref. \cite{Pauk:2015oaa}, we use $3\sigma$ bands around the curves, with the experimental resolution $\sigma=7\times 10^{-4}$. One sees from this plot, that the inclusion of radiative corrections is indispensable, since the ratio of cross sections, defined in Eq. \eqref{ratio-def}, is shifted to higher values by more than the $3\sigma$ band. The statement that lepton universality can be tested with a $3\sigma$ confidence level remains true if one adds radiative corrections as can be seen in Fig. \ref{Ratio1}.

\begin{figure}
\begin{center}
	\includegraphics[scale=0.95]{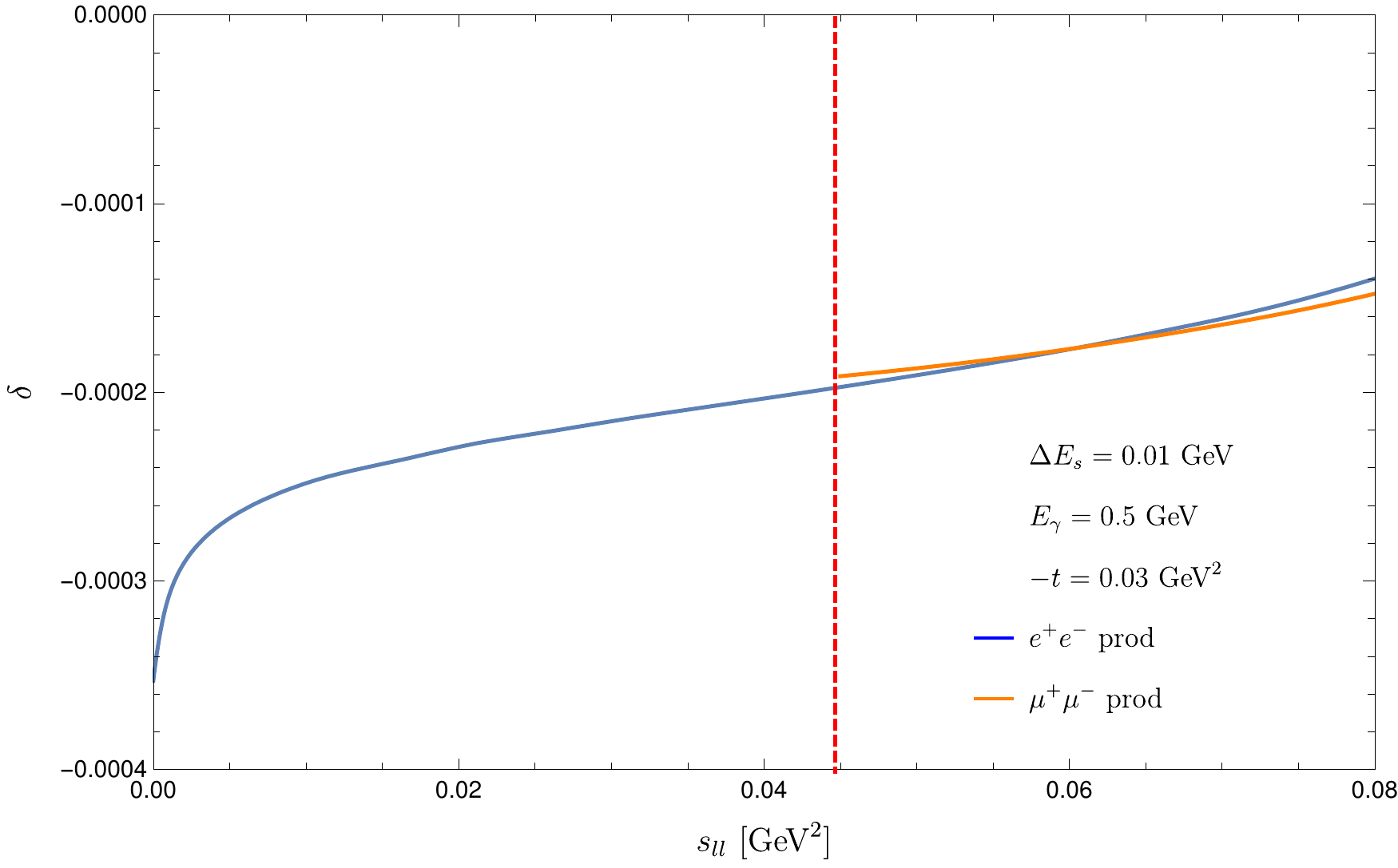}\hspace{0.cm}
\end{center}
	\caption{First-order radiative correction factor $\delta$, due to radiative corrections on the proton side. The vertical dashed red line indicates the muon-pair production threshold at $s_{ll} \approx 0.045~\mathrm{GeV}^2$.\label{proton_radiation1}}
\end{figure}
\begin{figure}
\begin{center}
	\includegraphics[scale=0.95]{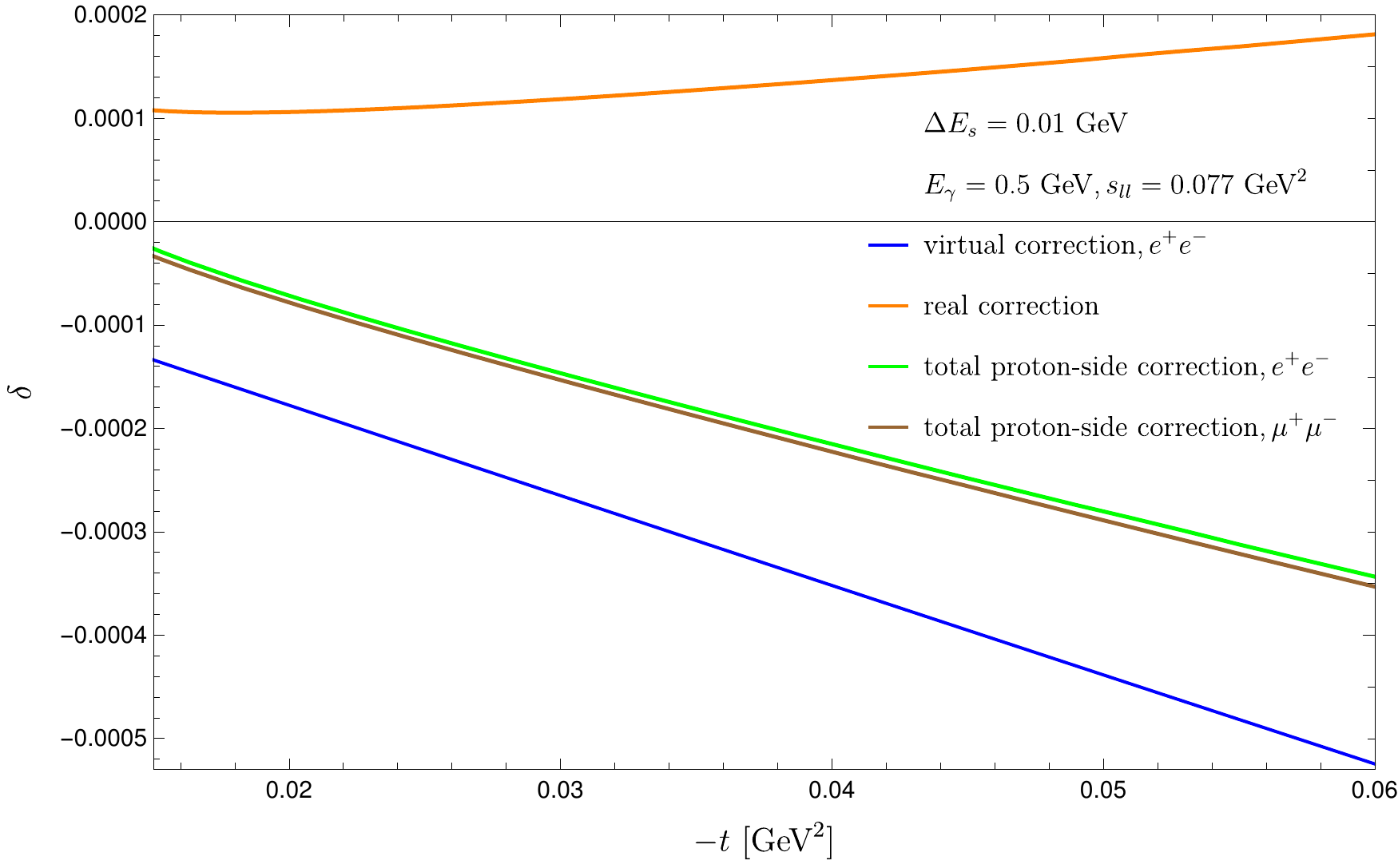}\hspace{0.cm}
\end{center}
	\caption{First-order radiative correction factor $\delta$ on the proton side as a function of $t$, with $ \Delta E_s = 0.01~\mathrm{GeV}$. \label{proton_radiation2}}
\end{figure}

We estimate the resulting correction on the proton side as a sum of soft bremsstrahlung (see Sec. \ref{proton_correction}) and renormalized vertex correction for the point-like particles with the on-shell form factors of Eqs. (\ref{renormalized_QED_F2}, \ref{renormalized_QED_F1}).  We choose the detector resolution $ \Delta E_s = 0.01~\mathrm{GeV}$. Both real and virtual contributions are almost independent of the lepton mass and $s_{ll}$ in the region $0.045~\mathrm{GeV}^2 \leq s_{ll} \leq 0.08~\mathrm{GeV}^2$. For the kinematics shown in Fig. \ref{proton_radiation1} (with $ t = -0.03~\mathrm{GeV}^2$), the total hadronic radiative correction amounts to $\delta\approx -(1.5-2)\times 10^{-4}$ in that range. The hadronic correction factor depends significantly on the momentum transfer $t$. We show this dependence for $s_{ll} = 0.077~\mathrm{GeV}^2$ and the photon beam energy $E_\gamma = 0.5~\mathrm{GeV}$ in Fig. \ref{proton_radiation2}. The resulting correction to the cross section is up to $ - 3\times10^{-4} \lesssim \delta \lesssim 0$. We see from Fig. \ref{proton_radiation2} that the proton vertex correction to the unpolarized cross section and the soft bremsstrahlung contributions from the proton line are both of order $10^{-4}$, but with opposite signs, resulting in an even smaller correction to the ratio of pair-production cross sections. Consequently, our approximation of the proton as a point QED particle is reliable for the envisaged precision $\approx 7\times 10^{-4}$ on the ratio of cross sections, which is required for a test of lepton universality.

\section{Conclusion}
\label{conclusion}
In this paper we calculated the first-order QED corrections to the Bethe-Heitler process in the $\gamma p\rightarrow l^+l^-p$ reaction, keeping the full lepton mass dependence. This reaction may serve as a test for lepton universality as the authors in Ref. \cite{Pauk:2015oaa} pointed out, and such experiment is presently in the planning stage at MAMI. 

The ratio of di-electron production cross sections above and below $\mu^+\mu^-$ threshold shows a sensitivity of $0.2\%$ when the difference between the larger proton charge radius from electron scattering is used versus the smaller proton radius which results from the muonic Hydrogen spectroscopy. Since the full one-loop radiative effects induce a correction around $1\%$ on this same ratio, its inclusion is indispensable in this comparison.

The calculation was done in two independent setups, whose results were found to be in perfect agreement. Furthermore, our calculation reproduces the correct leading logarithms of the soft-photon approximation, and is in agreement with Ref.\cite{Huld:1968zz} in the limit of small lepton masses.

We also showed, that hadronic corrections are negligible at the required level of precision, due to the cancellation of box type graphs after integration over the lepton angles. 

As a next step, we plan to extend our study to the reaction with an off-shell photon in the initial state, since the anticipated experiment at MAMI is designed to use a virtual photon.

\section*{Acknowledgements}

We  would like to thank Dr. Aleksandrs Aleksejevs for useful discussions.
This work was supported by the Deutsche Forschungsgemeinschaft (DFG, German Research Foundation), in part through the Collaborative Research Center [The Low-Energy Frontier of the Standard Model, Projektnummer 204404729 - SFB 1044], and in part through the Cluster of Excellence [Precision Physics, Fundamental Interactions, and Structure of Matter] (PRISMA$^+$ EXC 2118/1) within the German Excellence Strategy (Project ID 39083149). Matthias Heller was supported in part by GRK Symmetry Breaking (DFG/GRK 1581). Oleksandr Tomalak was supported in part by a NIST precision measurement grant and by the U. S. Department of Energy, Office of Science, Office of High Energy Physics, under Award Number DE-SC0019095. Shihao Wu was funded by the National Science and Engineering Research Council (NSERC) of Canada.  Our figures were generated using \texttt{Jaxodraw} \cite{Binosi:2003yf}, based on \texttt{AxoDraw} \cite{Vermaseren:1994je}. For the \texttt{Mathematica} plots we use the package \texttt{MaTeX}\footnote{https://github.com/szhorvat/MaTeX}.

Parts of this research were conducted using the supercomputer Mogon and/or advisory services offered by Johannes Gutenberg University Mainz (hpc.uni-mainz.de), which is a member of the AHRP (Alliance for High Performance Computing in Rhineland Palatinate,  www.ahrp.info) and the Gauss Alliance e.V.
\newpage
\appendix
\section{Contraction of the hadronic tensor $H_{\mu\nu}$ with the tree-level leptonic tensor $L_{0}^{\mu\nu}$}
\label{treecontraction}
In this appendix we give the explicit expression for the contraction between the hadronic tensor $H_{\mu\nu}$ and the tree-level leptonic tensor $L_0^{\mu\nu}$. It is given by:

\begin{align}\label{DirectTreeContraction}
H_{\mu\nu}L_{0,d}^{\mu\nu}&=\frac{4 H_1}{(m^2-t_{ll})^2(m^2-s_{ll}+t-t_{ll})}\left\{3 m^4 s_{ll}-m^4 t_{ll}+3 m^2 t s_{ll}-m^2 s_{ll} t_{ll}-m^2 s_{ll}^2-m^2 t_{ll}^2\right.\nonumber\\
&\left.+3 m^2 t t_{ll}+2 s_{ll} t_{ll}^2+s_{ll}^2
   t_{ll}-t s_{ll} t_{ll}+t^2 t_{ll}+t_{ll}^3-2 t t_{ll}^2+m^6-5 m^4 t-3 m^2 t^2\right\}\nonumber\\
&+\frac{H_2}{2(m^2-t_{ll})^2(m^2-s_{ll}+t-t_{ll})} \left\{ 16 M^2 m^6-12 s m^6+20 t m^6-12 s_{ll} m^6+18 M^4 m^4\right.\nonumber\\
&\left.+10 s^2 m^4-16 t^2 m^4-s_{ll}^2 m^4-28 M^2 s m^4+63 M^2 t m^4-23 s t m^4-8 M^2 u m^4+8 s u m^4\right.\nonumber\\
&\left.-32 t u m^4-49 M^2 s_{ll} m^4+21 s s_{ll} m^4+11 t s_{ll} m^4+20 u s_{ll} m^4-28 M^2 t_{ll} m^4+16 s t_{ll} m^4\right.\nonumber\\
&\left.-16 t t_{ll} m^4+4 s_{ll} t_{ll} m^4-11 M^2 t^2 m^2+12 s t^2 m^2+12 t u^2 m^2+2 M^2 s_{ll}^2 m^2+s s_{ll}^2 m^2\right.\nonumber\\
&\left.+u s_{ll}^2 m^2+16 M^2 t_{ll}^2 m^2-4 s t_{ll}^2 m^2+4 t t_{ll}^2 m^2+46 M^4 t m^2+12 s^2 t m^2-46 M^2 s t m^2\right.\nonumber\\
&\left.+11 t^2 u m^2-46 M^2 t u m^2+22 s t u m^2-36 M^4 s_{ll} m^2-10 s^2 s_{ll} m^2-8 u^2 s_{ll} m^2+38 M^2 s s_{ll} m^2\right.\nonumber\\
&\left.+7 M^2 t s_{ll} m^2-11 s t s_{ll} m^2+34 M^2 u s_{ll} m^2-18 s u s_{ll} m^2-8 t u s_{ll} m^2-20 M^4 t_{ll} m^2\right.\nonumber\\
&\left.-12 s^2 t_{ll} m^2+11 t^2 t_{ll} m^2+s_{ll}^2 t_{ll} m^2+32 M^2 s t_{ll} m^2-44 M^2 t t_{ll} m^2+8 s t t_{ll} m^2+8 M^2 u t_{ll} m^2\right.\nonumber\\
&\left.-8 s u t_{ll} m^2+20 t u t_{ll} m^2+24 M^2 s_{ll} t_{ll} m^2-4 s s_{ll} t_{ll} m^2-4 t s_{ll} t_{ll} m^2-4 u s_{ll} t_{ll} m^2-4 M^2 t_{ll}^3\right.\nonumber\\
&\left.+2 M^4 t_{ll}^2+2 s^2 t_{ll}^2-3 t^2 t_{ll}^2-4 M^2 s t_{ll}^2+13 M^2 t t_{ll}^2-s t t_{ll}^2-4 t u t_{ll}^2-7 M^2 s_{ll} t_{ll}^2-s s_{ll} t_{ll}^2+t s_{ll} t_{ll}^2\right.\nonumber\\
&\left.+3 M^2 t^2 t_{ll}-4 s t^2 t_{ll}-4 t u^2 t_{ll}-2 M^2 s_{ll}^2 t_{ll}-s s_{ll}^2 t_{ll}-u s_{ll}^2 t_{ll}-14 M^4 t t_{ll}-4 s^2 t t_{ll}+14 M^2 s t t_{ll}\right.\nonumber\\
&\left.-3 t^2 u t_{ll}+14 M^2 t u t_{ll}-6 s t u t_{ll}+4 M^4 s_{ll} t_{ll}+2 s^2 s_{ll} t_{ll}-6 M^2 s s_{ll} t_{ll}+M^2 t s_{ll} t_{ll}+3 s t s_{ll} t_{ll}\right.\nonumber\\
&\left.-2 M^2 u s_{ll} t_{ll}+2 s u s_{ll} t_{ll}\right\}.
\end{align}
The full tree-level contribution $H_{\mu\nu}L_{0}^{\mu\nu}$ is given by the sum of the two contractions of the direct lepton-tensor $H_{\mu\nu} L_{0,d}^{\mu\nu}$ and the crossed lepton-tensor $H_{\mu\nu}L_{0,c}^{\mu\nu}$ with the hadronic tensor $H_{\mu\nu}$, which can be derived by making the replacements $t_{ll}\rightarrow u_{ll}$ and $u \rightarrow -u - s - t + 3M^2+2m^2$ in $H_{\mu\nu}L_{0,d}^{\mu\nu}$.

\section{Master integrals} \label{master_integrals}
Here we give analytical expressions for all scalar master integrals which are needed for the calculation. The integrals $A_0(m^2)$ and $B_0\left(s_{ll},m^2,m^2\right)$ are needed up to order $\epsilon$, since they get multiplied by a factor proportional to $\frac{1}{\epsilon}$ stemming from the IBP identities. All other integrals are needed only up to order  $\epsilon^0$.

All integrals, except for $C_0(t_{ll},t,m^2,0,m^2,m^2)$, can be found in the literature, e.g., in \url{http://qcdloop.fnal.gov/}. The analytic expression for  $C_0(t_{ll},t,m^2,0,m^2,m^2)$ is to our knowledge a new result of this work.

We give the analytical results in the physical region:
\begin{align}
s_{ll}&>4m^2,\\
t&<0,\\
t_{ll}&<0,
\end{align}
in terms of real-valued logarithms and dilogarithms in this region.
We use the kinematical quantities:
\begin{align}
\beta_x&=\sqrt{1-\frac{4m^2}{x}},\\
\lambda&=\sqrt{-2 t_{ll} \left(m^2+t\right)+t_{ll}^2+\left(m^2-t\right)^2}.
\end{align}
Note that the $A$, $B$ and $C$ functions used in sections \ref{SectionSelfEnergy} and \ref{SectionVertex} always refer to the finite part, i.e. the coefficient in front of $\epsilon^0$ of the expansion, given in this appendix.

In the following, we use the notation:
\begin{equation}
    \frac{1}{\bar{\epsilon}}\equiv\frac{1}{\epsilon}-\gamma_E+\ln 4\pi,
\end{equation}
with $\epsilon=2-D/2$.

The tadpole up to order $\epsilon$ is given by:
\begin{align}
A_0\left(m^2\right)&=m^2\left\{\frac{1}{\bar{\epsilon}}+\left[1-\text{ln}\left(\frac{m^2}{\mu^2}\right)\right]+\left[-1-\frac{\pi^2}{6}+\text{ln}\left(\frac{m^2}{\mu^2}\right)-\frac{1}{2}\text{ln}^2\left(\frac{m^2}{\mu^2}\right)\right]\epsilon\right\}.
\end{align}

The two-point function are given by:
\begin{align}
B_0\left(t_{ll},0,m^2\right)&=\frac{1}{\bar{\epsilon}}+2-\ln\left(\frac{m^2}{\mu^2}\right)+\frac{t_{ll} -m^2}{t_{ll}}\ln\left(\frac{m^2}{m^2 - t_{ll}}\right),\\
B_0\left(s_{ll},m^2,m^2\right)&=\frac{1}{\bar{\epsilon}}+\left\{2-\text{ln}\left(\frac{m^2}{\mu^2}\right)+\beta_{s_{ll}}\left[i \pi- \ln\left(\frac{1+\beta_{s_{ll}}}{1-\beta_{s_{ll}}}\right)\right]\right\}\nonumber\\
&+\epsilon\left\{4+\frac{\pi^2}{6}-2\ln\left(\frac{m^2}{\mu^2}\right)+\frac{1}{2}\ln^2\left(\frac{m^2}{\mu^2}\right)+\beta_{s_{ll}}\left[-\frac{2}{3}\pi^2+\ln^2\left(\frac{1-\beta_{s_{ll}}}{2\beta_{s_{ll}}}\right)\nonumber\right.\right.\\
&\left.\left.+\ln\left(\frac{m^2}{\mu^2}\right)\ln\left(\frac{1+\beta_{s_{ll}}}{1-\beta_{s_{ll}}}\right)-\frac{1}{2}\ln^2\left(\frac{1+\beta_{s_{ll}}}{1-\beta_{s_{ll}}}\right)+2\text{Li}_2\left(\frac{\beta_{s_{ll}}-1}{2\beta_{s_{ll}}}\right)\right]\right.\nonumber\\
&\left.+i\pi\beta_{s_{ll}}\left[2-\ln\left(\frac{m^2}{\mu^2}\right)+\ln\left(\frac{1-\beta_{s_{ll}}}{1+\beta_{s_{ll}}}\right)+2\ln\left(\frac{1-\beta_{s_{ll}}}{2\beta_{s_{ll}}}\right)\right]\right\},\\
B_0\left(t,m^2,m^2\right)&=\frac{1}{\bar{\epsilon}}+\left\{2-\text{ln}\left(\frac{m^2}{\mu^2}\right)+\beta_{t}\ln\left(\frac{\beta_{t}-1}{\beta_{t}+1}\right)\right\}.
\end{align}

The three-point functions are given by:
\begin{align}
&C_0(t,s_{ll},0,m^2,m^2,m^2)=\frac{1}{2}\frac{1}{s_{ll}-t}\left\{\left[i\pi+\ln\left(\frac{1-\beta_{s_{ll}}}{1+\beta_{s_{ll}}}\right)\right]^2+\ln^2\left(\frac{\beta_t-1}{\beta_t+1}\right)\right\},\\
&C_0(0,m^2,t_{ll},m^2,m^2,0)=\frac{-1}{m^2-t_{ll}}\left\{\frac{\pi^2}{6}-\text{Li}_2\left(\frac{t_{ll}}{m^2}\right)\right\},\\
&C_0(t_{ll},t,m^2,0,m^2,m^2)\ \ \;=\frac{1}{\lambda}\left\{\frac{1}{2}\ln^2\left(\frac{-t_{ll}}{m^2-t-t_{ll}+\lambda}\right)+\ln^2\left(\frac{-t_{ll}}{m^2-t+t_{ll}+\lambda}\right)\right.\nonumber\\
&\left.-\frac{1}{2}\ln^2\left(\frac{-t_{ll}}{-m^2+t-t_{ll}+\lambda}\right)-\frac{1}{2}\ln^2\left(\frac{m^2-t-t_{ll}+\lambda}{m^2-t+t_{ll}+\lambda}\right)-\ln\left(\frac{m^2-t_{ll}}{m^2}\right)\ln\left(\frac{-m^2+t-t_{ll}+\lambda}{m^2-t+t_{ll}+\lambda}\right)\right.\nonumber\\
&\left.+\ln(2)\ln\left(\frac{2\;t_{ll}^2\;(-m^2+t-t_{ll}+\lambda)}{(m^2-t+t_{ll}+\lambda)^2(m^2-t-t_{ll}+\lambda)}\right)-\frac{1}{2} \ln^2\left(\frac{-m^2 t_{ll}-t t_{ll} \beta _t+t t_{ll}+m^4+\lambda  m^2-m^2 t}{-t_{ll} \left(2 m^2+t \beta _t-t\right)}\right)\right.\nonumber\\
&\left.-\frac{1}{2} \ln^2\left(\frac{-m^2 t_{ll}+t t_{ll} \beta _t+t t_{ll}+m^4+\lambda  m^2-m^2 t}{-t_{ll} \left(2 m^2-t \beta _t-t\right)}\right)+\text{Li}_2\left(\frac{m^2-t+\lambda +t_{ll}}{2 t_{ll}}\right)\right.\nonumber\\
&\left.+\text{Li}_2\left(\frac{-m^2+t+\lambda +t_{ll}}{2 t_{ll}}\right)-\text{Li}_2\left(\frac{t_{ll}\left(2 m^2-t-t \beta _t\right)}{-m^4+t m^2-\lambda  m^2+t_{ll} m^2-t t_{ll}-t t_{ll} \beta _t}\right)\right.\nonumber\\
&\left.-\text{Li}_2\left(\frac{-m^4+t m^2+\lambda  m^2+t_{ll} m^2-t t_{ll}-t t_{ll} \beta _t}{t_{ll} \left(2 m^2-t-t \beta _t\right)}\right)-\text{Li}_2\left(\frac{t_{ll} \left(2 m^2-t+t \beta _t\right)}{-m^4+t m^2-\lambda m^2+t_{ll} m^2-t t_{ll}+t t_{ll} \beta _t}\right)\right.\nonumber\\
&\left.-\text{Li}_2\left(\frac{-m^4+t m^2+\lambda  m^2+t_{ll} m^2-t t_{ll}+t t_{ll} \beta_t}{t_{ll} \left(2 m^2-t+t \beta _t\right)}\right)\right\}.
\end{align}

We also need the following four-point function:
\begin{align}
&D_0(m^2,0,t,m^2,t_{ll},s_{ll},0,m^2,m^2,m^2)=\frac{1}{s_{ll}\beta_{s_{ll}}(t_{ll}-m^2)}\left\{\frac{1}{\bar{\epsilon}}\left[i\pi+\ln\left(\frac{1-\beta_{s_{ll}}}{1+\beta_{s_{ll}}}\right)\right]\right.\nonumber\\
&\left.-\frac{7}{6}\pi^2-2\ln\left(\frac{m^2-t_{ll}}{m^2}\right)\left[i\pi+\ln\left(\frac{1-\beta_{s_{ll}}}{1+\beta_{s_{ll}}}\right)\right]-2\ln\left(\frac{4\beta_{s_{ll}}}{(1+\beta_{s_{ll}})^2}\right)\left[i\pi+\ln\left(\frac{1-\beta_{s_{ll}}}{1+\beta_{s_{ll}}}\right)\right]\right.\nonumber\\
&-\left.\ln\left(\frac{m^2}{\mu^2}\right)\left[i\pi+\ln\left(\frac{1-\beta_{s_{ll}}}{1+\beta_{s_{ll}}}\right)\right]+\ln^2 \left(\frac{2 \left(\beta _t-\beta_{s_{ll}}\right)}{\left(\beta _t-1\right) \left(1+\beta_{s_{ll}}\right)}\right)-\ln^2 \left(\frac{\beta _t-1}{\beta _t+1}\right)\right.\nonumber\\
   &\left.+2 \ln \left(\frac{2 \left(\beta _t-\beta_{s_{ll}}\right)}{\left(\beta _t-1\right) \left(1+\beta_{s_{ll}}\right)}\right)\left[-\ln \left(\frac{\left(\beta _t+1\right) \left(1-\beta_{s_{ll}}\right)}{\left(\beta _t-1\right) \left(1+\beta_{s_{ll}}\right)}\right)+\ln \left(\frac{\beta _t+1}{\beta _t-1}\right)+i \pi \right]\right.\nonumber\\
   &\left.+2 \ln \left(\frac{1-\beta_{s_{ll}}}{1+\beta_{s_{ll}}}\right) \left[\ln \left(\frac{2 \left(\beta _t-\beta_{s_{ll}}\right)}{\left(\beta _t-1\right) \left(\beta _{s_{ll}}+1\right)}\right)+\ln \left(\frac{2 \left(\beta_{s_{ll}}+\beta _t\right)}{\left(\beta _t+1\right) \left(1+\beta_{s_{ll}}\right)}\right)\right]+2 \ln \left(\frac{\beta _t-1}{\beta _t+1}\right)\right.\nonumber\\
 &\times\left. \ln \left(\frac{2 \left(\beta _{s_{ll}}+\beta _t\right)}{\left(\beta _t+1\right) \left(\beta _{s_{ll}}+1\right)}\right)-2 \ln \left(\frac{2\left(\beta _{s_{ll}}+\beta _t\right)}{\left(\beta _t+1\right) \left(\beta _{s_{ll}}+1\right)}\right)\left[\ln \left(\frac{\left(\beta _t-1\right) \left(1-\beta _{s_{ll}}\right)}{\left(\beta _t+1\right) \left(1+\beta _{s_{ll}}\right)}\right)-i \pi \right]\right.\nonumber\\
 &\left.+\ln^2 \left(\frac{2 \left(\beta _{s_{ll}}+\beta _t\right)}{\left(\beta _t+1\right) \left(1+\beta _{s_{ll}}\right)}\right)-\text{Li}_2\left(\frac{\left(1-\beta _{s_{ll}}\right)^2}{\left(1+\beta _{s_{ll}}\right)^2}\right)+2 \text{Li}_2\left(\frac{\left(\beta _t-1\right) \left(\beta_{s_{ll}}+1\right)}{2 \left(\beta _t-\beta _{s_{ll}}\right)}\right)\right.\nonumber\\
 &\left.+2 \text{Li}_2\left(\frac{\left(\beta _t+1\right) \left(\beta _{s_{ll}}+1\right)}{2 \left(\beta_t+\beta _{s_{ll}}\right)}\right)\right\}.
\end{align}
\section{Master integrals for small lepton mass} \label{master_integrals_massles}
Here we give the master integrals in the expansion for small $m^2$, keeping only terms propotional to $\ln(m^2)$.

The two-point functiosn are given by:
\begin{align}
B_0(t_{ll},0,m^2)&=\frac{1}{\bar{\epsilon}}+2-\ln\left(\frac{-t_{ll}}{m^2}\right)-\ln\left(\frac{m^2}{\mu^2}\right),\\
B_0(s_{ll},m^2,m^2)&=\frac{1}{\bar{\epsilon}}+\left[2+i\pi-\ln\left(\frac{s_{ll}}{m^2}\right)-\ln\left(\frac{m^2}{\mu^2}\right)\right]+\epsilon\left[4-2\ln\left(\frac{m^2}{\mu^2}\right)+\frac{1}{2}\ln^2\left(\frac{m^2}{\mu^2}\right)\right.\nonumber\\
&\left.+2i\pi-\frac{\pi^2}{2}+(-2-i\pi)\ln\left(\frac{s_{ll}}{m^2}\right)+\frac{1}{2}\ln^2 \frac{s_{ll}}{m^2}\right].
\end{align}

The three-point functions are given by:
\begin{align}
C_0(0,t,s_{ll},m^2,m^2,m^2)&=\frac{1}{2(t-s_{ll})}\left[\pi^2 + 2i\pi\ln \frac{s_{ll}}{m^2} - \ln^2 \frac{s_{ll}}{m^2} + \ln^2 \frac{-t}{m^2}\right],\\
C_0(0,m^2,t_{ll},m^2,m^2,0)&=\frac{1}{t_{ll}}\left[\frac{\pi^2}{3}+\frac{1}{2}\ln^2 \frac{-t_{ll}}{m^2}\right],\\
C_0(t,m^2,t_{ll},m^2,m^2,0)&=\frac{1}{2(t-t_{ll})}\left[\ln^2 \frac{-t}{m^2}-\ln^2 \frac{-t_{ll}}{m^2}-4\text{Li}_2\left(\frac{t_{ll}-t}{t_{ll}}\right)\right].
\end{align}

The four-point function for small lepton mass is given by:
\begin{align}
&D_0(m^2,0,t,t_{ll},s_{ll},0,m^2,m^2,m^2)=\frac{1}{\bar{\epsilon}}\frac{1}{s_{ll}t_{ll}}\left[i\pi-\ln\left(\frac{s_{ll}}{m^2}\right)\right]\nonumber\\
&+\frac{1}{s_{ll}t_{ll}}\left\{\ln \left(\frac{s_{ll}}{m^2}\right) \left[-2 \ln
   \left(\frac{s_{ll}-t}{s_{ll}}\right)+2 \ln
   \left(-\frac{t_{ll}}{s_{ll}}\right)+\ln \left(\frac{m^2}{\mu ^2}\right)-2 i
   \pi \right]\right.\nonumber\\
   &\left.+\ln \left(\frac{s_{ll}-t}{s_{ll}}\right) \left[-2 \ln
   \left(-\frac{t}{s_{ll}}\right)+2 \ln \left(-\frac{t}{m^2}\right)+2 i \pi \right]+2 \ln ^2\left(\frac{s_{ll}}{m^2}\right)+2 \text{Li}_2\left(\frac{s_{ll}}{s_{ll}-t}\right)\right.\nonumber\\
   &\left.+\ln^2\left(\frac{s_{ll}-t}{s_{ll}}\right)-2 i \pi  \ln\left(-\frac{t_{ll}}{s_{ll}}\right)-i \pi  \ln \left(\frac{m^2}{\mu^2}\right)-\ln ^2\left(-\frac{t}{m^2}\right)-\frac{5 \pi ^2}{6}\right\}
\end{align}

\section{Additional form factor $F_3$\label{F3}}
The third pair of half-off-shell form factors entering Eq. \eqref{VertexDec}, which do not contribute to physical quantities, are given by:
\begin{align}
F_{3^+}&(s',q^2)=\frac{\alpha}{4\pi}\times\frac{1}{s' \left(m^4-2 m^2 s'-2 m^2 q^2+s'^2-2 s' q^2+q^4\right)^2}\nonumber\\
&\times\left\{-4 m^2 s' \left(m^2-s'\right) \left(q^2 \left(m^2-5 s'\right)+4 \left(m^2-s'\right)^2+q^4\right) B_0\left(q^2,m^2,m^2\right)\right.\nonumber\\
&\left.+2 m^2 \left[\vphantom{\frac{1}{2}}m^8-m^6 \left(10
   s'+3 q^2\right)+m^4 \left(4 s'+q^2\right) \left(s'+3 q^2\right)+m^2 \left(18 s'^3-s'^2 q^2-6 s' q^4-q^6\right)\right.\right.\nonumber\\
&\left.\left.-s'\left(s'-q^2\right) \left(13 s'^2-10 s' q^2+3 q^4\right)\vphantom{\frac{1}{2}}\right] B_0\left(s',0,m^2\right)+4 m^2 s' \left(m^2-s'\right) \left[\vphantom{\frac{1}{2}}8 m^6-m^4
   \left(14 s'+5 q^2\right)\right.\right.\nonumber\\
&\left.\left.+m^2 \left(4 s'^2-6 s' q^2+4 q^4\right)+\left(s'-q^2\right)^2 \left(2 s'-q^2\right)\vphantom{\frac{1}{2}}\right]
   C_0\left(m^2,q^2,s',0,m^2,m^2\right)\right.\nonumber\\
&\left.-2  \left[\vphantom{\frac{1}{2}}m^8-3 m^6 \left(6 s'+q^2\right)+m^4 \left(28 s'^2+11 s' q^2+3 q^4\right)-m^2 \left(6
   s'^3-11 s'^2 q^2+8 s' q^4+q^6\right)\right.\right.\nonumber\\
&\left.\left.-s' \left(5 s'-3 q^2\right) \left(s'-q^2\right)^2\vphantom{\frac{1}{2}}\right]A_0\left(m^2\right)\right.\nonumber\\
&\left.+4 m^2 s' \left(9 m^6-m^4 \left(17
   s'+7 q^2\right)+m^2 \left(7 s'^2-6 s' q^2+5 q^4\right)+\left(s'-q^2\right)^3\right)\right\}.\label{F3+}
\end{align}

\begin{align}
F_{3^-}&(s',q^2)=\frac{\alpha}{4\pi}\times\frac{1}{s' \left(m^4-2 m^2 s'-2 m^2 q^2+s'^2-2 s' q^2+q^4\right)^2}\nonumber\\
&\times\left\{-4 m^2 s' \left[\vphantom{\frac{1}{2}}2 m^6+2 m^4 \left(s'+7 q^2\right)+m^2 \left(-10 s'^2+8 s' q^2-13 q^4\right)+3 \left(s'-q^2\right) \left(2
   s'^2-q^4\right)\right]\right.\nonumber\\
&\left.\times B_0\left(q^2,m^2,m^2\right)+2 m^2 \left[\vphantom{\frac{1}{2}}m^8-m^6 \left(8 s'+3 q^2\right)+m^4 \left(-14 s'^2+5 s' q^2+3 q^4\right)\right.\right.\nonumber\\
&\left.\left.-m^2 q^2 \left(29
   s'^2-4 s' q^2+q^4\right)+s' \left(s'-q^2\right) \left(21 s'^2-16 s' q^2+q^4\right)\vphantom{\frac{1}{2}}\right] B_0\left(s',0,m^2\right)\right.\nonumber\\
&\left.+4 m^2 s'
   \left[\vphantom{\frac{1}{2}}6 m^8+m^6 \left(4 s'+13 q^2\right)+m^4 \left(-24 s'^2+3 s' q^2-22 q^4\right)\right.\right.\nonumber\\
&\left.\left.+m^2 \left(12 s'^3-13 s'^2 q^2-8 s' q^4+11
   q^6\right)+\left(s'-q^2\right)^2 \left(2 s'^2+s' q^2-2 q^4\right)\vphantom{\frac{1}{2}}\right] C_0\left(m^2,q^2,s',0,m^2,m^2\right)\right.\nonumber\\
&\left.+2  \left[\vphantom{\frac{1}{2}}-m^8+3 m^6 \left(4
   s'+q^2\right)+m^4 \left(18 s'^2+23 s' q^2-3 q^4\right)\right.\right.\nonumber\\
&\left.\left.+m^2 \left(-20 s'^3+45 s'^2 q^2-30 s' q^4+q^6\right)-s' \left(9 s'-7
   q^2\right) \left(s'-q^2\right)^2\vphantom{\frac{1}{2}}\right]A_0\left(m^2\right)\right.\nonumber\\
&\left.+8 m^2 s' \left(5 m^6+m^4 s'+m^2 \left(-5 s'^2+5 s' q^2-3
   q^4\right)-\left(s'-q^2\right)^3\right)\vphantom{\frac{1}{2}}\right\}.\label{F3-}
\end{align}
\section{Soft-bremsstrahlung integral}
\label{soft_photon_integral}

The soft-bremsstrahlung integral $I$ of Eq. (\ref{equation_soft_bremsstrahlung}) can be expressed as \cite{Vanderhaeghen:2000ws}
\ber
I &=& -\frac{ 1- \beta_{p'} \beta_{p} \cos \theta_{p p'} }{2} \int \limits^{+1}_{-1} \frac{ \mathrm{d} y}{\beta_y\left(1-\beta_y^2\right)} \ln \frac{1-\beta_y}{1+\beta_y}, \label{equation_I2}
\eer
with the following notation:
\ber
\vec{\beta}_y &=& \frac{1+y}{2} \vec{\beta}_{p'} +  \frac{1-y}{2} \vec{\beta}_{p},
\eer
where the relative angle between the initial and final protons $\theta_{p' p}$ in the di-lepton pair rest frame is given by
\ber
\cos \theta_{p' p} &=& \frac{1}{\beta_{p'} \beta_{p}} \left(1-\frac{\sqrt{1 - \beta_p^2}\sqrt{1 - \beta_p'^2}}{\sqrt{1-v^2}} \right).
\eer

The integration in Eq. (\ref{equation_I2}) was performed in Ref. \cite{Vanderhaeghen:2000ws}. The resulting integral can be expressed as
\ber
I & = & -\frac{1}{2} \frac{ 1- \beta_{p'} \beta_{p} \cos \theta_{p p'} }{|\vec{\beta}_{p} - \vec{\beta}_{p'} |\tanh \alpha} \left( \left( -2 \ln 2 + \frac{1}{2} \ln \left( \sinh^2 \alpha - \sinh^2 \phi_1 \right) \right) \ln \frac{\sinh \alpha + \sinh \phi_1}{\sinh \alpha - \sinh \phi_1}   \right.  \nonumber \\
& - &  \left. \ln  \left( \sinh \alpha + \sinh \phi_1 \right) \ln \frac{\sinh \alpha - \sinh \phi_1}{4 \sinh^2 \alpha } +  2   \ln \left ( e^{-\alpha} \frac{e^{\alpha} + e^{\phi_1}}{e^{-\alpha} + e^{\phi_1}} \right ) \ln \frac{\cosh \alpha + \cosh \phi_1}{\cosh \alpha - \cosh \phi_1} \right.\nonumber \\
& - & \left. 2  \mathrm{Li}_2 \left( \frac{\sinh \alpha + \sinh \phi_1}{2 \sinh \alpha }\right) + \mathrm{Li}_2 \left[ \left( \frac{e^{\alpha} - e^{\phi_1}}{e^{\alpha} + e^{\phi_1}}\right)^2 \right]  - \mathrm{Li}_2 \left[ \left(  \frac{e^{-\alpha} - e^{\phi_1}}{e^{-\alpha} + e^{\phi_1}} \right)^2 \right]  - \left[ \phi_1 \leftrightarrow \phi_2 \right] \right), \nonumber \\
\eer
with
\ber
\cosh \alpha = \frac{|\vec{\beta}_{p} - \vec{\beta}_{p'} |}{\beta_{p} \beta_{p'} \sin \theta_{p p'}}, \quad \sinh \phi_1 = \frac{ \beta^2_{p'} -\beta_{p} \beta_{p'} \cos \theta_{p p'} }{\beta_{p} \beta_{p'} \sin \theta_{p p'}}, \quad \sinh \phi_2 = \frac{ - \beta^2_{p} +\beta_{p} \beta_{p'} \cos \theta_{p p'} }{\beta_{p} \beta_{p'} \sin \theta_{p p'}}. \nonumber \\
\eer


\begin{thebibliography}{99}
\bibitem{Bernauer:2010wm} 
  J.~C.~Bernauer {\it et al.} [A1 Collaboration],
  Phys.\ Rev.\ Lett.\  {\bf 105}, 242001 (2010).



\bibitem{Bernauer:2013tpr} 
  J.~C.~Bernauer {\it et al.} [A1 Collaboration],
  Phys.\ Rev.\ C {\bf 90}, no. 1, 015206 (2014).



\bibitem{Pohl:2010zza} 
  R.~Pohl {\it et al.},
  Nature {\bf 466}, 213 (2010).



\bibitem{Antognini:1900ns} 
  A.~Antognini {\it et al.},
  Science {\bf 339}, 417 (2013).

\bibitem{Beyer:2017}
A. Beyer {\it et al.}, Science {\bf 358}, 79 (2017). 


\bibitem{Fleurbaey:2018fih} 
  H.~Fleurbaey {\it et al.},
  Phys.\ Rev.\ Lett.\  {\bf 120}, no. 18, 183001 (2018).

\bibitem{Mihovilovic:2019jiz} 
  M.~Mihovilovic {\it et al.},
  arXiv:1905.11182 [nucl-ex].

\bibitem{Vutha2012}  
A. Vutha {\it et al.}, 2012, BAPS.DAMOP.D1.138.


\bibitem{Gasparian:2014rna} 
  A.~Gasparian [PRad at JLab Collaboration],
  EPJ Web Conf.\  {\bf 73}, 07006 (2014).
  
  
  
\bibitem{Lorenz:2012tm} 
  I.~T.~Lorenz, H.-W.~Hammer and U.~G.~Meissner,
  Eur.\ Phys.\ J.\ A {\bf 48}, 151 (2012).



\bibitem{Lorenz:2014vha} 
  I.~T.~Lorenz and U.~G.~Meißner,
  Phys.\ Lett.\ B {\bf 737}, 57 (2014).



\bibitem{Lorenz:2014yda} 
  I.~T.~Lorenz, U.~G.~Meißner, H.-W.~Hammer and Y.-B.~Dong,
  Phys.\ Rev.\ D {\bf 91}, no. 1, 014023 (2015).



\bibitem{Lee:2015jqa} 
  G.~Lee, J.~R.~Arrington and R.~J.~Hill,
  Phys.\ Rev.\ D {\bf 92}, no. 1, 013013 (2015).



\bibitem{Arrington:2015ria} 
  J.~Arrington and I.~Sick,
  J.\ Phys.\ Chem.\ Ref.\ Data {\bf 44}, 031204 (2015).



\bibitem{Arrington:2015yxa} 
  J.~Arrington,
  J.\ Phys.\ Chem.\ Ref.\ Data {\bf 44}, 031203 (2015).



\bibitem{Griffioen:2015hta} 
  K.~Griffioen, C.~Carlson and S.~Maddox,
  Phys.\ Rev.\ C {\bf 93}, no. 6, 065207 (2016).



\bibitem{Higinbotham:2015rja} 
  D.~W.~Higinbotham, A.~A.~Kabir, V.~Lin, D.~Meekins, B.~Norum and B.~Sawatzky,
  Phys.\ Rev.\ C {\bf 93}, no. 5, 055207 (2016).

\bibitem{Alarcon:2018zbz} 
  J.~M.~Alarcón, D.~W.~Higinbotham, C.~Weiss and Z.~Ye,
  Phys.\ Rev.\ C {\bf 99}, no. 4, 044303 (2019).

\bibitem{TuckerSmith:2010ra} 
  D.~Tucker-Smith and I.~Yavin,
  Phys.\ Rev.\ D {\bf 83}, 101702 (2011).



\bibitem{Barger:2010aj} 
  V.~Barger, C.~W.~Chiang, W.~Y.~Keung and D.~Marfatia,
  Phys.\ Rev.\ Lett.\  {\bf 106}, 153001 (2011).



\bibitem{Barger:2011mt} 
  V.~Barger, C.~W.~Chiang, W.~Y.~Keung and D.~Marfatia,
  Phys.\ Rev.\ Lett.\  {\bf 108}, 081802 (2012).



\bibitem{Batell:2011qq} 
  B.~Batell, D.~McKeen and M.~Pospelov,
  Phys.\ Rev.\ Lett.\  {\bf 107}, 011803 (2011).

\bibitem{Brax:2010gp} 
  P.~Brax and C.~Burrage,
  Phys.\ Rev.\ D {\bf 83}, 035020 (2011).



\bibitem{Jentschura:2010ha} 
  U.~D.~Jentschura,
  Annals Phys.\  {\bf 326}, 516 (2011).


\bibitem{Carlson:2012pc} 
  C.~E.~Carlson and B.~C.~Rislow,
  Phys.\ Rev.\ D {\bf 86}, 035013 (2012).



\bibitem{Wang:2013fma} 
  L.~B.~Wang and W.~T.~Ni,
  Mod.\ Phys.\ Lett.\ A {\bf 28}, 1350094 (2013).



\bibitem{Onofrio:2013fea} 
  R.~Onofrio,
  EPL {\bf 104}, no. 2, 20002 (2013).



\bibitem{Karshenboim:2014tka} 
  S.~G.~Karshenboim, D.~McKeen and M.~Pospelov,
  Phys.\ Rev.\ D {\bf 90}, no. 7, 073004 (2014)
  Addendum: [Phys.\ Rev.\ D {\bf 90}, no. 7, 079905 (2014)].

\bibitem{Carlson:2015jba} 
  C.~E.~Carlson,
  Prog.\ Part.\ Nucl.\ Phys.\  {\bf 82}, 59 (2015). 

\bibitem{Gilman:2013eiv} 
  R.~Gilman {\it et al.} [MUSE Collaboration],
  arXiv:1303.2160 [nucl-ex].



\bibitem{Gilman:2017hdr} 
  R.~Gilman {\it et al.} [MUSE Collaboration],
  arXiv:1709.09753 [physics.ins-det].

\bibitem{Denisov:2018unj} 
  B.~Adams {\it et al.},
  arXiv:1808.00848 [hep-ex].
  
  
\bibitem{Drell:1963ej} 
  S.~D.~Drell and J.~D.~Walecka,
  Annals Phys.\  {\bf 28}, 18 (1964).
  
\bibitem{Pauk:2015oaa} 
  V.~Pauk and M.~Vanderhaeghen,
  Phys.\ Rev.\ Lett.\  {\bf 115}, no. 22, 221804 (2015).


\bibitem{MAMI_photopr} 
   A2~Collaboration at MAMI, Letter of intent,
  \url{https://www.blogs.uni-mainz.de/fb08-mami-experiments/files/2016/07/A2-LOI2016-1.pdf}

\bibitem{Heller:2018ypa} 
  M.~Heller, O.~Tomalak and M.~Vanderhaeghen,
  Phys.\ Rev.\ D {\bf 97}, no. 7, 076012 (2018).


\bibitem{Vanderhaeghen:2000ws} 
  M.~Vanderhaeghen, J.~M.~Friedrich, D.~Lhuillier, D.~Marchand, L.~Van Hoorebeke and J.~Van de Wiele,
  Phys.\ Rev.\ C {\bf 62}, 025501 (2000).



\bibitem{tHooft:1978jhc} 
  G.~'t Hooft and M.~J.~G.~Veltman,
  Nucl.\ Phys.\ B {\bf 153}, 365 (1979).
  

\bibitem{Yennie:1961ad} 
  D.~R.~Yennie, S.~C.~Frautschi and H.~Suura,
  Annals Phys.\  {\bf 13}, 379 (1961).
  
\bibitem{Nogueira:1991ex} 
  P.~Nogueira,
  J.\ Comput.\ Phys.\  {\bf 105}, 279 (1993).
  
\bibitem{Kuipers:2012rf}
  J.~Kuipers, T.~Ueda, J.~A.~M.~Vermaseren and J.~Vollinga,
  Comput.\ Phys.\ Commun.\  {\bf 184} (2013) 1453
  doi:10.1016/j.cpc.2012.12.028
  [arXiv:1203.6543 [cs.SC]].
  
\bibitem{vonManteuffel:2012np} 
  A.~von Manteuffel and C.~Studerus,
  arXiv:1201.4330 [hep-ph].
  
\bibitem{Mertig:1990an} 
  R.~Mertig, M.~Bohm and A.~Denner,
  Comput.\ Phys.\ Commun.\  {\bf 64}, 345 (1991).


\bibitem{Hahn:2010zi} 
  T.~Hahn,
  PoS ACAT {\bf 2010}, 078 (2010).

  
\bibitem{Binosi:2003yf}
  D.~Binosi and L.~Theussl,
  Comput.\ Phys.\ Commun.\  {\bf 161} (2004) 76.

\bibitem{Vermaseren:1994je}
  J.~A.~M.~Vermaseren,
  Comput.\ Phys.\ Commun.\  {\bf 83} (1994) 45.

\bibitem{Bjorken:1958zz} 
  J.~D.~Bjorken, S.~D.~Drell and S.~C.~Frautschi,
  Phys.\ Rev.\  {\bf 112}, 1409 (1958).

\bibitem{Fomin:1959zz} 
  Ya.~Guzenko and P.~I.~Fomin,
  JETP\  {\bf 8}, 491 (1959).
  
\bibitem{Parsons:1966jw} 
  R.~G.~Parsons,
  Phys.\ Rev.\  {\bf 150}, 1165 (1966).
 
\bibitem{Huld:1968zz} 
  B.~Huld,
  Phys.\ Rev.\  {\bf 168}, 1782 (1968).

\bibitem{Arbuzov:2007ct} 
  A.~B.~Arbuzov,
  JHEP {\bf 0801}, 031 (2008).
  
\end{thebibliography}
\end{document}